                      \def\version{10 October, 2006}               %

\documentclass[reqno,twoside,11pt]{amsart}
\usepackage{amsmath}
\usepackage{amssymb}
 
\def\G{\Gamma} 
\def\d{\delta} 
 
\def\l{\lambda} 

\def\L{\Lambda} 

\newfam\Bbbfam 
\font\tenBbb=msbm10 
\font\sevenBbb=msbm7 
\font\fiveBbb=msbm5 
\textfont\Bbbfam=\tenBbb 
\scriptfont\Bbbfam=\sevenBbb 
\scriptscriptfont\Bbbfam=\fiveBbb

\newcommand{\R}     {\mathbb{R}} 
\newcommand{\Z}     {\mathbb{Z}} 
\newcommand{\N}     {\mathbb{N}} 
\renewcommand{\P}   {\mathbb{P}} 
 
\newcommand{\E}     {\mathbb{E}} 
 
 \newcommand{\floor}[1]{\left\lfloor #1 \right\rfloor}

\newcommand{\Pmf}{\mathfrak{P}}
\def\1{{\mathchoice {1\mskip-4mu\mathrm l}      
{1\mskip-4mu\mathrm l} 
{1\mskip-4.5mu\mathrm l} {1\mskip-5mu\mathrm l}}} 
\newcommand{\ssup}[1] {{\scriptscriptstyle{({#1}})}} 
\def\comment#1{} 
\newtheoremstyle{thm}{2ex}{2ex}{\itshape\rmfamily}{} 
{\bfseries\rmfamily}{}{1.7ex}{} 
 
\newtheoremstyle{rem}{1.3ex}{1.3ex}{\rmfamily}{} 
{\itshape\rmfamily}{}{1.5ex}{} 
 
\newenvironment{proofsect}[1] 
{\vskip0.1cm\noindent{\bf #1.}\hskip0.5cm}

\newtheorem{theorem}{Theorem}[section] 
\newtheorem{lemma}[theorem]{Lemma} 
\newtheorem{prop}[theorem] {Proposition} 
 
\newtheorem{remark}[theorem]  {Remark}

\theoremstyle{definition}
\newtheorem{example}[theorem] {Example}

 \newcommand{\s}{\sigma}

\renewcommand{\section}{\secdef\sct\sect} 
\newcommand{\sct}[2][default]{\refstepcounter{section} 
\vspace{0.8cm} 
\setcounter{equation}{0} 
\centerline{ 
\large\scshape \arabic{section}.\ #1} 
\vspace{0.2cm}} 
\newcommand{\sect}[1]{ 
\vspace{0.8cm} 
\centerline{\large\scshape #1} 
\vspace{0.2cm}} 
 
\renewcommand{\subsection}{\secdef \subsct\sbsect} 
\newcommand{\subsct}[2][default]{\refstepcounter{subsection} 
\nopagebreak 
\vspace{0.5\baselineskip} 
{\flushleft\bf \arabic{section}.\arabic{subsection}~\bf #1  } 
\nopagebreak} 
\newcommand{\sbsect}[1]{\vspace{0.1cm}\noindent 
{\bf #1}\vspace{0.1cm}} 
\newcommand{\heap}[2]{\genfrac{}{}{0pt}{}{#1}{#2}} 
 
\renewcommand{\subsubsection}{%
\secdef \subsubsect\sbsbsect} 
\newcommand{\subsubsect}[2][default]{%
\refstepcounter{subsubsection} 
\nopagebreak 
\vspace{0.1\baselineskip} 
\nopagebreak 
{\flushleft 
\sffamily\slshape 
\arabic{section}.\arabic{subsection}.\arabic{subsubsection} 
\ %
\sffamily #1\/.}\ } 
\newcommand{\sbsbsect}[1]{\vspace{0.1cm}\noindent 
{\bf #1}\ } 
 
\newcommand{\cou}{{\rm Cou}}
 
\renewcommand{\d}{{\rm d}}

\newcommand{\eps}{\varepsilon} 
\newcommand{\Sym}{\mathfrak{S}}

\newcommand{\supp}{{\operatorname {supp}}} 
\newcommand{\dist}{{\operatorname {dist}}}

\newcommand{\tr}{{\operatorname {Tr}\,}}
 
\newcommand{\Bcal}   {{\mathcal B }} 
\newcommand{\Ccal}   {{\mathcal C }}

\newcommand{\Hcal}   {{\mathcal H }}

\newcommand{\Mcal}   {{\mathcal M }} 
 
\newcommand{\Ocal}   {{\mathcal O }}

\newcommand{\Xcal}   {{\mathcal X }}

\begin{document}

\title[Symmetrisation for large systems of random walks]{\large Asymptotic Feynman-Kac formulae for large symmetrised systems of random walks}

\author[Stefan Adams and Tony Dorlas]{} 

\date{}

\maketitle

\thispagestyle{empty}
\vspace{0.2cm}
\centerline {\sc By Stefan Adams\footnote{Max-Planck Institute for Mathematics in the Sciences, Inselstra{\ss}e 22-26, D-04103 Leipzig, Germany, Dublin Institute for Advanced Studies, School of Theoretical Physics, 10, Burlington Road, Dublin 4, Ireland,{\tt adams@mis.mpg.de}} \/ and  Tony Dorlas\footnote{Dublin Institute for Advanced Studies, School of Theoretical Physics, 10, Burlington Road, Dublin 4, Ireland, {\tt dorlas@stp.dias.ie}}}
\vspace{0.4cm}
\renewcommand{\thefootnote}{}
 \footnote{
Partially supported by DFG grant AD 194/1-3\rq}

\vspace{0.4cm}

\vspace{0.2cm}

\centerline{\small(\version)}
\vspace{.3cm}
\begin{quote}
{\small {\bf Abstract:}} 
We study large deviations principles for $ N $ random processes on the lattice $ \Z^d $ with finite time horizon $ [0,\beta] $ under a symmetrised measure where all initial and terminal points are uniformly given by a random permutation. That is, given a permutation $ \sigma $ of $ N $ elements and a vector $ (x_1,\ldots,x_N) $ of $ N $ initial points we let the random processes terminate in the points $ (x_{\sigma(1)},\ldots,x_{\sigma(N)}) $ and then sum over all possible permutations and initial points, weighted with an initial distribution. There is a two-level random mechanism and we prove two-level large deviations principles for the mean of empirical path measures, for the mean of paths and for the mean of occupation local times under this symmetrised measure. The  symmetrised measure cannot be written as any product of single random process distributions. We show a couple of important applications of these results in quantum statistical mechanics using the Feynman-Kac formulae representing traces of certain trace class operators. In particular we prove a non-commutative Varadhan Lemma for quantum spin systems with Bose-Einstein statistics and mean field interactions. 

A special case of our large deviations principle for the mean of occupation local times of $ N $ simple random walks has the Donsker-Varadhan rate function as the rate function for the limit $ N\to\infty $ but for finite time $ \beta $. We give an interpretation in quantum statistical mechanics for this surprising result.

\end{quote}


\eject

\setcounter{section}{0}
\section{Introduction}

Let $ N $ random processes in continuous time on the lattice $ \Z^d $ with initial distribution $ m\in\Pmf(\Z^d) $, where $ \Pmf(\Z^d) $ is the set of probability measures on $ \Z^d $, be given. We fix the time horizon as $ [0,\beta] $.
In this paper we study large deviations for different functionals of the $ N $ random processes for large $ N $ under the symmetrised distribution
\begin{equation}\label{defsymP}
\P^{\ssup{\rm sym}}_{N,\beta}=\frac{1}{N!}\sum\limits_{\sigma\in\Sym_N}\sum\limits_{x_1\in\Z^d}\cdots\sum\limits_{x_N\in\Z^d}\bigotimes\limits_{i=1}^Nm(x_i)\P^{\beta}_{x_i,x_{\s(i)}}.
\end{equation}
Here $ \Sym_N $ is the set of all permutations of $ N $ elements  and the measure $ \P^{\beta}_{x_i,x_{\s(i)}} $ is defined for any $ \s\in\Sym_N $ and $ x_i\in\Z^d, 1\le i\le N, $ as
the conditional probability measure for the $i$-th random process starting at $ x_i $ with terminal location $ x_{\s(i)} $. 
That is, in \eqref{defsymP} we have two mechanisms. First we draw uniformly a permutation and after that we pick $ N$ initial points which are permuted according to the chosen permutation to obtain $ N $ terminal points. Then these $ N $ initial and terminal points determine the $ N $ random processes. Finally we average over all permutations and initial points which are weighted with the given initial distribution $ m $. Hence we prove two-level large deviations principles for the symmetrised distribution $ \P^{\ssup{\rm sym}}_{N,\beta} $. The symmetrised measure $ \P^{\ssup{\rm sym}}_{N,\beta} $ itself is of interest because of the following reasons.

The symmetrisation in \eqref{defsymP} is described by the set of $ N $ pairs $ (x_1,\ldots,x_N;x_{\sigma(1)},\ldots,x_{\sigma(N)})$ for any permutation $\sigma\in\Sym_N $ and any $ x\in\Z^d $. The mixing procedure for the second entry in these pairs has been studied both in \cite{DZ92} and \cite{T02}, which were motivated from asymptotic questions about exchangeable vectors of random variables.  \cite{DZ92} studies large deviations for the empirical measures $\frac 1N\sum_{i=1}^N\delta_{Y_i}$, where $Y_1,\dots,Y_N$ have distribution $\int_\Theta \mu(\d \theta)\, P_N^{\ssup \theta}$ for some distribution $\mu$ on some compact space $\Theta$, and the empirical measures are assumed to satisfy a large deviation principle under $P_N^{\ssup \theta}$ for each $\theta$. In \cite{T02}, a similar problem is studied: given a sequence of random vectors $(Y_1^{\ssup N},\dots,Y_N^{\ssup N})$ such that the empirical measures $\frac 1N\sum_{i=1}^N\delta_{Y_i^{\ssup N}}$ satisfy a large deviation principle, another principle is established for the process of empirical measures $\frac 1N\sum_{i=1}^{\lfloor tN\rfloor}\delta_{X_{i}^{\ssup N}}$, where 
$$
\big(X_1^{\ssup N},\dots,X_N^{\ssup N}\big)=\frac1{N!}\sum_{\s\in\Sym_N}\big(Y_{\s(1)}^{\ssup N},\dots,Y_{\s(N)}^{\ssup N}\big).
$$

Our second main motivation for studying the symmetrised distribution $ \P_{N,\beta}^{{\ssup{\rm sym}}} $ stems from the application of Feynman-Kac formulae to express thermodynamic functions in quantum statistical mechanics. These thermodynamic functions are given as traces over exponentials of the Hamilton operator describing the quantum system. There exist two kinds of elementary particles in nature, the {\it Fermions\/} and the {\it Bosons\/}. The state of a system of $ N $ {\it Bosons\/} is described by a symmetrisation procedure like in \eqref{defsymP}, whereas the state for {\it Fermions\/} is given with the corresponding anti-symmetrisation procedure. Thus one is lead to employ large deviation technique to study the large $ N$-limit for expectations with respect to the symmetrised distribution. We apply our main large deviation results in Section~\ref{ncV-sec} and Section~\ref{speccase-sec} to systems of {\it Bosons\/} in quantum statistical mechanics.

We derive large deviations principles under the symmetrised distribution $ \P_{N,\beta}^{\ssup{\rm sym}} $ for the {\it empirical path measure\/} $ L_N $, for the {\it mean path\/} $ Y_N $ and for the {\it mean of occupation measures\/} $ Z_N $, all defined as functions of the $ N $ random paths $ \xi^{\ssup{1}},\ldots,\xi^{\ssup{N}}\colon [0,\beta]\to\Z^d $, which are elements of the space $ D_{\beta}=D([0,\beta];\Z^d) $ of all functions $ \omega:[0,\beta]\to\Z^d $, which are right continuous with left limits. 
The {\it empirical path measures}
\begin{equation}
L_N=\frac{1}{N}\sum\limits_{i=1}^N\delta_{\xi^{\ssup{i}}}
\end{equation}
are random elements in the set $ \Pmf(D_\beta) $ of probability measures on $ D_\beta $, the {\it mean path}
\begin{equation}
Y_N=\frac{1}{N}\sum\limits_{i=1}^N\xi^{\ssup{i}}
\end{equation}
is a random element in $ D_\beta([0,\beta];\R^d) $ whereas the {\it mean of the normalised occupation local times}
\begin{equation}
Z_N=\frac{1}{N}\sum\limits_{i=1}^N l_{\beta}^{\ssup{i}},
\end{equation}
is a random probability measure on $ \Z^d $, where the normalised occupation local times are defined as
\begin{equation}
l^{\ssup{i}}_{\beta}(z)=\frac{1}{\beta}\int_0^{\beta} \1_{\{\xi^{\ssup{i}}_s=z\}} \d s, \qquad i=1,\ldots,N, z\in\Z^d,
\end{equation} which represent the relative time the $i$-th random process spends up to time $ \beta $ in the state $ z $.

Our large-deviation rate functions for the three principles are explicit in terms of variational problems involving an entropy term (describing the large deviations of the permutations) and a certain Legendre transform (describing the large deviations of $L_N, Y_N $ and $Z_N$, respectively, for a fixed permutation). These two parts in the variational formula for the rate function are due to the two-level large deviations, which has something in common with the multilevel large deviations studied in \cite{DG94}. We draw a number of conclusions about variants of the principles, laws of large numbers and asymptotic independence. Let us remark that all our large deviations results maybe obtained for random processes (Markovian or not) on any connected graph with finite or enumerable vertices.

A first application of our large deviation results is given in Subsection~\ref{ncV-sec}, where we use the Feynman-Kac formula to represent the trace of any trace class operator restricted to the symmetric subspace of the $ N$-th tensor product of $n\times n $ complex matrix algebras as an expectation with a measure for $ N$ Markov processes on the index set $ \{1,\ldots m\} $. The trace class operator here is given by the {\it Boltzmann factor\/} $ {\rm e}^{-\beta h} $ for any self-adjoint matrix $ h $ representing in quantum mechanics the Hamilton operator for  a system of $ N $ quantum spins (lattice systems) for the inverse temperature $ \beta $, i.e. the time horizon of our random processes is given by the inverse temperature. In particular we derive the thermodynamic limit of the free energy, which is the trace of the Boltzmann factor,  for a general class of mean-field interactions and thus we get a non-commutative version of Varadhan's Lemma with {\it Bose-Einstein statistics\/}, i.e. where the trace is restricted to the symmetric subspace. Here, an analysis of the variational formula for the rate function is achieved with $ L^2 $ techniques for the mean paths, which are embedding in the corresponding $ L^2 $ space. Non-commutative Varadhan's Lemmas have been studied in \cite{CLR88} and \cite{PRV89}. In \cite{Dor96} a non-commutative central limit theorem under Bose-Einstein statistics has been proved for the case $ n=2 $. Hence our results complement and extend these results.

If we consider $ N $ simple random walks on $ \Z^d $ conditioned to stay within a finite set $ \L $ and replace the initial probability distribution $ m $ by the counting measure in $ \L $ we are able to show that the corresponding rate function for the large deviations principle for the mean of occupation local times of the $ N $ random walks under the symmetrised measure $ \mu_{N,\beta}^{\ssup{\rm sym}} $ \eqref{symmeasure} is given by the well-known {\it Donsker-Varadhan\/} rate function. The latter governs the large deviations principle for the occupation local time of a single random walk but for the limit $ \beta\to\infty $. This remarkable result has an interpretation for the cycle-structure given by our symmetrisation procedure, i.e. the appearance of cycles whose lengths grow like some potential of $ N $. This long cycles are considered  to be an order parameter for the occurrence of the {\it Bose-Einstein condensation\/} (BEC), a quantum phase transition solely driven by the symmetrisation procedure (\cite{BCMP05},\cite{DMP05}). Details of this interpretation are given at the end of Subsection~\ref{speccase-sec}.

We consider this as a first step towards a rigorous understanding of large Boson systems at positive temperature $ \beta $, because the time horizon $ \beta $ represents the inverse temperature for the Feynman-Kac formulae. Future work will be devoted to the mutually interacting case. Interacting Brownian motions in trap potentials so far have been analysed without symmetrisation, in particular, systems for vanishing temperature in \cite{ABK04} and large systems of interacting motions for fixed positive temperature in \cite{ABK05}. In \cite{AK06} some results for Brownian motions under the symmetrised distribution are obtained, which are not so general and which cannot be applied to mean field models.

Let us make some remarks on related literature. We found a most interesting old work \cite{S31} by Schr\"odinger, which is related to the pair probability method we applied in our large deviations principle. In \cite{S31} Schr\"odinger raised the question of the most probable behaviour of a large system of diffusion particles in thermal equilibrium. F\"ollmer \cite{F88} gave a mathematical formulation of these ideas in terms of large deviations. He applied Sanov's theorem to obtain a large deviations principle for $ L_N $ when $ B^{\ssup{1}},B^{\ssup{2}},\ldots $ are i.i.d.~Brownian motions with initial distribution $ m$ and no condition at time $\beta$. The rate function is the relative entropy with respect to $ \int_{\R^d} m(\d x)\,\P_{x}\circ B^{-1} $, where the motions start in $x$ under $\P_x$. Then Schr\"odinger's question amounts to identifying the minimiser of that rate function under given fixed independent initial and terminal distributions. It turns out that the unique minimiser is of the form $\int_{\R^d}\int_{\R^d}\d x\d y\,f(x) g(y)\,\P_{x,y}^\beta\circ B^{-1}$, i.e. a Brownian bridge with independent initial and terminal distributions. The probability densities $f$ and $g$ are characterised by a pair of dual variational equations, originally appearing in \cite{S31} for the special case that both given initial and terminal measures are the Lebesgue measure. 


An important work combining combinatorics and large deviations for symmetrised measures is \cite{Toth90}. T\'oth \cite{Toth90} considers $ N $ continuous-time simple random walks on a complete graph with $ \rho N $ vertices, where $ \rho\in (0,1) $ is fixed. He looks at the symmetrised distribution as in \eqref{defsymP} and adds an exclusion constraint: there is no collision of any two particles during the time interval $ [0,\beta] $. The combinatorial structure of this model enabled him to express the free energy in terms of a cleverly chosen Markov process on $ \N_0 $. Using Freidlin-Wentzell theory, he derives an explicit formula for the large-$ N $ asymptotic of the free energy; in particular he obtains a phase-transition, called {\it Bose-Einstein-condensation\/}, for large $ \beta $ and sufficiently large $ \rho $.
T\'oth's work inspired partly our approach and future work will be contributed to question of application of our main results to this setting and the large deviations for appearance of long cycles. Large deviations for integer partitions and cycle structures, where the random walk bridges with different time horizon are weighted, are obtained in  \cite{A06}.

The structure of the paper is as follows. In Section~\ref{Intro} we present all our results. In Subsection~\ref{LDP-sec} we describe  our main large deviation results and a couple of conclusions. The application to general quantum spin models and to a non-commutative version of Varadhan's Lemma is given in Subsection~\ref{ncV-sec}. In Subsection~\ref{speccase-sec} we study the special case for simple random walks on a finite set and in Subsection~\ref{pre-sec} we provide some basic facts about the space $ D_\beta $ and large deviations theory. The Section~\ref{Proofsec} is devoted to the proofs of our main results. In the appendix in Section~\ref{appendix-sec} we prove a lemma on pair probability measures and an entropy estimation which we use in our proofs.

\section{The Results}\label{Intro}
In this section we are going to formulate our main results. In Subsection~\ref{LDP-sec} we present the large deviations results and important conclusions. Following  in Subsection~\ref{ncV-sec} we apply this to a non-commutative version of Varadhan's Lemma and give an application of our large deviations result for quantum spin models and in Subsection~\ref{speccase-sec}
we study the very important case when for finite time the large time rate function, the {\it Donsker-Varadhan\/} rate function, is the rate function for the large $ N $-limit under the symmetrised distribution. At the end in Subsection~\ref{pre-sec} we provide some preliminaries about the topology in $ D_\beta $ and some notion on large deviations theory.

\subsection{Large deviations for symmetrised distributions}\label{LDP-sec}
We fix throughout the paper $ \beta>0 $. We equip the space $ D_\beta $ with the Skorokhod metric. 
Our main aim is to encode the combinatorics for the sum over permutations for the symmetrised measure \eqref{defsymP} with a sum over pair probability measures with equal marginals. In order to formulate the large deviations results we introduce the following notations. By $ \Pmf(\Z^d\times\Z^d) $ be denote the set of pair probability measures on $ \Z^d\times\Z^d $ and we let 
$$ \widetilde{\Pmf}(\Z^d\times\Z^d)=\{Q\in\Pmf(\Z^d\times\Z^d)\colon Q^{\ssup{1}}=Q^{\ssup{2}}\} $$ 
be the set of pair probability measures on $ \Z^d\times\Z^d $ with equal first and second marginal, respectively $ Q^{\ssup{1}}(x)=\sum_{y\in\Z^d}Q(x,y) , x\in\Z^d,$ and $ Q^{\ssup{2}}(y)=\sum_{x\in\Z^d}Q(x,y) , y\in\Z^d$. 
The relative entropy of the pair probability measure $ Q \in\Pmf(\Z^d\times\Z^d) $ with respect to the product $ Q^{\ssup{1}}\otimes m $ is given by 
\begin{equation}
H(Q|Q^{\ssup{1}}\otimes m)=\sum\limits_{x,y\in\Z^d}Q(x,y)\log\frac{Q(x,y)}{Q^{\ssup{1}}(x)m(y)}.
\end{equation}
Note that $ Q\mapsto H(Q|Q^{\ssup{1}}\otimes m) $ is strictly convex. 
All the rate functions of our large deviations principles include this relative entropy of pair probability measures as the part coming from the combinatorics of the symmetrised measure $ \P_{N,\beta}^{\ssup{\rm sym}} $. The other part of the rate functions comes into play in each of the principles from large deviations principles for a product of not necessarily identical distributed objects. Thus the encoding of the sum over permutations with a sum over pair probability measures represents a certain two-level large deviations principle. This is seen in the definition \eqref{defsymP} of the symmetrised measure $ \P_{N,\beta}^{\ssup{\rm sym}}$, where permutations are sampled uniformly and for each permutation there is a product of not necessarily identical distributions of single random walks with initial and terminal condition. 

On the level of path measures we define the following functional on the space of probability measures on the set $ D_{\beta} $ of path as
\begin{equation}\label{ratefunctionsym1}
I^{\ssup{\rm sym}}_{\beta}(\mu)=\inf\limits_{Q\in\widetilde{\Pmf}(\Z^d\times\Z^d)}\Bigl\{H(Q|Q^{\ssup{1}}\otimes m)+ I^{\ssup{Q}}_{\beta}(\mu)\Bigr\}\quad\mbox{ for }\mu\in\Pmf(D_{\beta}),
\end{equation} where the functional $ I^{\ssup{Q}}_{\beta} $ is given by
\begin{equation}\label{variationallevel2}
I^{\ssup{Q}}_{\beta}(\mu)=\sup\limits_{F\in\Ccal_{\rm b}(D_{\beta})}\Bigl\{\langle F,\mu\rangle-\sum\limits_{x,y\in\Z^d}Q(x,y)\log\E^{\beta}_{x,y}\Bigl({\rm e}^{\langle F,\delta_{\xi}\rangle}\Bigr)\Bigr\}\quad \mbox{ for }\mu\in\Pmf(D_\beta),
\end{equation} where we write $ \xi $ for $ \xi^{\ssup{1}} $ and $ \langle F,\delta_{\xi}\rangle=\int_{D_{\beta}}F(\omega)\delta_{\xi}(\d \omega)=F(\xi). $
Clearly, $ I^{\ssup{Q}}_{\beta} $ is a Legendre-Fenchel transform, but {\it not} one of a logarithmic moment generating function of a random variable, hence there seems to be no way to represent this functional as the relative entropy of $ \mu $ with respect to any measure. $ I^{\ssup{Q}}_{\beta} $ and $ I^{\ssup{\rm sym}}_{\beta} $ are nonnegative, and $ I^{\ssup{Q}}_{\beta} $ is convex as a supremum of linear functions. 

Let $ \pi_s\colon D_\beta\to\R^d $ be the projection $ \pi_s(\omega)=\omega_s $ for any $ s\in[0,\beta] $ and $ \omega\in D_\beta $. We denote the marginal measure of $ \mu\in\Pmf(D_\beta) $ on $ \Z^d $ by $ \mu_s=\mu\circ\pi_s^{-1} \in\Pmf(\Z^d)$, and analogously we write $ \mu_{0,\beta}=\mu\circ(\pi_0,\pi_\beta)^{-1}\in\Pmf(\Z^{d}\times\Z^{d}) $ for the joint distribution of the initial and the terminal point of a random process with distribution $ \mu $. If we restrict the supremum in \eqref{variationallevel2} over all $ F\in\Ccal_{\rm b}(D_\beta) $ to all functions of the form  $ \omega\mapsto g(\omega_0,\omega_\beta) $ with $ g\in\Ccal_{\rm b}(\R^d\times\R^d) $, we see that $ Q=\mu_{0,\beta} $ if $ I^{\ssup{Q}}_{\beta}(\mu)<\infty $. Indeed,
$$
\begin{aligned}
\sup_{g\in \Ccal_{\rm b}(\R^d\times\R^d)}&\Big\{\sum_{x,y\in\Z^d}g(x,y)\big(\mu_{0,\beta}(x,y)-Q(x,y)\big)\Big\}\\
&=\sup_{g\in \Ccal_{\rm b}(\R^d\times\R^d)}\Big\{\sum_{x,y\in\Z^d}g(x,y)\mu_{0,\beta}(x,y)-\sum_{x,y\in\Z^d}Q(x,y)\log\E_{x,y}\Big({\rm e}^{g(\xi_0,\xi_\beta)}\Big)\Big\}\\
& \le I^{\ssup{Q}}_{\beta}(\mu) <\infty,
\end{aligned}
$$
which implies that $ \mu_{0,\beta}=Q $. Therefore the infimum in \eqref{ratefunctionsym1} is uniquely attained at this pair probability measure $ Q $, i.e.
\begin{equation}\label{ratefunctioninf}
I^{\ssup{\rm sym}}_{\beta}(\mu)=\left\{\begin{array}{r@{\;\;}l}
H(\mu_{0,\beta}|\mu_0\otimes m)+\sup\limits_{F\in\Ccal_{\rm b}(D_{\beta})}\big\langle \mu,F-\log\E^\beta_{\pi_0,\pi_\beta}\big({\rm e}^{F(\xi)}\big)\big\rangle & \mbox{ if } \mu_0=\mu_\beta,\\
+\infty & \;\mbox{otherwise}.
\end{array}\right.
\end{equation}
In particular, $ I^{\ssup{\rm sym}}_{\beta} $ is convex.

\begin{theorem}[{\bf LDP for the mean of path measures $ L_N $}]\label{mainthm1}
Under the symmetrised measure $ \P^{\ssup{\rm sym}}_{N,\beta} $ the empirical path measures $ (L_N)_{N\ge 1} $ satisfy a large deviations principle on $ \Pmf(D_{\beta}) $ with speed $ N $ and rate function $ I^{\ssup{\rm sym}}_{\beta} $.
\end{theorem}

The proof of Theorem~\ref{mainthm1} is in Section~\ref{ProofTH1}. The proof does not rely on any Markov property of the $ N $ random processes, hence this assumption can be dropped.

We also have a large deviations principle for the mean path level. Note that any mean path is an element in the space $ D_\beta([0,\beta];\R^d) $ due to the averaging of paths with values in the lattice $ \Z^d $. For the path level we consider the continuous embedding of the space $ D([0,\beta];\R^d) $ into $ L^2([0,\beta];\R^d) $ (see \cite[Lemma~2.3]{Dor96} for details).  The scalar product for the latter space is defined by
\begin{equation}\label{scalarproduct}
\langle \xi,\omega\rangle=\int_0^\beta\d s\langle\xi(s),\omega(s)\rangle_{\R^d}
\end{equation}
for $ \xi,\omega\colon[0,\beta]\to\R^d $, where $ \langle\cdot,\cdot\rangle_{\R^d} $ is the scalar product on $ \R^d $. In the following we also write $ \xi_s $ for $ \xi(s) $.
We define the following functional on $ L^2([0,\beta];\R^d) $ as 
\begin{equation}
\widetilde I^{\ssup{\rm sym}}_\beta(\omega)=\inf_{Q\in\widetilde{\Pmf}(\Z^d\times\Z^d)}\Bigl\{H(Q|Q^{\ssup{1}}\otimes m)+\widetilde I^{\ssup{Q}}_\beta(\omega)\Big\}\quad\mbox{ for }\omega\in  L^2([0,\beta];\R^d),
\end{equation}
where
\begin{equation}
\widetilde I^{\ssup{Q}}_\beta(\omega)=\sup_{f\in L^2([0,\beta];\R^d)}\Big\{\langle f,\omega\rangle-\sum\limits_{x,y\in\Z^d}Q(x,y)\log\E_{x,y}^{\beta}\Big({\rm e}^{\langle \omega,\xi\rangle}\Big)\Big\}\quad\mbox{ for } \omega\in  L^2([0,\beta];\R^d).
\end{equation}

Then the large deviations principle for the mean path reads as 
\begin{theorem}[{\bf LDP for the mean of paths $ Y_N $}]\label{mainthm3}
Under the symmetrised measure $ \P^{\ssup{\rm sym}}_{N,\beta} $ the mean  $ (Y_N)_{N\ge 1} $ of the paths satisfies a large deviations principle on  $ L^2([0,\beta];\R^d) $ with speed $ N $ and rate function $ \widetilde I^{\ssup{\rm sym}}_{\beta} $.
\end{theorem}

The proof of Theorem~\ref{mainthm3} is in Subsection~\ref{main3}. The contraction principle \cite[Th.~4.2.1]{DZ98} yields a large deviations principle for the mean path from the one for the mean of path measures. However, the identification of that rate function from the contraction principle with the one in Theorem~\ref{mainthm3} seems to be rather difficult task from a technical point of view. Luckily, our proof of Theorem~\ref{mainthm1} is so general that it can be slightly modified to give  the proof for the large deviations principle for the mean path. For details see Subsection~\ref{main3}.

Denote by $ \Bcal(\Z^d) $ all bounded function $ f\colon\Z^d\to\R $.
On the level of probability measures on $ \Z^d $ we define the functional $ J^{\ssup{\rm sym}}_{\beta} $  on the set $ \Pmf(\Z^d) $ of probability measures on $ \Z^d $ as
\begin{equation}
J^{\ssup{\rm sym}}_{\beta}(p)=\inf\limits_{Q\in\widetilde{\Pmf}(\Z^d\times\Z^d)}\Bigl\{H(Q|Q^{\ssup{1}}\otimes m)+ J_{\beta}^{\ssup{Q}}(p)\Bigr\}\quad\mbox{ for } p\in\Pmf(\Z^d),
\end{equation}
where
\begin{equation}
J_{\beta}^{\ssup{Q}}(p)=\sup\limits_{f\in \Bcal(\Z^d)}\Bigl\{\beta\sum\limits_{x\in\Z^d} f(x)p(x) -\sum\limits_{x,y\in\Z^d}Q(x,y)\log\E_{x,y}^{\beta}\Bigl({\rm e}^{\beta\langle f,l_{\beta}\rangle}\Bigr)\Bigr\}.
\end{equation}

\begin{theorem}[{\bf LDP for the mean of normalised occupation local times $ Z_N $}]\label{mainthm2}
Under the symmetrised measure $ \P^{\ssup{\rm sym}}_{N,\beta} $ the mean $ (Z_N)_{N\ge 1} $ of the normalised occupation measures satisfy a large deviations principle on $ \Pmf(\Z^d) $ with speed $ N $ and rate function $  J^{\ssup{\rm sym}}_{\beta} $.
\end{theorem}

The proof of Theorem~\ref{mainthm2} is in Subsection~\ref{main2}. Here, the same remarks as for the proof of Theorem~\ref{mainthm3} concerning the contraction principle apply. For details see Subsection~\ref{main2}.

In the following remark we compare the symmetrised distribution with the i.i.d. case.
\begin{remark}
For i.i.d. random walks with initial distribution $ m $, the empirical path measure $ (L_N)_{N\ge 1} $ satisfies a large deviations principle with speed $ N $ and rate function
$$
I_{\beta,m}(\mu)=\sup\limits_{F\in\Ccal_{\rm b}(D_{\beta})}\Bigl\{\langle F,\mu\rangle -\log\sum_{x,y\in\Z^d}m(x)\E_{x,x}^\beta\Bigl({\rm e}^{F(\xi)}\Bigr)\Bigr\}.
$$
This is an application of Cramer's theorem \cite[Th.~6.1.3]{DZ98} for the mean of the independent identically distributed random walks with initial distribution $ m $.
Note that $ I_{\beta}^{\ssup{Q}}\ge I_{\beta,m} $ for the pair measure $ Q $ defined as $ Q(x,y)=m(x)\delta_x(y)$ for $ x,y\in\Z^d$, since
\begin{eqnarray}
\begin{aligned}
-\sum\limits_{x,y\in\Z^d}Q(x,y)\log\E^{\beta}_{x,y}({\rm e}^{F(\xi)})&\ge -\log \sum_{x,y\in\Z^d}m(x)\E_{x,x}^\beta(\E_{\xi_0,\xi_{\beta}}^{\beta}({\rm e}^{F(\xi)}))\\
& =-\log\sum_{x,y\in\Z^d}m(x)\E_{x,x}^\beta({\rm e}^{F(\xi)}). 
\end{aligned}
\end{eqnarray} In particular $ I^{\ssup{\rm sym}}_{\beta}\ge I_{\beta} $.
\hfill $ \Diamond $
\end{remark}

Our large deviations Theorems~\ref{mainthm1}-\ref{mainthm2} may be extended by considering a finite and positive measure $ m $ not necessarily normalised to one. However, more interesting is the question if we replace the conditional probability measure $ \P_{x,y}^\beta $ by the measure $ \mu_{x,y}^\beta(\cdot)=\P_x^\beta(\cdot\1\{\xi_{\beta}=y\}) $ for any $ x,y\in\Z^d $. This is included in the following proposition.

\begin{prop}\label{prop1}
Let $ m $ be a positive finite measure on $ \Z^d $ and let $ g\colon\R^d\times\R^d\to\R_+ $ be a bounded and continuous strictly positive function. Replace $ \P_{x,y}^\beta $ by $g(x,y)\P_{x,y}^\beta $ in the definition of the symmetrised measure $ \P^{\ssup{\rm sym}}_{N,\beta}$ in \eqref{defsymP}. Then
\begin{enumerate}
\item[(i)] Theorem~\ref{mainthm1} remains true with the rate function replaced by 
$$ 
\mu\mapsto I^{\ssup{\rm sym}}_{\beta}(\mu)-\sum_{x,y\in\Z^d}\mu_{0,\beta}(x,y)\log g(x,y).
$$
\item[(ii)] Theorem~\ref{mainthm3} remains true with the rate function replaced by
$$
\omega\mapsto \inf_{Q\in\widetilde{\Pmf}(\Z^d\times\Z^d)}\Bigl\{H(Q|Q^{\ssup{1}}\otimes m)+\widetilde I^{\ssup{Q}}_\beta(\omega)-\sum_{x,y\in\Z^d}Q(x,y)\log g(x,y)\Big\}.
$$
\item[(iii)] Theorem~\ref{mainthm2} remains true with the rate function replaced by
$$
p\mapsto \inf_{Q\in\widetilde{\Pmf}(\Z^d\times\Z^d)}\Big\{H(Q|Q^{\ssup{1}}\otimes m)+J_{\beta}^{\ssup{Q}}(p) -\sum_{x,y\in\Z^d}Q(x,y)\log g(x,y)\Big\}.
$$
\end{enumerate}
\end{prop}

\begin{proofsect}{Proof}
We will prove (i). Define the function $ F_g(\omega)=\log g(\omega_0,\omega_\beta) $ for any path $ \omega\in D_\beta([0,\beta];\Z^d) $. Then with probability one with respect to $ \bigotimes_{i=1}^N\P_{x_i,x_{\sigma(i)}}^\beta $,
$$
\prod_{i=1}^Ng(x_i,x_{\sigma(i)})=\exp\Big(\sum_{i=1}^N\log g(\xi_0^{\ssup{i}},\xi_\beta^{\ssup{i}})\Big)=\exp\Big(N\langle L_N,F_g\rangle\Big).
$$
Clearly, $ \omega\mapsto F_g(\omega) $ is continuous and bounded. Hence the large deviations principle follows from \cite[Th.~III.17]{dH00} for any $ Q\in \widetilde{\Pmf}(\Z^d\times\Z^d) $. The rate function follows as $ \mu\mapsto I^{\ssup{\rm sym}}_{\beta}(\mu)-\langle\mu,F_g\rangle $ which together with \eqref{ratefunctioninf} gives the proof.
The proof of (ii) and (iii) follows analogously with (i) and the proofs of Theorem~\ref{mainthm3} and Theorem~\ref{mainthm2} respectively. Note for (ii) that
$$
\begin{aligned}
\omega\mapsto &\inf_{\mu\in\Pmf(D_\beta)\colon \Psi(\mu)=\omega}\Big\{I_\beta^{\ssup{\rm sym}}(\mu)-\langle\mu_{0,\beta},\log g\rangle\Big\}\\
& =\inf_{Q\in \widetilde{\Pmf}(\Z^d\times\Z^d)}\Big\{\widetilde I_\beta^{\ssup{\rm sym}}(\omega)-\sum_{x,y\in\Z^d}Q(x,y)\log g(x,y)\Big\}\quad\mbox{ for }\omega\in L^2([0,\beta];\Z^d).
\end{aligned}
$$ where $ \Psi\colon\Pmf(D_\beta)\to D([0,\beta];\R^d) $ is the continuous mapping for the contraction principle, compare the proof of Theorem~\ref{mainthm3} in Section~\ref{main3}. Note that we used the fact that $ Q=\mu_{0,\beta} $ if $ I_\beta^{\ssup{Q}}(\mu)<\infty $.
\qed
\end{proofsect}

From the previous proposition we get.

\begin{prop}
Let $ m $ be a positive finite measure on $ \Z^d $ and let $ g\colon\R^d\times\R^d\to\R_+ $ be a bounded and continuous strictly positive function. 
\begin{enumerate}
\item[(i)] 
\begin{equation}\label{comblemma}
\begin{aligned}
\lim_{N\to\infty}\frac{1}{N}&\log\Big(\frac{1}{N!}\sum_{\sigma\in\Sym_N}\sum_{x_i\in\Z^d,1\le i\le N}\prod_{i=1}^Nm(x_i)\prod_{i=1}^Ng(x_i,x_{\sigma(i)})\Big)\\
&=-\inf_{Q\in\widetilde{\Pmf}(\Z^d\times\Z^d)}\Big\{H(Q|Q^{\ssup{1}}\otimes m)-\sum_{x,y\in\Z^d}Q(x,y)\log g(x,y)\Big\}.
\end{aligned}
\end{equation}
\item[(ii)]
The unique minimiser of the rate function $ \mu\mapsto I^{\ssup{\rm sym}}_{\beta}(\mu)-\sum_{x,y\in\Z^d}\mu_{0,\beta}(x,y)\log g(x,y) 
$ is given by
\begin{equation}
\mu^0=\sum_{x,y\in\Z^d}Q^0(x,y)\P_{x,y}^\beta\circ \xi^{-1},
\end{equation}
where $ Q^0\in\widetilde{\Pmf}(\Z^d\times\Z^d) $ is the unique minimiser on the right hand side of \eqref{comblemma}. Under the symmetrised measure $ \P_{N,\beta}^{\ssup{\rm sym}} $ the sequence $ (L_N)_{N\in\N} $ converges in distribution to the measure $ \mu^0 $ as $ N\to\infty $.
\end{enumerate}
\end{prop}

\begin{proofsect}{Proof}
(i) Proposition~\ref{prop1} gives that the left hand side of \eqref{comblemma} equals $ -\inf_{\mu\in\Pmf(D_\beta)}I_\beta^{\ssup{\rm ysm}}(\mu)-\langle \mu_{0,\beta},\log g\rangle $. If we use \eqref{ratefunctioninf} and substitute $ Q=\mu_{0,\beta} $ we get
$$
\begin{aligned}
-\inf_{Q\in\widetilde{\Pmf}(\Z^d\times\Z^d)}&\Big\{H(Q|Q^{\ssup{1}}\otimes m)-\sum_{x,y\in\Z^d}Q(x,y)\log g(x,y)\\ &\qquad +\inf_{\mu\in\Pmf(D_\beta)\colon Q=\mu_{0,\beta}}\sup_{F\in\Ccal_{\rm b}(D_\beta)}\Big\{\langle\mu,F-\log\E_{\pi_0,\pi_\beta}^\beta\big({\rm e}^{F(\xi)}\big)\Big\}\Big\}.
\end{aligned}
$$
The latter infimum over $ \mu $ is equal to zero. To see this, pick $ F=0 $ to get the lower bound. To get the corresponding upper bound take $ \mu=\sum_{x,y\in\Z^d}Q(x,y)\P_{x,y}^\beta\circ\xi^{-1} $ and use Jensen's  inequality to get
$$
-\langle\mu,\log\E_{\pi_0,\pi_\beta}^\beta\big({\rm e}^{F(\xi)}\big)\rangle\le -\sum_{x,y\in\Z^d}Q(x,y)\E_{x,y}^\beta(F).
$$

\noindent We are going to prove that $ \mu^0 $ is the unique minimiser of the rate function $ \mu\mapsto I^{\ssup{\rm sym}}_{\beta}(\mu)-\sum_{x,y\in\Z^d}\mu_{0,\beta}(x,y)\log g(x,y) $. This proves then both (ii) and (iii). For that, let $ \mu\in\Pmf(D_\beta) $ be a zero of $ I^{\ssup{\rm sym}}_{\beta} $. As the relative entropy has compact level sets, there is a $ Q^0 \in\widetilde{\Pmf}(\Z^d\times\Z^d) $ that minimises the formula on the right hand side of \eqref{comblemma}. As $ I^{\ssup{Q^0}}_\beta(\mu)<\infty$, we have $ \mu_{0,\beta}=Q^0 $ and hence
$$
0=I^{\ssup{Q^0}}_\beta(\mu)=\sup_{F\in\Ccal_{\rm b}(D_\beta)}\Big\{\langle \mu,F-\log\E_{\pi_0,\pi_\beta}^\beta\big({\rm e}^{F(\xi)}\big)\rangle\Big\}.
$$
Clearly, $ F=0 $ is optimal and the Euler-Lagrange equations yield, for any $ g\in\Ccal_{\rm b}(D_\beta) $,
$$
\langle \mu,g\rangle=\langle \mu,\E_{\pi_0,\pi_\beta}^\beta\big(h(\xi)\big)\rangle,
$$
which identifies $ \mu $ as $ \mu^0 $.
\qed
\end{proofsect}

\subsection{Non-commutative Varadhan's Lemma with Bose-Einstein Statistics}\label{ncV-sec}
In this section we use our large deviations principle for the mean paths under the symmetrised distribution $ \P_{N,\beta}^{\ssup{\rm sym}} $ to derive a non-commutative version of Varadhan's Lemma with Bose-Einstein statistics. Let $ h $ be a self-adjoint $ n\times n $ matrix with $ h_{x,y}\le 0 $ for all $ x\not= y, x,y\in G=\{1,\ldots,n\} $.
We define a Markov process on the finite index set $ G $ with transition probabilities
\begin{equation}\label{transitionprobab}
\P(\xi(t+\delta t)=y|\xi(t)=x)=\left\{\begin{array}{r@{\;,\;}l} -h_{y,x}\delta t & \mbox{ if } \, y\not=x\\ 1+\sum_{z\not= x}h_{z,x}\delta t & \mbox{ if }\, y=x\end{array} ,\right.
\end{equation}
analogously
$$
p_{t-s}(x,y)=\P(\xi(t^{\prime})=y|\xi(s)=x)=\Big({\rm e}^{-(t-s)\widetilde h}\Big)_{y,x}\quad \mbox{ for } x,y\in G,
$$
where the matrix $ \widetilde h $ is defined by $ \widetilde h_{y,x}=h_{y,x} $ for $ y\not=x $ and $ \widetilde h_{x,x}=-\sum_{z\not= x}h_{z,x} $. 
For later convenience we let $ \lambda\colon[-n,n]\to \R $ a continuous function such that $ \lambda(x)=\lambda_x $ for $ x\in G $, and we denote by $ h_D $ a continuous function $ h_D(x)=\widetilde h_{x,x}-h_{x,x} $ for each $ x\in G $.

We let $N$ Markov processes $ \xi^{\ssup{1}},\ldots,\xi^{\ssup{N}} $ with transition probabilities \eqref{transitionprobab} and time horizon $ [0,\beta] $ be given. 
Let $ \P_{x,y}^{\beta,h} $ denote the conditional probability measure with density $ {\rm e}^{\int_0^\beta h_D(\omega(s))\d s} $ starting in $ x\in G $ conditioned to terminate in $ y\in G $. We write $ \E_{x,y}^{\beta,h} $ for the expectation with respect to the bridge probability measure $ \P_{x,y}^{\beta,h} $.

We derive a large deviations principle for the mean path $ Y_N $ under the symmetrised distribution (compare \eqref{defsymP})
$$
\P^{\ssup{{\rm sym},h}}_{N,\beta}=\frac{1}{N!}\sum_{\sigma\in\Sym_N}\sum_{x_1\in G}\cdots\sum_{x_N\in G}\bigotimes_{i=1}^Nm(x_i)\P_{x_i,x_{\sigma(i)}}^{\beta,h},
$$
where the initial distribution $ m $ is defined  by $ m(x)=\frac{1}{m} $ for $ x\in G $.

We consider 
mean-field type interactions for the $ N $ Markov processes $ \xi^{\ssup{1}},\ldots,\xi^{\ssup{N}} $ of the form $ N\int_0^\beta f(\frac{1}{N}\sum_{i=1}^N \xi^{\ssup{i}}_s)\d s $ for some bounded continuous function $ f\colon \R\to\R $. In the following we write $ \E^{\ssup{\rm sym,h}}_N $ for the expectation with respect to the symmetrised distribution.

\begin{theorem}\label{THMmain4}
Fix $ \beta >0 $ and $ n\in\N $. Let $ h $ be a selfadjoint $n\times n $-matrix with $ h_x,y\le 0 $ for $ x\not=y $ for all $ x,g\in G $.
\begin{enumerate}
\item[(a)]  The mean paths, $ Y_N $, under the symmetrised distribution $ \P^{\ssup{{\rm sym},h}}_{N,\beta} $ satisfy, as $ N\to \infty $, a large deviations principle on $ L^2([0,\beta];\R) $ with rate function $ I^{\ssup{{\rm sym},h}}_\beta $ defined by
\begin{equation}
I^{\ssup{{\rm sym},h}}_\beta(\omega)=\inf_{Q\in \widetilde \Pmf(G\times G)}\Big\{H(Q|Q^{\ssup{1}}\otimes m)+I_\beta^{\ssup{Q,h}}(\omega) \Big\},
\end{equation} for $ \omega\in L^2([0,\beta];\R) $,
where
$$
I_\beta^{\ssup{Q,h}}(\omega)=\sup_{g\in L^2([0,\beta];\R)}\Big\{\langle g,\omega\rangle -\sum_{x,y\in G}Q(x,y)\log\E_{x,y}^{\beta,h}\Big({\rm e}^{\int_0^\beta g(s)\xi(s) \d s}\Big)\Big\}.
$$

\item[(b)] 
\begin{equation}\label{varformulava}
\lim_{N\to\infty}\frac{1}{N}\log \E_N^{\ssup{{\rm sym},h}}\Big({\rm e}^{N\int_0^\beta f(\frac{1}{N}\sum_{i=1}^N \xi^{\ssup{i}}_s)\d s}\Big)=\sup_{\omega\in  L^2([0,\beta];\R)}\Big\{\int_0^\beta f(\omega(s))\d s- I^{\ssup{{\rm sym},h}}_\beta(\omega)\Big\}.
\end{equation}
\end{enumerate}

\end{theorem}

\begin{proofsect}{Proof}
(a) is a direct application of our main Theorem~\ref{mainthm3}, and (b) is an application of Varadhan's Lemma \cite[Th.~4.3.1]{DZ98}.
\qed
\end{proofsect}

We outline how this large deviations principle gives a non-commutative version of Varadhan's Lemma under Bose-Einstein statistics. Let $ \rho $ be the state on the algebra $ \Mcal $ of all complex $ n\times n $ matrices given by $ \rho(A)=\tr({\rm e}^{-\beta h}A) $ for $ A\in\Mcal $ for the given matrix $ h $, and let $ \tr {\rm e}^{-\beta h}=1 $.  We fix a self-adjoint element $ x\in\Mcal $ and some continuous bounded function $ f\colon \R\to\R $. This self-adjoint element $ x $ describes a mean-field interaction expressed through the mean matrix
$$
x^{\ssup{N}}=\frac{1}{N}\sum_{i=1}^Nx_i, 
$$ where $ x_i , i=1,\ldots,N $, is a copy of the matrix $ x $. Further, let $ h^{\ssup{N}}=\frac{1}{N}\sum_{i=1}^N h_i$, where $ h_i, i=1,\ldots,N $, is a copy of the matrix $ h $. Hence $ h^{\ssup{N}} $ and $ x^{\ssup{N}} $ both act on the $N$-th tensor product of the $n$-dimensional single variable space. The symbol $\tr_+ $ denotes the trace restricted to the subspace of all symmetric $N$-variables with respect to any permutation of their single indices. The restriction to these symmetric variables is called {\it Bose-Einstein statistics}.

We shall calculate the trace $ $ via the Feynman-Kac formula and our previous results in Theorem~\ref{THMmain4}. For this we need the following path measure 
\begin{equation}
\mu_{x,y}^{\beta,h}(\cdot) =\P_x^{\beta,h}(\cdot{\rm e}^{\int_0^\beta h_D(\omega(s))\d s}\1\{\xi_\beta=y\}),
\end{equation}
which is the probability for the Markov process to start in $ x\in G $ and to terminate in $ y\in G $. Note that this measure can be normalised with the function $ g_\beta(x,y)=\P_x^{\beta,h}({\rm e}^{\int_0^\beta h_D(\omega(s))\d s}\1\{\xi_\beta=y\}) $ to obtain the conditional probability measure $ \P_{x,y}^{\beta,h} $. Here we apply our Proposition~\ref{prop1} in combination with Theorem~\ref{THMmain4}. Note that this results in substituting the symmetrised distribution $ \P_{N,\beta}^{\ssup{{\rm sym},h}} $ with the symmetrised measure
\begin{equation}\label{symmeasure0}
\mu_{N}^{\ssup{{\rm sym},h}}=\frac{1}{N!}\sum_{\sigma\in\Sym_N}\sum_{x_1\in\L}\cdots\sum_{x_N\in\L}\bigotimes_{i=1}^N\mu_{x_i,x_{\sigma(i)}}^{\beta,h}.
\end{equation}

The Feynman-Kac formula gives 
\begin{equation}\label{PRVthm}
\begin{aligned}
\lim_{N\to\infty}\frac{1}{N}\log \tr_+ \big({\rm e}^{-N(h^{\ssup{N}}-f(x^{\ssup{N}}))}\big)&=\lim_{N\to\infty}\frac{1}{N}\log \mu^{\ssup{{\rm sym},h}}_{N,\beta}\big({\rm e}^{N\int_0^\beta f(\frac{1}{N}\sum_{i=1}^N \xi^{\ssup{i}}_s)\d s}\big)\\
&=\sup_{\omega\in  L^2([0,\beta];\R)}\Big\{\int_0^\beta f(\omega(s))\d s- I^{\ssup{{\rm sym},h}}_\beta(\omega)+\langle Q,g_\beta\rangle\Big\},
\end{aligned}
\end{equation}
where we replaced $ m $ in Theorem~\ref{THMmain4} by the counting measure $ \cou $, and where $ \langle Q,g_\beta\rangle $ denotes the expectation of $ g_\beta $ with respect to the pair probability measure $ Q $.

The analysis of the variational formula of the right hand side of \eqref{PRVthm} gives the following theorem.

\begin{theorem}[{\bf Mean-field interaction with Bose-Einstein statistics}]\label{NoncommutativeTHM}
\begin{equation}\label{PRVthm1}
\begin{aligned}
\lim_{N\to\infty}&\frac{1}{N}\log \tr_+ \big({\rm e}^{-N(h^{\ssup{N}}-f(x^{\ssup{N}}))}\big)=
\sup_{\omega\in  L^2([0,\beta];\R)}\Big\{\int_0^\beta f(\omega(s))\d s- I^{\ssup{{\rm sym},h}}_\beta(\omega)+\langle Q,g_\beta\rangle\Big\}\\
&=\beta\sup_{u\in\R}\;\inf_{Q\in\widetilde \Pmf(G\times G)}\Big\{ f(u)-I^{\ssup{Q,h}}_\beta(u)-H(Q|Q^{\ssup{1}}\otimes \cou)
\Big\},
\end{aligned}
\end{equation}
where
\begin{equation}
I^{\ssup{Q,h}}_\beta(u)=\sup_{a\in\R} \Big\{au-\sum_{x,y\in G}Q(x,y)\log\E_{x}^{\beta,h}\big({\rm e}^{a\int_0^\beta\xi(s)\d s}\1\{\xi_\beta=y\}\big)\Big\}\quad\mbox{ for } u\in\R.
\end{equation}
\end{theorem}

Trace formulas like \eqref{PRVthm} and \eqref{PRVthm1} go back to the work of Cegla, Lewis and Raggio \cite{CLR88} in which the authors use a combination of large deviations theory and group representation to derive a variational formula for the free energy of mean-field quantum spin systems. Inspired by their work, Petz, Raggio and Verbeure \cite{PRV89} derived a non-commutative version of Varadhan's theorem using $ C^*$-algebraic methods. We thus have in our Theorem~\ref{THMmain4} and Theorem~\ref{NoncommutativeTHM} derived a non-commutative version of Varadhan's Lemma and hence  a variational formula for the free energy of mean-field quantum spin systems under symmetrised distributions, i.e. a version with Bose-Einstein statistics.

\begin{proofsect}{Proof of Theorem~\ref{NoncommutativeTHM}}
The proof follows from Theorem~\ref{THMmain4} and \eqref{PRVthm1} and the analysis of the variational problems, which is done in the following Lemma~\ref{prooflemma} and Proposition~\ref{proofprop}.
\qed
\end{proofsect}

\begin{lemma}\label{prooflemma}
Let $ Q\in\widetilde \Pmf(G\times G) $. If $ \omega\in L^2([0,\beta];\R) $ is a constant function then the supremum in
$$
\sup_{g\in L^2([0,\beta];\R)}\Big\{\langle g,\omega\rangle -\sum_{x,y\in G}Q(x,y)\log\E_{x}^{\beta,h}\big({\rm e}^{\int_0^\beta g(s)\xi(s) \d s}\1\{\xi_\beta=y\}\big)\Big\}
$$
is attained at a constant function $ g $.
\end{lemma}

\begin{proofsect}{Proof}
Fix any $ Q\in \widetilde \Pmf(G\times G) $. Clearly 
$$ g\mapsto \sum_{x,y\in G}Q(x,y)\log\E_{x}^{\beta,h}\Big({\rm e}^{\int_0^\beta g(s)\xi(s) \d s}\1\{\xi_\beta=y\}\Big) $$ 
is as a convex combination of logarithmic moment generating functions convex and continuous. We introduce the Haar basis $ \{h_i\}_{i\ge 0 } $ for $ L^2([0,\beta];\R) $ consisting of the functions $ h_i $ defined by $ h_0(s)=1 $  for $ s\in[0,\beta] $ and if $ 2^m\le i\le 2^{m+1}-1 $,
$$
h_i(s)=\left\{\begin{array}{r@{\;,\;}l}
2^{m/2} & \mbox{ if } \beta(i2^{-m}-1)\le s < \beta((i+\frac{1}{2})2^{-m}-1);\\
-2^{m/2} & \mbox{ if } \beta((i+\frac{1}{2})2^{-m}-1)\le s \le \beta((i+1)2^{-m}-1);\\
0 & \mbox{ otherwise } 
\end{array}\right..
$$
Now for every $ \eps>0 $ there is a $ m\in\N $ and a function $ g $ in the space $ \Hcal_m $ spanned by the basis functions $ h_0,h_1,\ldots,h_{2^m-1} $ such that $ I_\beta^{\ssup{Q,h}}(\omega)<\langle g,\omega\rangle-\sum_{x,y\in G}Q(x,y)\log\E_{x}^{\beta,h}\big({\rm e}^{\int_0^\beta g(s)\xi(s) \d s}\1\{\xi_\beta=y\}\big)+\eps $. If $ \omega(s)=u $ for all $ s\in[0,\beta] $ we get the Euler-Lagrange equations for the finite dimensional variational problem as
\begin{equation}\label{EL-finite}
\begin {aligned}
u &=\sum_{x,y\in\Z^d}Q(x,y)\frac{\E_{x}^\beta\big(\langle h_0,\xi\rangle {\rm e}^{\langle g,\xi\rangle}\1\{\xi_\beta=y\}\big)}{\E_{x}^\beta\big({\rm e}^{\langle g,\xi\rangle}\1\{\xi_\beta=y\}\big)}\\
0&=\sum_{x,y\in\Z^d}Q(x,y)\frac{\E_{x}^\beta\big(\langle h_i,\xi\rangle {\rm e}^{\langle g,\xi\rangle}\1\{\xi_\beta=y\}\big)}{\E_{x}^\beta\big({\rm e}^{\langle g,\xi\rangle}\1\{\xi_\beta=y\}\big)}\quad\mbox{ for } 1\le i\le 2^m-1.
\end {aligned}
\end{equation}
We observe that 
$$
\sum_{x,y\in\Z^d}Q(x,y)\frac{\E_{x}^\beta\big(\cdot {\rm e}^{\langle g,\xi\rangle}\1\{\xi_\beta=y\}\big)}{\E_{x}^\beta\big({\rm e}^{\langle g,\xi\rangle}\1\{\xi_\beta=y\}\big)}
$$
is symmetric if the function $ g $ is constant. This is easily seen by
$$
\begin{aligned}
\sum_{x,y\in\Z^d}Q(x,y)&\frac{\E_{x}^\beta\big(\cdot {\rm e}^{\langle g,\xi\rangle}\1\{\xi_\beta=y\}\big)}{\E_{x}^\beta\big({\rm e}^{\langle g,\xi\rangle}\big)}\le \sum_{x,y\in\Z^d}\frac{\E_{x}^\beta\big(\cdot {\rm e}^{\langle g,\xi\rangle}\1\{\xi_\beta=y\}\big)}{\E_{x}^\beta\big({\rm e}^{\langle g,\xi\rangle}\1\{\xi_\beta=y\}\big)}
\end{aligned}
$$
and a complementary lower bound with $ \min_{x,y\in\L} Q(x,y)> 0 $ (indices $ x $ and $ y $ with $ Q(x,y)=0 $ do not contribute at all). Hence, the equations in \eqref{EL-finite} for $  1\le i\le 2^m-1 $ are trivially solved for any constant function $ g $. The constant is determined through the first equation of \eqref{EL-finite}.
\qed
\end{proofsect}

\begin{prop}\label{proofprop}
The supremum in the variational formula \eqref{varformulava} with the mean-field energy $ f $ is attained at a constant function $ \omega(s)=u\in\R $ for all $ s\in[0,\beta] $.
\end{prop}
 \begin{proofsect}{Proof}
We split the proof into two steps. In the second step we will show that
\begin{equation}\label{step3}
I_\beta^{\ssup{{\rm sym},h}}(\omega)\ge\int_0^\beta I_\beta^{\ssup{{\rm sym},h}}(\widehat \omega_s)\d s,
\end{equation}
where $ \widehat \omega_s(t)=\omega(s) , t\in[0,\beta],$ is the constant function with value $ \omega(s) $. 

\noindent {\bf Step 1}
From \eqref{step3} we get
\begin{equation}
\begin{aligned}
-\int_0^\beta f(\omega(s))\d s+ I_\beta^{\ssup{{\rm sym},h}}(\omega)&\ge \int_0^\beta\big(-f(\omega(s))+I_\beta^{\ssup{{\rm sym},h}}(\widehat \omega_s)\big)\d s\\
&\ge \beta \inf_{u\in\R} \big\{-f(u)+I_\beta^{\ssup{{\rm sym},h}}(\widehat u)\big\},
\end{aligned}
\end{equation}
and hence
\begin{equation}
\sup_{\omega\in L^2([0,\beta];\R)}\Big\{\int_0^\beta f(\omega(s))\d s- I_\beta^{\ssup{{\rm sym},h}}(\omega)\}\le \beta\sup_{u\in\R} \big\{f(u)-I_\beta^{\ssup{{\rm sym},h}}(\widehat u)\big\}
\end{equation}

\noindent {\bf Step 2} We now prove \eqref{step3}.
We only need to show that 
\begin{equation}\label{toshowstep2}
I^{\ssup{Q}}_\beta(\omega)\ge\int_0^\beta I_\beta^{\ssup{Q}}(\widehat \omega_s)\d s
\end{equation}
for any $ Q\in\widetilde \Pmf(\G^2) $.
We fix $ Q\in \widetilde \Pmf(\G^2) $. For any $ u\in\R $ we get from Lemma~\ref{prooflemma} that
\begin{equation}\label{Ifunctionreal}
I^{\ssup{Q}}_\beta(u)=\sup_{a\in\R}\Big\{au-\sum_{x,y\in G}Q(x,y)\log\E_{x}^{\beta,h}\Big({\rm e}^{a\int_0^\beta\xi(s)\d s}\1\{\xi_\beta=y\}\Big)\Big\}.
\end{equation}
The function $ \sum_{x,y\in G}Q(x,y)\log\E_{x}^{\beta,h}\big({\rm e}^{a\int_0^\beta\xi(s)\d s}\1\{\xi_\beta=y\}\big)$ is clearly convex and infinitely differentiable with increasing first derivative. From this we conclude that there exists, for any $ \eps>0 $, a function $ g\in L^\infty([0,\beta];\R) $ such that 
\begin{equation}\label{inequalityae}
I^{\ssup{Q}}_\beta(\widehat \omega_t) < g(t)\omega(t)-\sum_{x,y\in G}Q(x,y)\log\E_{x}^{\beta,h}\Big({\rm e}^{\widehat g_t\int_0^\beta\xi(s)\d s}\1\{\xi_\beta=y\}\Big)+\eps\quad\mbox{ for } a.e. t\in[0,\beta].
\end{equation}
As $ g\in L^2([0,\beta];[0,m]) $ we get the lower bound
\begin{equation}\label{laststeplower}
I^{\ssup{Q}}_\beta(\omega)\ge \langle g,\omega\rangle-\sum_{x,y\in G}Q(x,y)\log\E_{x}^{\beta,h}\Big({\rm e}^{\int_0^\beta g(s)\xi(s)\d s}\1\{\xi_\beta=y\}\Big).
\end{equation}
Therefore we are finished with our proof if we show that
\begin{equation}\label{laststeptoshow1}
\E_{x}^{\beta,h}\Big({\rm e}^{\int_0^\beta g(s)\xi(s)\d s}\1\{\xi_\beta=y\}\Big)\le \int_0^\beta\E_{x}^{\beta,h}\Big({\rm e}^{\widetilde g_t\int_0^\beta \xi(s)\d s}\1\{\xi_\beta=y\}\Big)\d t. 
\end{equation}

By continuity, we may assume that $ g\in\Hcal_m $ for some $ m\in\N $. Then $ g $ can be written as $ g=\sum_{k=1}^{2^m}\langle h_k,g\rangle\1\{[\beta(k-1)2^{-m},\beta k2^{-m}]\} $. Hence, we need to show that
\begin{equation}\label{laststeptoshow2}
\E_{x}^{\beta,h}\Big({\rm e}^{\int_0^\beta g(s)\xi(s)\d s}\1\{\xi_\beta=y\}\Big)\le 2^{-m}\sum_{k=1}^{2^m}\E_{x}^{\beta,h}\Big({\rm e}^{\langle h_k,g\rangle \int_0^\beta\xi(s)\d s}\1\{\xi_\beta=y\}\Big).
\end{equation}
But \eqref{laststeptoshow2} is an application of the H\"older inequality for random walk expectations. Hence, \eqref{laststeptoshow1} follows. Now, \eqref{laststeptoshow1} implies, together with \eqref{inequalityae} and \eqref{laststeplower}, the inequality \eqref{toshowstep2}, and finishes the proof.
\qed
 \end{proofsect}

\begin{example}[{\bf Quantum-Spin-$1/2 $ variables and Telegraph process}]
We shall apply our results for the mean field free energy in Theorem~\ref{PRVthm1} to the quantum spin $1/2 $ model introduced in \cite{Dor96}. In \cite{Dor96} the non-commutative central limit theorem for the following model was studied. We show in this example the extension to the non-commutative Varadhan Lemma to obtain the mean mean-field free energy.
In the setting of Theorem~\ref{THMmain4} and Theorem~\ref{PRVthm1} we consider the  set $ G=\{-1,+1\} $ of possible spin values and the following process on $ G $ with transition probabilities
$$
\P(\xi(t+\delta t)=y|\xi(t)=x)=\left\{\begin{array}{r@{\;,\;}l} \frac{1}{2}\delta t & \mbox{ if } \, y\not=x\\ 1-\frac{1}{2}\delta t & \mbox{ if }\, y=x\end{array} ,\right.
$$
equivalently
$$
p_{t}(x,y)=\frac{1}{2}\big(1+xy{\rm e}^{-t}\big)\quad ,x,y\in G.
$$

From Theorem~\ref{THMmain4} and Proposition~\ref{proofprop} we get for this process and any continuous bounded function $ f\colon[-1,1]\to\R $ the result
\begin{equation}
\begin{aligned}
\lim_{N\to\infty}\frac{1}{N}\log\mu_{N,\beta}^{\ssup{\rm sym,t}}\Big({\rm e}^{\frac{1}{N}\sum_{i=1}^Nf(\xi^{\ssup{i}})}\Big)&=\sup_{\omega\in L^2([0,\beta];\R)}\Big\{\int_0^\beta f(\omega(s))\d s - I_\beta^{\ssup{\rm sym,t}}(\omega)\Big\}\\
& =\sup_{u\in\R}\big\{f(u)-I_\beta^{\ssup{\rm sym,t}}(u)\big\},
\end{aligned}
\end{equation} where
\begin{equation}\label{ratetelegraph}
I_\beta^{\ssup{\rm sym,t}}(u)=\inf_{Q\in\widetilde \Pmf(G^2)}\Big\{ H(Q|Q^{\ssup{1}}\otimes \cou)+\sup_{a\in\R}\Big\{ua-\sum_{x,y\in G}Q(x,y)\log \E_{x}^\beta\Big({\rm e}^{a\int_0^\beta \xi(s)\d s}\1\{\xi_\beta=y\}\Big)\Big\}\Big\}.
\end{equation}

To analyse \eqref{ratetelegraph} we have to calculate the expectations $ \E_{x}^\beta\big({\rm e}^{a\int_0^\beta \xi(s)\d s}\1\{\xi_\beta=y\}\big) $ for any $ x,y\in G $. This can be done by simple matrix calculations as follows.
\begin{equation}\label{calexp}
\begin{aligned}
\E_{x}^\beta\Big({\rm e}^{a\int_0^\beta \xi(s)\d s}\1\{\xi_\beta=y\}\Big)&=\big\langle y|{\rm e }^{\beta(a\sigma_z-\frac{1}{2}(\1-\sigma_x))}|x\big\rangle,
\end{aligned}
\end{equation}
where 
$$ \sigma_z=\left(\begin{array}{cc} 1 & 0\\0 & -1\end{array}\right) \mbox{  and }  \sigma_x=\left(\begin{array}{cc} 0 & 1\\ 1 & 0\end{array}\right) 
$$ are the two-dimensional Pauli matrices. The eigenvalues of the matrix $  (a\sigma_z-\frac{1}{2}(\1-\sigma_x)) $ are $ \lambda_{\pm 1}=-\frac{1}{2}\pm\sqrt{\frac{1}{4}+a^2} $. In principle, direct calculations of  all the expectations in \eqref{calexp} lead to an evaluation of \eqref{ratetelegraph}. The resulting mean mean-field free energy  is therefore given by
\begin{equation}
\begin{aligned}
\lim_{N\to\infty}\frac{1}{N}\log\mu_{N,\beta}^{\ssup{\rm sym,t}}\Big({\rm e}^{\frac{1}{N}\sum_{i=1}^Nf(\xi^{\ssup{i}})}\Big)=\beta \sup_{u\in\R} \{f(u)-\frac{1}{2}(1-\sqrt{1-u^2})\big\},
\end{aligned}
\end{equation}
compare \cite{D02}.
As an alternative to the tedious calculations we immediately get a lower bound via the following considerations. Let $ a\in\R $ and let $ u_a $ denote the eigenvector for the eigenvalue $ \l_1 $ of the matrix $ (a\sigma_z-\frac{1}{2}(\1-\sigma_x)) $. Then
$$
M_\beta^{\ssup{a}}={\rm e}^{a\int_0^\beta\xi(s)\d s}{\rm e}^{-\beta\l_1}\frac{u_a(\xi_\beta)}{u_a(\xi_0)}
$$
defines a martingale for the telegraph process. We insert $ M_\beta^{\ssup{a}} $ in the expectation on the right hand side of \eqref{ratetelegraph} and obtain
$$
\sup_{a\in\R}\Big\{\beta(au-\l_1)-\sum_{x,y\in G}Q(x,y)\log\E_{x}^\beta\big(M_\beta^{\ssup{a}}\1\{\xi_\beta=y\})\Big\}.
$$
As $ \E_x(M_\beta^{\ssup{a}}\1\{\xi_\beta=y\}) $ is a probability measure for all $ x\in G $ due to the martingale property we can combine the logarithm term with the one from the entropy to get
$$
I_\beta^{\ssup{\rm sym,t}}(u)=\inf_{Q\in\widetilde \Pmf(G^2)}\Big\{\beta(au-\l_1)+\sum_{x,y\in G}Q(x,y)\log\frac{Q(x,y}{Q^{\ssup{1}}\E_x(M_\beta^{\ssup{a}}\1\{\xi_\beta=y\})}\Big\}.
$$
Hence, as the relative entropy of probability measures is positive, we get as a lower the Legendre-Fenchel transform of $ \l_1 $, which is given by
$ \beta\frac{1}{2}(1-\sqrt{1-u^2})$.

\end{example}

\subsection{A special case for the rate function}\label{speccase-sec}
In this subsection we study motivated by the results for the quantum spin models in the previous section a very important special case for the rate function for the mean of the normalised occupation local times when our $ N $ random processes are simple random walks on $ \Z^d $ whose generator is given by the discrete Laplacian. We let $ \L\subset\Z^d $ be a finite box and we put $ m $ equal the counting measure $ \cou_\L $ on $ \L $. Instead of the symmetrised probability measure $ \P_{N,\beta}^{\ssup{\rm sym}} $ in \eqref{defsymP} we consider the symmetrised measure 
\begin{equation}\label{symmeasure}
\mu_{\L,N}^{\ssup{\rm sym}}=\frac{1}{N!}\sum_{\sigma\in\Sym_N}\sum_{x_1\in\L}\cdots\sum_{x_N\in\L}\bigotimes_{i=1}^N\mu_{x_i,x_{\sigma(i)}}^\beta.
\end{equation}

Proposition~\ref{prop1}(iii) gives a large deviations principle for the mean of the normalised occupation local times under the measure $ \mu_{\L,N}^{\ssup{\rm sym}} $, i.e. we have
$$
\lim_{N\to\infty}\frac{1}{N}\log (\mu_{\L,N}^{\ssup{\rm sym}}\circ Z_N^{-1})=-\inf_{p\in\Pmf(\Z^d)}J^{\ssup{\rm sym}}_{\beta,\L}(p),
$$
where
\begin{equation}\label{JsymLambdadef}
J^{\ssup{\rm sym}}_{\beta,\L}(p)=\inf_{Q\in\widetilde \Pmf(\Z^d\times\Z^d)}\big\{H(Q|Q^{\ssup{1}}\otimes\cou_\L)+J^{\ssup{Q}}_{\beta,\L}(p)\big\},\quad p\in\Pmf(\Z^d),
\end{equation}
with
\begin{equation}\label{JqLambdadef}
J^{\ssup{Q}}_{\beta,\L}(p)=\sup_{f\in\Bcal(\Z^d)}\Big\{\beta\sum_{x\in\Z^d}f(x)p(x)-\sum_{x,y\in\Z^d}Q(x,y)\log\E_x\Big({\rm e}^{\int_0^\beta f(\xi_s)\d s}\1\{\xi_\beta=y\}\Big)\Big\}.
\end{equation}

Our goal is to identify the rate function $ $ in much easier and more familiar terms. It turns out that this rate function is the {\it Donsker-Varadhan\/} rate function $ I_\L $,  defined as
\begin{equation}\label{defDVratef}
I_\L(p)=\left\{\begin{array}{r@{\;,\;}l} \frac{1}{2}\sum_{\heap{x,y\in\L\colon}{|x-y|=1}}\big(\sqrt{p(x)}-\sqrt{p(y)}\big)^2 & \mbox{if } \supp(p)\subset\L\\ +\infty & \mbox{otherwise }
\end{array}\right.,
\end{equation}
where $ |x-y|=\max_{1\le i\le d}|x_i-y_i| , x,y\in\Z^d $, is the lattice distance.
This is the  rate function for the large deviations principle for the normalised occupation local time $ l_\beta $ as $ \beta\to\infty $. More precisely, denote by $ \xi_{[0,\beta]} $ the path of the random walk and define the sub probability measure $ P_x^{\ssup{\beta}}=\P_x(\cdot|\supp\,(l_\beta)\subset\L)=\P_x(\cdot|\xi_{[0,\beta]}\subset\L) $. The the normalised occupation local time satisfies as $ \beta\to\infty $ a large deviations principle on $ \Pmf(\L) $ with speed $ \beta $ and rate function $ I_\L -C_\L$ \cite{DV75-83}, where $ C_\L $ is just the normalisation $ \inf_{p\in\Pmf(\Z^d)}\frac{1}{2}\sum_{\heap{x,y\in\L\colon}{|x-y|=1}}\big(\sqrt{p(x)}-\sqrt{p(y)}\big)^2 $ for the rate function. 

The surprising result is that the rate function $ J^{\ssup{\rm sym}}_{\beta,\L} $ for finite $ \beta $ but for large $ N $ under the symmetrised measure equals the {\it Donsker-Varadhan\/} rate function. 

\begin{theorem}\label{Jident} Let $ \L\subset\Z^d $ be a finite set. Then $ J^{\ssup{\rm sym}}_{\beta,\L}(p)= \beta I_{\L}(p)$ for any $p\in\Pmf(\Z^d)$ with $ \supp\,(p)=\L $.
\end{theorem}

For the proof of the theorem we need the following lemma.

\begin{lemma}\label{Jrestr} Fix $\beta\in(0,\infty)$ and a finite set  $\L\subset\Z^d$. Then, for all $ p\in\Pmf(\Z^d) $ having support in $\L$,
\begin{equation}\label{JqLambdaident}
J^{\ssup{\rm sym}}_{\beta,\L}(p)=
\inf_{Q\in\widetilde \Pmf(\Z^d\times\Z^d)}\Big\{H(Q|Q^{\ssup{1}}\otimes\cou_\L)+J^{\ssup Q}_{\beta,\L}(p)\Big\},
\end{equation}
where
$$
J^{\ssup Q}_{\beta,\L}(p)=\sup_{f\in\R^\L}\Bigl\{\beta\sum_{x\in\L}f(x)p(x)-\sum_{x,y\in\L}Q(x,y)\log\E_x\big( {\rm e}^{\int_0^{\beta}f(\xi_s)\,\d s}\1\{\xi_{[0,\beta]}\subset\L\}\1\{\xi_\beta=y\}\Big)\Bigr\}.
$$
\end{lemma}

\begin{proofsect}{Proof} Note that $H(Q|Q^{\ssup{1}}\otimes\cou_\L)=\infty$ if the support of $Q$ is not contained in $\L\times\L$. Hence, in \eqref{JsymLambdadef} we need to take the infimum over pair probability measures $Q$ only on the set $  \widetilde \Pmf(\L\times\L) $.
From an inspection of the right hand side of \eqref{JqLambdadef} it follows that the function (vector) $f$ in the supremum can be taken arbitrarily negative outside $\L$ to approximate the supremum. Hence, we may add in the expectation the indicator on the event that the random walk does not leave $\L$ by time $\beta$. But then the values of $f$ outside $\L$ do not contribute. This shows that we need to consider only functions $f$ that are defined on $\L$; in other words,  \eqref{JqLambdaident} holds.
\qed
\end{proofsect}

\begin{proofsect}{Proof of Theorem~\ref{Jident}}
We proceed in two steps. First we show that $ J^{\ssup{\rm sym}}_{\beta,\L}(p)\ge \beta I_\L(p) $ for any $ p\in\Pmf(\Z^d) $ with support in $ \L $. In a second step we show the complementary inequality for any $ p\in\Pmf(\Z^d) $ with $ \supp\,(p)=\L $. We start from \eqref{JqLambdaident}.

Denote by $ \Delta_\L $ the restriction of the discrete Laplacian to the set $ \L $ with zero boundary conditions, i.e. for any $ f\in\R^\L $,
\begin{equation}
\Delta_\L f(x)=\sum_{\heap{x,y\in\L\colon}{|x-y|=1}}[f(x)-f(y)], \quad x\in\L,  f\in\R^{\Z^d},\supp\,(f)\subset\L.
\end{equation}
For $ f\in l^2(\L)=\R^\L $ let $ u_f $ be the unique positive eigenfunction for the operator $ \Delta_\L +f $ for the eigenvalue $ \l(f) $ given by
\begin{equation}\label{eigenvdefi}
\begin{aligned}
\l_\L(f)&=\sup_{g\in l^2(\L)\colon ||g||_2=1}\langle (\Delta_\L+f)g,g\rangle=-\inf_{\heap{g\in l^2(\Z^d)\colon}{\supp(g)\subset\L,||g||_2=1}}\Big\{\frac{1}{2}\sum_{\heap{x,y\in\Z^d\colon}{|x-y|=1}}(g(x)-g(y))^2-\langle f,g\rangle^2\Big\}.
\end{aligned}
\end{equation}
To check the positivity and uniqueness of the eigenfunction consider the operator $ \Delta_\L+c\1+f $, which is non-negative and irreducible  for some $ c>0 $. The theorem of Perron-Frobenius (see \cite{Se81}) gives that $ \Delta_\L+c\1+f $ has a unique simple eigenvalue and that the corresponding eigenfunction is positive in $ \L $ and unique up to constants. Clearly, $ \Delta_\L+c\1+f  $ and $ \Delta_\L+f $ have the same eigenfunction, and the corresponding eigenvalues differ by the constant $ c $. Hence, $\l(f) $ is simple and the eigenfunction $ u_f $  is unique and positive.

We now introduce a martingale which permits a transformation of the random walk in the expectation of the variational formula in \eqref{JqLambdaident}. The expression
\begin{equation}\label{densitygirsanov}
M_{\beta}^{\ssup{f}}:={\rm e}^{\int_0^{\beta}f(\xi_s)\,\d s}{\rm e}^{-\beta\l_{\L}(f)}\1\{\xi_{[0,\beta]}\subset\L\}\frac{u_f(\xi_{\beta})}{u_f(\xi_{0})}
\end{equation} 
defines a martingale $ (M_{\beta}^{\ssup{f}})_{\beta\ge 0} $ under $\P_x $ for any $x\in\L$ with respect to the canonical filtration (see the Markov process variant of \cite[Prop. VIII.3.1]{RY99} or \cite{Kal01} for martingale theory for Markov processes).

We insert now  $M_{\beta}^{\ssup{f}}$ on the right hand side of \eqref{JqLambdaident}, obtain  an extra $ \beta\l_\L(f) $  and use the marginal property of the pair probability measure $ Q $, which gives
$ \sum_{x,y}Q(x,y)\log \frac{u_f(y)}{u_f(x)}=0 $. Thus  we see that
$$
J^{\ssup{Q}}_{\beta,\L}(p)=\sup_{f\in\R^\L}\Bigl\{\beta(\langle f,p\rangle - \l_{\L}(f))-\sum_{x,y\in\L}Q(x,y)\log \E_{x}\big(M_{\beta}^{\ssup{f}}\1\{\xi_\beta=y\}\big)\Bigr\},
$$ 
where $ \langle f,p\rangle=\sum_{x\in\L}f(x)p(x) $ is the scalar product in $ \L $.
Here $\E_x$ denotes expectation for the simple random walk with generator $\Delta$ starting at $x$. Note that by the martingale property of $ (M_{\beta}^{\ssup{f}})_{\beta\ge 0} $ the measure $ \E_x(M_{\beta}^{\ssup{f}}\1\{\xi_\beta=\cdot\}) $ is a probability measure on $ \L $ for any $ x \in\L $.
Substituting this in \eqref{JqLambdaident} and recalling  the definition of the relative entropy of the pair measures, we obtain that
\begin{equation}\label{infqsupf}
J^{\ssup{\rm sym}}_{\beta,\L}(p)=\inf_{Q\in\widetilde \Pmf(\Z^d\times\Z^d)}\sup_{f\in\R^\L}\Bigl\{\beta(\langle f,p\rangle - \l_{\L}(f))+\sum_{x,y\in\L}Q(x,y)\log\frac{Q(x,y)}{Q^{\ssup{1}}(x)\E_{x}\big(M_{\beta}^{\ssup{f}}\1\{\xi_\beta=y\}\big)}\Bigr\}.
\end{equation}
The double sum in \eqref{infqsupf} is, because $ \E_x(M_{\beta}^{\ssup{f}}\1\{\xi_\beta=\cdot\}) $ is a probability measure, an entropy between probability measures and therefore nonnegative.

Therefore we get that
\begin{equation}\label{JsymLambdaident}
J^{\ssup{\rm sym}}_{\L}(p)\ge \beta \sup_{f\in\R^\L}\Bigl\{\langle f,p\rangle-\l_{\L}(f)\Bigr\}.
\end{equation}
Note that the map $f\mapsto \l_{\L}(f) $ is the Legendre-Fenchel transform of  $ I_\L $, as is seen from the Rayleigh-Ritz principle in \eqref{eigenvdefi}. According to the Duality Lemma \cite[Lemma~4.5.8]{DZ98}, the r.h.s.\ of \eqref{JsymLambdaident} is therefore equal to $ \beta I_\L(p) $ since it is equal to the Legendre-Fenchel transform of $\l_{\L}$. Hence, we have shown that $ J^{\ssup{\rm sym}}_{\L}(p)\ge \beta I_\L(p)$ for any $ p\in\Pmf(\Z^d)$ with support in $ \L$.

In our second step we prove the complementary inequality. For that we construct a pair probability measure to get an upper for \eqref{infqsupf}. The resulting upper bound is given by the supremum over  any function $ f\in\R^\L $. Thus we are left to solve this variational problem for the upper bound. Let's turn to the details.

For $ p\in\Pmf(\Z^d) $ with $ \supp\,(p)=\L $ define the function $ u\in\R^\L $ by $ u(x)=\sqrt{p(x)} $ and the function $ f^*\in\R^\L $ by 
$$ 
f^*(x)=-\frac{\Delta_\L u(x)}{u(x)}\quad\mbox{ for } x\in\L.
$$
Then $ u=u_{f^*} $ is the unique positive eigenfunction for the operator $ \Delta_\L+f^* $ with eigenvalue $ \l_\L(f^*)=0 $. Given these objects we define the function $ Q^*\colon\L\times\L\to [0,1] $ by
\begin{equation}
Q^*(x,y)=u(x)u(y)\E_x\Big({\rm e}^{\int_0^\beta f^*(\xi_s)\d s}\1\{\xi_{[0,\beta]}\subset\L\}\1\{\xi_\beta=y\}\Big),\quad x,y\in\L.
\end{equation}
This function is obviously symmetric and sums up to one. By the martingale property of $ (M_\beta^{\ssup{f^*}})_{\beta\ge 0} $ the first marginal is identified as 
\begin{equation}
Q^{*\ssup{1}}(x)=u(x)\E_x\Big({\rm e}^{\int_0^\beta f^*(\xi_s)\d s}\1\{\xi_{[0,\beta]}\subset\L\}u(\xi_\beta)\Big)=u(x)^2,\quad x\in\L.
\end{equation}
Therefore $ Q^*\in\widetilde \Pmf(\L\times\L) $, which gives the following upper bound for \eqref{infqsupf}, when we apply the marginal property of $ Q^* $ and separate the eigenvalue in the definition of $ M_\beta^{\ssup{f}} $. The marginal property gives
$$
\sum_{x,y\in\L}Q^*(x,y)\log\frac{u(x)u(y)}{Q^{*\ssup{1}}(x)}=0.
$$
Hence
\begin{equation}\label{Jsymupperboundf}
\begin{aligned}
J^{\ssup{\rm sym}}_{\beta,\L}(p)&\le \sup_{f\in\R^\L}\Bigl\{\beta\langle f,p\rangle +\sum_{x,y\in\L}Q^*(x,y)\log \frac{\E_x\Big({\rm e}^{\int_0^\beta f^*(\xi_s)\d s}\1\{\xi_{[0,\beta]}\subset\L\}\1\{\xi_\beta=y\}\Big)}{\E_x\Big({\rm e}^{\int_0^\beta f(\xi_s)\d s}\1\{\xi_{[0,\beta]}\subset\L\}\1\{\xi_\beta=y\}\Big)}\Big\}.
\end{aligned}
\end{equation}

We will show that the variational problem on the right hand side of \eqref{Jsymupperboundf} is solved for $f=f^* $. Due to the strict concavity we need to get the solution for the Euler-Lagrange equation, which reads  
\begin{equation}\label{EL}
\beta\langle v,p\rangle=\sum_{x,y\in\L}Q^*(x,y)\frac{\E_x\Big(\big(\int_0^\beta v(\xi_s)\d s\big) {\rm e}^{\int_0^\beta f(\xi_s)\d s}\1\{\xi_{[0,\beta]}\subset\L\}\1\{\xi_\beta=y\}\Big)}{\E_x\Big({\rm e}^{\int_0^\beta f(\xi_s)\d s}\1\{\xi_{[0,\beta]}\subset\L\}\1\{\xi_\beta=y\}\Big)}
\end{equation}
for all $ v\in\R^\L $. The right hand side of \eqref{EL} for $ f=f^* $ reads 
\begin{equation}
\begin{aligned}
\sum_{x,y\in\L}u(x)u(y)&\E_x\Big(\big(\int_0^\beta v(\xi_s)\d s\big) {\rm e}^{\int_0^\beta f(\xi_s)\d s}\1\{\xi_{[0,\beta]}\subset\L\}\1\{\xi_\beta=y\}\Big)\\
&=\sum_{x\in\L}u(x)^2\E_x\Big(M_\beta^{\ssup{f^*}}\int_0^\beta v(\xi_s)\d s\Big)=\int_0^\beta\d s\widetilde \E^{\ssup{f^*}}\big(v(\xi_s)\big),
\end{aligned}
\end{equation}
where $ \E^{\ssup{f^*}} $ is the expectation with respect to the  transform with the martingale $ (M_\beta^{\ssup{f^*}})_{\beta\ge 0} $ (compare remark following Prop.~VIII.3.9 in \cite{RY99}), starting at its invariant measure $ u^2(x)=p(x), x\in\L $. Therefore $ \widetilde \E^{\ssup{f^*}}\big(v(\xi_s)\big)=\langle v,p\rangle $, because the transformed random walk does not leave the set $ \L $ and is stationary when started in its invariant measure. Thus $f=f^* $ solves the variational problem on the right hand side of \eqref{EL} and gives finally after short calculations
\begin{equation}
\begin{aligned}
J_{\beta,\L}^{\ssup{\rm sym}}(p)&\le \beta \langle f^*,p\rangle=-\beta\big\langle \frac{\Delta_\L u}{u},u\big\rangle=-\beta\sum_{x\in\L}u(x)\Delta_\L u(x)=
\beta\Big(2d-\sum_{\heap{x,y\in\L\colon}{|x-y|=1}}u(x)u(y)\Big)\\ &=\beta\frac{1}{2}\sum_{\heap{x,y\in\L\colon}{|x-y|=1}}\big(\sqrt{p(x)}-\sqrt{p(y)}\big)^2=\beta I_\L(p).
\end{aligned}
\end{equation}

\qed
\end{proofsect}

We will give a heuristic interpretation in terms of the cycle structure. The measure $\mu_{\Lambda,N}^{\ssup{{\rm sym}}}$ in \eqref{symmeasure} admits a representation which goes back to Feynman 1953 \cite{F53}; in fact he considered Brownian motions instead of random walks on $ \Z^d $. 
Every permutation $\sigma\in\Sym_N$ can be written as a concatenation of cycles. Given a cycle $(i,\s(i),\s^2(i),\dots,\s^{k-1}(i))$ with $\s^k(i)=i$ and precisely $k$ distinct indices, the contribution coming from this cycle is independent of all the other indices. Furthermore, by the fact that $\mu^{\beta}_{x_i,x_{\s(i)}}$ is the conditional distribution given that the random walk ends in $x_{\s(i)}$, this contribution (also executing the $k$ integrals over $x_{\s^l(i)}\in\Lambda$ for $l=k-1,k-2,\dots,0$) turns the corresponding $k$ random walk bridges of length $\beta$ into one random walk  bridge of length $k\beta$, starting and ending in the same point $x_i\in\Lambda$ and visiting $\L$ at the times $\beta,2\beta,\dots,(k-1)\beta$. Hence, 
$$
\mu_{\Lambda,N}^{\ssup{{\rm sym}}}=\frac{1}{N!}\sum_{\s\in\Sym_N}\;\bigotimes_{k\in \N}\Big(\int_\Lambda \d y_k\,\mu^{k,\beta,\L}_{y_k,y_k}\Big)^{\otimes f_k(\sigma)},
$$
where $f_k(\s)$ denotes the number of cycles in $\s$ of length precisely equal to $k$, and $\mu^{k,\beta,\L}_{x,y}$ is the random walk  bridge measure $\mu^{k\beta}_{x,y}=\P_x(\cdot\1\{\xi_\beta=y\}) $ , restricted to the event $\bigcap_{l=1}^k\{\xi_{l\beta}\in\L\}$. (See \cite[Lemma 2.1]{Gin71} for related combinatorial considerations.) If $f_N(\s)=1$ (i.e., if $\s$ is a cycle), then we are considering just one random walk bridge $\xi$ of length $N\beta$, with uniform initial measure on $\L$, on the event $\bigcap_{l=1}^N\{\xi_{l\beta}\in\L\}$. Furthermore, $Y_N$ is equal to the normalised occupation measure of this random walk. For such a $\s$, the limit $N\to\infty$ turns into a limit for diverging time, and the corresponding large-deviation principle of Donsker and Varadhan formally applies. In \cite{A06} large deviation results are obtained for the cycle representation for integer partitions.

It has been proved by S\"ut\"o \cite{S93,S02} that the non-vanishing of the probability for long cycles is equivalent with {\it Bose-Einstein condensation\/} for non-interacting Bosons. \cite{DMP05} studied the density of long cycles for mean-field interactions and \cite{BCMP05} study large deviations for different cycles statistics in the grandcanonical ensemble.  

If a permutation $\s$ does not contain a cycle of length $\approx N^\alpha$, for some $ \alpha>0 $, its contribution is quantified with a different rate, compare \cite{BCMP05} and for the canonical ensemble \cite{A06}. In this way, Theorem~\ref{Jident} says that the large-$N$ behaviour of $\mu_{\Lambda,N}^{\ssup{{\rm sym}}}\circ Y_N^{-1}$ is predominantly determined by all those permutations consisting of just one cycle of length $N$. 

\subsection{Preliminaries}\label{pre-sec}

In this subsection we provide some definitions and basic properties. A function $ \xi\colon[0,\beta]\to\R^d $ is an element of $ D_{\beta} $ if and only for any $t_0\in(0,\beta) $
\begin{equation}
\lim\limits_{t\downarrow t_0}\xi(t)=\xi(t_0)\;\mbox{ and }\;\lim\limits_{t\uparrow t_0}\xi(t)\;\mbox{ exists}, 
\end{equation} and, moreover,
$ \xi(0)=\lim_{t\downarrow 0} $ and $ \xi(\beta)=\lim_{t\uparrow \beta}\xi(t)$.
Every function in $ D_{\beta} $ has at most a countable number of points where it is not continuous (\cite{Par67}). On $ D_{\beta} $ one can introduce a metric, the Skorokhod metric, defined by
\begin{equation}
\rho(\xi,\xi^{\prime})=\inf\limits_{\lambda\in H}\Bigl\{\sup\limits_{t\in[0,\beta]} |\xi(\lambda(t))-\xi^{\prime}(t)|+\sup\limits_{t\in [0,\beta]}|\lambda(t)-t|\Bigr\},
\end{equation} where $ H $ is the space of increasing functions $ \lambda\colon [0,\beta]\to[0,\beta] $. The Skorokhod topology is the topology induced on $ D_{\beta} $ by the metric $ \rho$ and turns $ D_{\beta} $ into a Polish space(\cite{Par67},\cite{DS01}). One has the following characterisation of compact subsets of $ D_{\beta} $. 

\begin{lemma}[\cite{Par67}]
In order that a subset $ K\subset D_{\beta} $ be compact under the Skorokhod topology it is necessary and sufficient that $ K $ be bounded and closed and satisfy the condition
\begin{equation}
\lim\limits_{\delta\downarrow 0}\sup\limits_{\xi\in K} \omega_{\xi}(\delta)=0,
\end{equation} where $ \omega_{\xi}(\delta) $ is defined by 
\begin{equation}
\begin{aligned}
\omega_{\xi}(\delta)&=\sup\limits_{t-\delta\le t^{\prime}\le t\le t^{\prime\prime}\le t+\delta}\Bigl\{|\xi(t^{\prime})-\xi(t)|\wedge |\xi(t^{\prime\prime})-\xi(t)|\Bigr\} +\sup\limits_{0\le t\le \delta} |\xi(t)-\xi(0)|\\
&+\sup\limits_{\beta-\delta \le t \le\beta}|\xi(t)-\xi(\beta)|.
\end{aligned}
\end{equation}
\end{lemma}

For the convenience of our reader, we repeat the notion of a large-deviations principle and of the most important facts that are used in the present paper. See \cite{DZ98} and \cite{DS01} for a comprehensive treatment of this theory. 

Let $ \Xcal $ denote a topological vector space.  A lower semi-continuous function $ I\colon \Xcal\to [0,\infty] $ is called a {\it rate function\/} if  $ I $ is not identical $ \infty$ and  has compact level sets, i.e. if $ I^{-1}([0,c])=\{x\in\Xcal\colon I(x)\le c\} $ is compact for any $ c\ge 0 $. A sequence $(X_N)_{N\in\N}$ of $\Xcal$-valued random variables $X_N$  satisfies the {\it large-deviation upper bound\/} with {\it speed\/} $a_N$ and rate function $I$ if, for any closed subset $F$ of $\Xcal$,
\begin{equation}\label{LDPupper}
\limsup_{N\to\infty}\frac 1{a_N}\log \P(X_N\in F)\leq -\inf_{x\in F}I(x),
\end{equation}
and it satisfies the {\it large-deviation lower bound\/} if, for any open subset $G$ of $\Xcal$,
\begin{equation}\label{LDPlower}
\liminf_{N\to\infty}\frac 1{a_N}\log \P(X_N\in G)\leq -\inf_{x\in G}I(x).
\end{equation}
If both, upper and lower bound, are satisfied, one says that $(X_N)_N$ satisfies a {\it large-deviation principle}. The principle is called {\it weak\/} if the upper bound in \eqref{LDPupper} holds only for {\it compact\/} sets $F$. A weak principle can be strengthened to a full one by showing that the sequence of distributions of $X_N$ is {\it exponentially tight}, i.e. if for any $L>0$ there is a compact subset $K_L$ of $\Xcal$ such that $\P(X_N\in K_L^{\rm c})\leq {\rm e}^{-LN}$ for any $N\in\N$. 

One of the most important conclusions from a large deviation principle is {\it Varadhan's Lemma}, which says that, for any bounded and continuous function $F\colon \Xcal\to\R$,
$$
\lim_{N\to\infty}\frac 1N\log \int {\rm e}^{N F(X_N)}\,\d\P=-\inf_{x\in \Xcal}\big(I(x)-F(x)\big).
$$

One standard situation in which a large deviation principle holds is the case where $\P$ is a probability measure, and $X_N=\frac 1N(Y_1+\dots+Y_N)$ is the mean of $N$ i.i.d.~$\Xcal$-valued random variables $Y_i$ whose moment generating function 
$$ M(F)=\int {\rm e}^{F(Y_1)}\,\d\P$$ 
is finite for all elements $F$ of the topological dual space $\Xcal^*$ of $\Xcal$. In this case, the abstract Cram\'er theorem provides a weak large deviation principle for $(X_N)_{N\in\N}$ with rate function equal to the Legendre-Fenchel transform of $\log M$, i.e. $I(x)=\sup_{F\in \Xcal^*}(F(x)-\log M(F))$. An extension to independent, but not necessarily identically distributed random variables is provided by the abstract G\"artner-Ellis theorem.

For our main theorems Theorem~\ref{mainthm1}, Theorem~\ref{mainthm3} and Theorem \ref{mainthm2} we shall rely on the following conventions.
For the random probability measure in $ \Pmf(D_{\beta}) $ we conceive $ \Pmf(D_{\beta}) $ as a closed convex subset of $ \Mcal(D_{\beta}) $, the space of finite signed Borel measures on $ D_{\beta} $.  $ \Mcal(D_{\beta}) $ is a topological Hausdorff space, whose topology is induced by the set $ \Ccal_{\rm b}(D_{\beta}) $ of continuous bounded functions on $ D_{\beta} $. Here $ \Ccal_{\rm b}(D_{\beta}) $ is the topological dual of $ \Mcal(D_{\beta})$. The set $ \Pmf(D_{\beta}) $ of probability measures on $ D_{\beta} $ inherits its topology from $ \Mcal(D_{\beta}) $.  When we speak of a
large deviations principle of $\Pmf(D_{\beta})$-valued random variables, then we
mean a principle on $\Mcal(D_{\beta}) $ with a rate function that is tacitly
extended from $ \Pmf(D_{\beta})$ to $\Mcal(D_{\beta})$ with the value $+\infty$. Thus, in the variational formula \eqref{variationallevel2} the Legendre-Fenchel transform with respect to the set $ \Ccal_{\rm b}(D_{\beta}) $ appeared. For the mean path we embed the space $ D_\beta([0,\beta];\R^d) $ continuously into $ L^2([0,\beta];\R^d) $, hence the dual pairing is here given by the $ L^2$-scalar product.


\section{Proofs}\label{Proofsec} In this section we prove our main Theorems~\ref{mainthm1}, \ref{mainthm3} and \ref{mainthm2}. In the first subsection we prove the large deviations principle in Theorem~\ref{mainthm1} for the empirical path measures. In Subsection~\ref{main3} we prove Theorem~\ref{mainthm3}, in Subsection~\ref{main2} we prove Theorem~\ref{mainthm2} both on the basis of the proof for Theorem~\ref{mainthm1} and the contraction principle.
The last subsection, Subsection~\ref {productldp-proof}, is devoted to the proof of the exponential tightness of the distribution of the empirical path measure under the  symmetrised measure \eqref{defsymP} and the exponential tightness for certain products of not necessarily identical distributed objects coming form the two-level large deviation method occurring in the proof of our main Theorems~\ref{mainthm1} and \ref{mainthm2}.

\subsection{Proof of Theorem \ref{mainthm1}}\label{ProofTH1}

For the proof of Theorem~\ref{mainthm1} we have to show the following inequalities and properties,\\
(1) For any open set $ A\subset\Pmf(D_{\beta}) $,
\begin{equation}
\liminf\limits_{N\to\infty}\log\P^{\ssup{\rm sym}}_{N,\beta}(L_N\in A)\ge -\inf\limits_{\mu\in A} I^{\ssup{\rm sym}}_{\beta}(\mu),
\end{equation}
\noindent (2) for any compact set $ F \subset\Pmf(D_{\beta}) $,
\begin{equation}
\limsup\limits_{N\to\infty}\log\P^{\ssup{\rm sym}}_{N,\beta}(L_N\in F)\le -\inf\limits_{\mu\in F} I^{\ssup{\rm sym}}_{\beta}(\mu).
\end{equation} 
\noindent (3) The sequence of probability measures $ (\P^{\ssup{\rm sym}}_{N,\beta}\circ L_N^{-1})_{N\in\N} $ is exponentially tight.

The symmetrised measure $ \P^{\ssup{\rm sym}}_{N,\beta} $ is not a product of  i.i.d. measures, hence we first transform the probability measure into a probability measure which permits an application of a version of the G\"artner-Ellis theorem. This will be done in Section \ref{seclowerboundmain} for the lower bound and in Section~\ref{secupperboundmain} for the upper bound respectively. The technique applied here is a purely combinatoric one and it is inspired by the works of \cite{Gin71} and \cite{Toth90}. The counting arguments for the proposed combinatoric methods are given by the results in \cite{A01}, where combinatoric counting was used to evaluate partition functions in the microcanonical ensemble.
For the whole proof of Theorem \ref{mainthm1} the initial distribution measure $ m\in\Pmf(\Z^d) $ allows some kind of compactification. The exponential tightness will be proved in Subsection~\ref{productldp-proof}.

\subsubsection{Proof of the lower bound of Theorem \ref{mainthm1}}\label{seclowerboundmain}

Fix an open set $ A\in \Pmf(D_{\beta}) $ and a finite set $ \L\subset\Z^d$ with $ m(\L)>0$. We get the lower bound  
\begin{equation}\label{estlower1}
\P^{\ssup{\rm sym}}_{N,\beta}(L_N\in A)\ge m(\L)^N\frac{1}{N!}\sum\limits_{\s\in\Sym_N}\sum\limits_{x_1\in \L}\cdots\sum\limits_{x_N\in\L}\prod\limits_{i=1}^Nm_{\L}(x_i)\Bigl(\bigotimes\limits_{i=1}^N\P^{\beta}_{x_i,x_{\s(i)}}\Bigr)(L_N\in A),
\end{equation}
where $ m_{\L}(x):=m(x)/m(\L) $ for any $ x\in\L$ defines a probability measure on $ \L $. 
The terminal points $ x_{\s(i)} $ are in $ \L $ as well for any $ \sigma\in\Sym_N, x_i\in\L, 1\le i\le N$. 
We want to employ the following combinatoric scheme. We shall rewrite the sum over permutations with a sum on pair probability measures, where we are only asking for the frequency of transitions from an initial point in $ \L $ to a terminal point in $ \L $. This frequency can be expressed in an easy way by a pair probability measure, i.e. a probability measure on $ \L\times\L $. Let 
\begin{equation} 
\Pmf^{\ssup{N}}_{\L}=\widetilde{\Pmf}(\L\times\L)\cap\frac{1}{N}\N_0^{\L\times\L},
\end{equation} be the set of pair probability measures $ Q $ with equal marginals such that $ NQ(x,y)\in\N_0 $ for any $ Q\in \Pmf^{\ssup{N}}_{\L} $ and any $ x,y\in\L $. Note here, that at most $ N $ entries of $ (Q(x,y))_{x,y\in\L} $ are nonzero for any $ Q\in\Pmf^{\ssup{N}}_{\L} $.
Further, for a configuration $ \overline{x}=(x_1,\ldots,x_{N})\in \L^N $ and given pair probability measure $ Q\in\Pmf^{\ssup{N}}_{\L} $, let
\begin{equation}
\Sym_N(\overline{x},Q)=\{\s\in\Sym_N\colon\,\forall x,y\in\L\colon\sharp\{i\colon x_i=x, x_{\s(i)}=y\}=NQ(x,y)\},
\end{equation} be the set of permutations which are admissible with a given configuration $ \overline{x} $ and pair probability measure $ Q $.
Define the distribution
\begin{equation}
\P_{Q,N}^{\beta}:=\Bigl(\bigotimes\limits_{x,y\in\L}\Bigl(\P^{\beta}_{x,y}\Bigr)^{\otimes N Q(x,y)}\Bigr)
\end{equation}
for any pair probability measure $ Q\in\Pmf^{\ssup{N}}_{\L} $.
Note that
\begin{equation}\label{measurerelation}
\bigotimes_{i=1}^N\P_{x_i,x_{\sigma(i)}}^\beta=\P_{Q,N}^{\beta},\qquad Q\in\Pmf^{\ssup{N}}_{\L},\sigma\in\Sym_N(\overline{x},Q),\overline{x}\in\L^N.
\end{equation}
Clearly the measure in \eqref{measurerelation} does not depend on $ \sigma\in\Sym_N $ as long as $ \sigma\in \Sym_N(\overline{x},Q) $. Furthermore $ Q^{\ssup{1}} $ is the empirical measure of the configuration $ \overline{x}\in\L^N $, hence
$$
\prod_{i=1}^Nm_\L(x_i)=\prod_{x\in\L}m_\L(x)^{N Q^{\ssup{1}}(x)}.
$$
We insert a sum over $ Q\in\Pmf^{\ssup{N}}_{\L} $ in \eqref{estlower1} and continue the estimation in \eqref{estlower1}
\begin{equation}\label{estlower2}
\begin{aligned}
\P^{\ssup{\rm sym}}_{N,\beta}&(L_N\in A)\ge\frac{m(\L)^N}{N!}\sum\limits_{Q\in \Pmf^{\ssup{N}}_{\L}}\sum\limits_{\heap{x_i\in\L,}{1\le i\le N}}\prod\limits_{x\in\L} m_{\L}(x)^{N Q^{\ssup{1}}(x)}\sum\limits_{\s\in\Sym_N(\overline{x},Q)}\Bigl(\bigotimes\limits_{x,y\in\L}\Bigl(\P^{\beta}_{x,y}\Bigr)^{\otimes N Q(x,y)}\Bigr)(L_N\in A)\\
&=m(\L)^N\sum\limits_{Q\in \Pmf^{\ssup{N}}_{\L}}\prod\limits_{x\in\L} m_{\L}(x)^{NQ^{\ssup{1}}(x)} \P_{Q,N}^{\beta}(L_N\in A)\sum\limits_{\heap{x_i\in\L,}{1\le i\le N}}\frac{\sharp\Sym_N(\overline{x},Q)}{N!}.
\end{aligned}
\end{equation}
The measure $ \P_{Q,N}^{\beta} $ is a mixed product of single distributions of the random walks with given starting and terminal points. Recall that $ \P_{x,y}^{\beta} $ is the conditional probability measure for a random walk starting in $ x $ with terminal location $ y $.
Now we need to compute the counting term $$ \sum\limits_{\heap{x_i\in\L,}{1\le i\le N}}\frac{\sharp\Sym_N(\overline{x},Q)}{N!}, $$ 
because the remaining factors in (\ref{estlower2}) do not depend on  $ \overline{x}=(x_1,\ldots,x_N)\in \L^N $. We do this with the help of an additional sum over  configurations $ \omega\in\L^N $ with $ \overline{x}_{\s}=\omega $, where $ \overline{x}_{\s}=(x_{\s(1)},\ldots,x_{\s(N)})\in(\L)^N $ for any permutation $ \s\in\Sym_N $. Thus
\begin{equation}\label{counting}
\begin{aligned} 
\sum_{\overline{x}\in\L^N} \sum_{\sigma\in\Sym_N({\overline{x}},Q)}1
&=\sum\limits_{\overline{x}\in\L^N}\sum_{\omega\in\L^N}\sum_{\sigma\in\Sym_N}
\1_{\{{\overline{x}_{\s}}=\omega\}}(\sigma)\1_{\{\forall x,y\in\L\colon \#\{i\colon x_i=x,\omega_i=y\}=N Q(x,y)\}}(\omega)\\
&=\sum\limits_{\heap{\overline x\in\L^N\colon}{L(\overline{x})=Q^{\ssup{1}}}}\sum\limits_{\heap{\omega\in\L^N\colon}{L(\omega)= Q^{\ssup{1}}}}\sum\limits_{\s\in\Sym_N}\1_{\{{\overline{x}_{\s}}=\omega\}}(\sigma)\1_{\{\forall x,y\in\L\colon \#\{i\colon x_i=x,\omega_i=y\}=N Q(x,y)\}}(\omega),
\end{aligned}
\end{equation} where in the second line we took $ \overline{x}\in\L^N $ according to the marginal $ Q^{\ssup{1}} $, i.e. $ L(\overline{x})=Q^{\ssup{1}} $. Here, $ L(\overline{x})=Q^{\ssup{1}} $ means that the first marginal of $ Q $ equals the empirical measure of the configuration $ \overline{x}$, i.e. 
$$ 
Q^{\ssup{1}}(y)=\frac{1}{N}\sum\limits_{i=1}^N\1_{x_i}(y)\quad\mbox{ for }\;y\in\L.
$$
The same holds for the configurations $ \omega\in\L^N $, i.e. the sum is restricted to $ L(\omega)=Q^{\ssup{1}} $.  
Now we have to count the single terms in \eqref{counting}. We start with the sum over the permutations. For fixed $ \overline{x}\in\L^N $ and fixed $ \omega\in\L^N $ the number of permutations $ \sigma\in\Sym_N $ with $ \overline{x}_{\sigma}=\omega $ is exactly $ \prod_{x\in\L}(NQ^{\ssup{1}}(x))! $, which follows from the fact that for any $ x\in\L $ exactly $ (NQ^{\ssup{1}}(x)) $ times the $ x $ appears in the configuration $ \overline{x} $ and $ \omega $ giving for the first permutation $ NQ^{\ssup{1}}(x) $ possibilities, for the second one $ NQ^{\ssup{1}}(x)-1 $ and so on. The sum over the configurations $ \overline{x}\in\L^N $ with $ Q^{\ssup{1}}=L(\overline{x}) $ is just the multinomial distribution
$$
\binom{N}{NQ^{\ssup{1}}}:=\frac{N!}{\prod\limits_{x\in\L}(NQ^{\ssup{1}}(x))!},
$$ where $ \sum_{x\in\L}NQ^{\ssup{1}}(x)=N $. In \eqref{counting} the sum over the configurations $ \overline{x}\in\L^N $ gives the multinomial factor, because all the other terms do not depend on the configurations $ \overline{x} $ anymore.
Thus, together with the counting over the permutations we get
\begin{equation}
\sum_{\heap{x_i\in\L,}{1\le i\le N}}\sum_{\sigma\in\Sym_N({\overline{x}},Q)}1=N!\sum_{\heap{\omega\in\L^N}{L(\omega)=\overline{x}}}\1_{\{\forall x,y\in\L\colon 
\#\{i\colon x_i=x,\omega_i=y\}=NQ(x,y)\}}(\omega).
\end{equation}
It remains to count the configurations $ \omega $ relative to the given configuration $ \overline x $ and frequency $ NQ(x,y) $ for any $ x,y\in\L.$ This number corresponds to the number of Euler trails in a complete graph or the number of configurations in a microcanonical ensemble specified through the pair measure $ Q $ (cf. \cite{A01} and references therein), and equals
\begin{eqnarray}
\sum_{\heap{\omega\in\L^N}{L(\omega)=\overline{x}}}\1_{\{\forall x,y\in\L\colon 
\#\{i\colon x_i=x,\omega_i=y\}=N Q(x,y)\}}(\omega)=\frac{\prod_{x\in\L}(NQ^{\ssup{1}}(x))!}{\prod_{x,y\in\L} (N Q(x,y))!}.
\end{eqnarray}
Thus we get from (\ref{estlower2})

\begin{equation}\label{estlower4}
\begin{aligned}
\P^{\ssup{\rm sym}}_{N,\beta}(L_N\in A)&\\\ge & m(\L)^N\sum\limits_{Q\in\Pmf^{\ssup{N}}_{\L}}
\prod\limits_{x\in\L}m_{\L}(x)^{NQ^{\ssup{1}}(x)} \frac{\prod_{x\in\L} (NQ^{\ssup{1}}(x))!}{\prod_{x,y\in\L} (N Q(x,y))!} \P_{Q,N}^\beta(L_N\in A).
\end{aligned}
\end{equation}
Applying the Stirling formula for $ N!\approx N^{N+1/2}{\rm e}^{-N}\sqrt{2\pi} $, we get after some computation and estimation
$$
\prod\limits_{x\in\L} m_{\L}(x)^{NQ^{\ssup{1}}(x)}\frac{\prod_{x\in\L}(NQ^{\ssup{1}}(x))!}{\prod_{x,y\in\L} (N Q(x,y))!}\ge {\rm e}^{C|\L|^2\log N}{\rm e}^{-NH(Q|Q^{\ssup{1}}\otimes m_{\L})},
$$ 
where
$$
H(Q|Q^{\ssup{1}}\otimes m_{\L})=\sum\limits_{x,y\in\L}Q(x,y)\log\frac{Q(x,y)}{\overline{Q}(x) m(y)}
$$ is the relative entropy of the pair probability measure $ Q\in\widetilde{\Pmf}(\L\times\L) $ relative to the probability measure $ Q^{\ssup{1}}\otimes m_{\L}\in \widetilde{\Pmf}(\L\times\L)$, and
where $ C > 0 $ is an absolute constant.
Hence,
\begin{equation}\label{estlower3}
\P^{\ssup{\rm sym}}_{N,\beta}(L_N\in A)\ge m(\L)^N{\rm e}^{-C|\L|^2\log N}\sum\limits_{Q\in\Pmf^{\ssup{N}}_{\L}} {\rm e}^{-N H(Q|Q^{\ssup{1}}\otimes m_{\L})}\P_{Q,N}^{\beta}(L_N\in A).
\end{equation} 

In \eqref{estlower3} there is a sum over all pair measures in $ \Pmf^{\ssup{N}}_{\L} $. For any pair measure $ Q\in\widetilde \Pmf(\Z^d\times\Z^d) $ with finite relative entropy $ H(Q|\overline{Q}\otimes m) $ and finite functional $ I^{\ssup {Q}}_{\beta} $ we need an approximation with a sequence $ (Q_N)_{N\in\N} $ of pair probability  measures $ Q_N\in \Pmf^{\ssup{N}}_{\L} $ converging to $ Q $ as $ N\to\infty $ in an appropriate way. Additionally, we let $ (\L_N)_{N\in\N} $ be a sequence of sub boxes $ \L_N\subset\Z^d $ such that $ \L_N\uparrow \Z^d$ as $ N\to\infty$. 

\noindent This will be done in the proof of the following proposition.

\begin{prop}\label{mainlowerboundproposition}  Let $ (\L_N)_{N\in\N} $ be a sequence of sub boxes $ \L_N\subset\Z^d $ such that $ \L_N\uparrow \Z^d$ as $ N\to\infty $ and such that there is  $ \eps>0 $ with $ |\L_N|^3/N=N^{-\eps} $ and $ 4|\L_N|^2/N\le \frac{1}{2} $ for all $ N\in\N $. Then for any open set $ A\subset\Pmf(D_{\beta}) $ we have
\begin{equation}\label{proplower}
\liminf_{N\to\infty}\frac{1}{N}\log\sum\limits_{Q\in\Pmf^{\ssup{N}}_{\L_N}} {\rm e}^{-N H(Q|Q^{\ssup{1}}\otimes m_{\L_N})}\P_{Q,N}^{\beta}(L_N\in A)\ge -\inf\limits_{\mu\in A}\inf\limits_{Q\in\widetilde \Pmf (\Z^d\times\Z^d)}\Bigl\{H(Q|Q^{\ssup{1}}\otimes m)+I^{\ssup{Q}}_{\beta}(\mu)\Bigr\}.
\end{equation}
\end{prop}

\begin{proofsect}{Proof}
Let $ A\subset\Pmf(D_\beta) $ be an open set, and let $ \min A $ and $ Q\in\widetilde{\Pmf}(\Z^d\times\Z^d) $ be given. In the case $ H(Q|Q^{\ssup{1}}\otimes m)=+\infty $ or $ J^{\ssup{Q}}_{\beta}(\mu)=+\infty  $ the assertion \eqref{proplower} follows immediately.
We assume therefore that the relative entropy $ H(Q|Q^{\ssup{1}}\otimes m) $ and the functional $ J^{\ssup{Q}}_{\beta}(\mu) $ are finite. The first assumption implies that $ \supp\,(Q)\subset(\supp\,(m)\times\supp\,(m)) $, because 
\begin{equation}\label{rewriterelentropy}
H(Q|Q^{\ssup{1}}\otimes m)=H(Q|Q^{\ssup{1}}\otimes Q^{\ssup{1}})+H(Q^{\ssup{1}}|m),
\end{equation}
where $ H(Q^{\ssup{1}}|m)= \sum_{x\in\Z^d}Q^{\ssup{1}}(x)\log\frac{Q^{\ssup{1}}(x)}{m(x)}.$  Hence, $ Q\in\widetilde\Pmf(\supp\,(m)\times\supp\,(m)) $.

\noindent Our strategy is as follows: \\[1ex]
\noindent{\bf Step 1:} In the first step we construct a sequence $ (Q_N^{\ssup{N}})_{N\in\N} $ of pair probability measures $ Q_N^{\ssup{N}}\in\Pmf^{\ssup{N}}_{\L_N} $, such that $ Q_N^{\ssup{N}}\to Q $ weakly in the sense of probability measures as $ N\to\infty $ and such that the

\noindent {\bf Step 2:} lower bound 
\begin{equation}\label{assertionlower1}
\liminf\limits_{N\to\infty}\frac{1}{N}\log\P_{Q^{\ssup{N}}_{N},N}^{\beta}(L_N\in A)\ge - I^{\ssup{Q}}_{\beta}(\mu)
\end{equation}
for any $ \mu\in A $ follows from a lower bound of a large deviations principle for the distribution of $ L_N $ under the measure $ \P_{Q_N^{\ssup{N}},N}^\beta $, and that

\noindent {\bf Step 3:}
\begin{equation}\label{assertionlower2}
\liminf\limits_{N\to\infty}\frac{1}{N}\log\Big({\rm e}^{-N H(Q^{\ssup{N}}_N|Q^{\ssup{N,(1)}}_N\otimes m_{\L_N})}\Big)\ge -H(Q|Q^{\ssup{1}}\otimes m).
\end{equation}

We will start with step 1.

\noindent {\bf Step 1: Construction of pair probability measures $ \mathbf{Q_N^{\ssup{N}}\in\Pmf^{\ssup{N}}_{\L_N}}$}

This construction is splitted into two parts. First we construct for a given pair probability measure on $ \Z^d\times\Z^d $ with equal marginals a pair probability measure on $ \L_N\times\L_N $ with equal marginals such that it converges weakly to the given pair probability measure on $ \Z^d\times\Z^d $. This is done in Lemma~\ref{marginalconstructionlemma}. Ones we have a sequence of  pair probability measures on $ \L_N\times\L_N $ with equal marginals we construct for this sequence a second sequence also with equal marginals but such that their single entries are integers when multiplied by $ N $. 

\begin{lemma}[{\bf Marginal construction}]\label{marginalconstructionlemma}
Let $ Q\in\widetilde\Pmf(\Z^d\times\Z^d) $, $ \L_N\subset\Z^d $ and $ x_0\in\L_N $ and $ (\eta_N)  $ a sequence in $ (0,1) $ with $ \eta_N\to 0 $ as $ N\to\infty $. Define the $ |\L_N|^2 $ entries of a function $ Q_N\colon\L_N\times\L_N\to[0,1] $ as 
\begin{equation}\label{defentriesmarginal} 
Q_N(x,y):=\left\{\begin{array}{l@{\,,\mbox{ if }  \,}l} Q(x,y) & x,y\in\L_N\setminus\{x_0\}\\[1.5ex]
Q^{\ssup{1}}(x)-\sum_{z\in\L_N\setminus\{x_0\}}Q_N(x,z) & x\in\L_N\setminus\{x_0\}, y=x_0\\[1.5ex]
Q^{\ssup{1}}(y)-\sum_{z\in\L_N\setminus\{x_0\}}Q_N(z,y) & y\in\L_N\setminus\{x_0\}, x=x_0\\[1.5ex]
1-\sum_{\heap{(x,y)\in\L_N^2\colon}{(x,y)\not=(x_0,x_0)}} Q_N(x,y) & (x,y)=(x_0,x_0).
                 \end{array}\right.
\end{equation}
Then the following holds
\begin{enumerate}
\item[(i)] $ Q_N\in\widetilde\Pmf(\L_N\times\L_N) $

\item[(ii)] Let $ (\L_N)_{N\in\N} $ be a sequence of boxes $ \L_N\subset\Z^d $ with $ \L_N\uparrow \Z^d $ as $ N\to\infty $. Then $ Q_N\to Q $ strongly as $ N\to\infty $.
\item[(iii)] $ Q_N(x_0,x_0) \ge \eta_N $ for all $ N\in\N $. 
\end{enumerate}
\end{lemma}

\begin{proofsect}{Proof}
(i) Clearly $ Q_N(x,y)\in[0,1] $ for all $ x,y\in\L_N\setminus\{x_0\} $, and $ Q_N(x,x_0)=\sum_{z\in\L_N^{\rm c}\cup\{x_0\}}Q(x,z)\in[0,1] $ and $ Q_N(x_0,x)=\sum_{z\in\L_N^{\rm c}\cup\{x_0\}}Q(z,x) \in[0,1] $ for all $ x\in\L_N\setminus\{x_0\} $. Also
$$
Q_N(x_0,x_0)=1-\Big(\sum_{x,y\in\L_N\setminus\{x_0\}}Q(x,y)+\sum_{x\in\L_N\setminus\{x_0\}}\sum_{y\in\L_N^{\rm c}\cup\{x_0\}}\big(Q(x,y)+Q(y,x)\big)\Big)\in[0,1]\quad,
$$ because the terms in the brackets are bounded by $ \sum_{x,y\in\Z^d}Q(x,y)$.
All $ |\L_N|^2 $ entries of the function $ Q_N $ sum up to one, because
$$
\begin{aligned}
\sum_{x,y\in\L_N}Q_N(x,y)&=\sum_{x,y\in\L_N\setminus\{x_0\}} Q(x,y)+\sum_{x\in\L_N\setminus\{x_0\}}\Big(2Q^{\ssup{1}}(x)-\sum_{y\in\L_N\setminus\{x_0\}}\big(Q(x,y)+Q(y,x)\big)\Big)\\& \quad +Q_N(x_0,x_0)=1.
\end{aligned}
$$

For any $ x\in\L_N\setminus\{x_0\} $ we get
$$
\begin{aligned}
Q^{\ssup{1}}_N(x)&=Q_N(x,x_0)+\sum_{y\in\L_N\setminus\{x_0\}}Q(x,y)=Q^{\ssup{1}}(x)-\sum_{y\in\L_N\setminus\{x_0\}}Q_N(x,y)+\sum_{y\in\L_N\setminus\{x_0\}}Q(x,y)\\
&=Q^{\ssup{1}}(x)=Q^{\ssup{2}}(x)=Q_N(x_0,x)+\sum_{y\in\L_N\setminus\{x_0\}}Q_N(y,x)\\
&=Q^{\ssup{2}}_N(x);
\end{aligned}
$$
and
$$
\begin{aligned}
Q^{\ssup{2}}_N(x_0)&=Q_N(x_0,x_0)+\sum_{y\in\L_N\setminus\{x_0\}}Q(y,x_0)=\sum_{y\in\L_N\setminus\{x_0\}}\Big(Q^{\ssup{1}}(y)-\sum_{z\in\L_N\setminus\{x_0\}}Q(y,z)\Big)\\ &\quad +Q_N(x_0,x_0)\\
&=1-\Big(\sum_{x,y\in\L_N\setminus\{x_0\}}Q(x,y)+\sum_{x\in\L_N\setminus\{x_0\}}\Big(2Q^{\ssup{1}}(x)-\sum_{y\in\L_N\setminus\{x_0\}}\big(Q(x,y)+Q(y,x)\big)\Big)\Big)\\&\quad+\sum_{y\in\L_N\setminus\{x_0\}}\Big(Q^{\ssup{1}}(y)-\sum_{z\in\L_N\setminus\{x_0\}}Q(y,z)\Big)\\
&=1-\sum_{x\in\L_N\setminus\{x_0\}}Q^{\ssup{1}}(x)=1-\sum_{x\in\L_N\setminus\{x_0\}}Q^{\ssup{1}}_N(x)=
Q^{\ssup{1}}_N(x_0).
\end{aligned}
$$
Hence $ Q_N\in\widetilde\Pmf(\L_N\times\L_N) $.

\noindent (ii)
First, a direct computation gives
$$
\begin{aligned}
Q_N(x_0,x_0)&=1-2\sum_{z\in\L_N\setminus\{x_0\}}Q^{\ssup{1}}(z)+\sum_{y,z\in\L_N\setminus\{x_0\}}Q(z,y)\\
&=\sum_{z\in\L_N^{\rm c}\cup\{x_0\}}Q^{\ssup{1}}(z)-\sum_{y\in\L_N\setminus\{x_0\}}Q^{\ssup{1}}(y)+\sum_{y,z\in\L_N\setminus\{x_0\}}Q(z,y)\\
&=\sum_{z\in\L_n^{\rm c}\cup\{x_0\}}Q^{\ssup{1}}(z)-\sum_{\heap{y\in\L_N\setminus\{x_0\}}{y\in\L_N^{\rm c}\cup\{x_0\}}} Q(y,z)\\
&=\sum_{y,z\in\L_N^{\rm c}\cup\{x_0\}}Q(x,y),
\end{aligned}
$$
this gives
\begin{equation}\label{estfixedpoint1}
|Q_N(x_0,x_0)-Q(x_0,x_0)|=\sum_{y,z\in\L_N^{\rm c}}Q(x,y),
\end{equation}
and hence
$$
\begin{aligned}
\sum_{x,y\in\Z^d}|Q(x,y)-Q_N(x,y)|&\le \sum_{(x,y)\in(\L_N\times\L_N)^{\rm c}}Q(x,y)+\sum_{x\in\L_N\setminus\{x_0\}}\sum_{y\in\L_N^{\rm c}} \big(Q(x,y)+Q(y,x)\big)\\
&\le 2  \sum_{(x,y)\in(\L_N\times\L_N)^{\rm c}}Q(x,y).
\end{aligned}
$$
The assertion then follows since for any $ \eps>0 $ there is $ N_0\in\N $ such that
$$
\sum_{x,y\in\Z^d}|Q(x,y)-Q_N(x,y)|<\eps\quad\mbox{for } N\ge N.
$$
was arbitrary.

\noindent (iii) If $ Q_N(x_0,x_0)=\sum_{y,z\in\L_N^{\rm c}\cup\{x_0\}}Q(x,y) <\eta_N $ for all $ N\in\N $ we will multiply all the entries $ Q_N(x,y) $ for $ (x,y)\in (\L_N\times\L_N)\setminus\{x_0,x_0\} $ with the factor $ \alpha_N=(1-\eta_N)/(1-Q_N(x_0,x_0)) $. The resulting entries are then denoted by $ \widetilde Q_N(x,y) $. Clearly, $ \widetilde Q_N(x_0,x_0)=1-\sum_{(x,y)\in\L_N^2\setminus\{x_0,x_0\}}\widetilde Q_N(x,y)\ge \eta_N $ and $ \alpha_N\to 0 $ as $ N\to\infty $. As
$$
\sum_{x,y\in\Z^d}|Q(x,y)-\widetilde Q_N(x,y)|\le  2  \sum_{(x,y)\in(\L_N\times\L_N)^{\rm c}}Q(x,y)+(1-\alpha_N),
$$
all requirements in (i)-(ii) are satisfied by $ \widetilde Q_N $.
\qed
\end{proofsect}

\noindent We now construct pair probability measures $ Q_N^{\ssup{N}}\in\Pmf^{\ssup{N}}_{\L_N} $. We apply the previous Lemma~\ref{marginalconstructionlemma} for the choice $\eta_N=(|\L_N|-1)^2/N $.
Fix $ N\in\N $ and denote the pair probability measure $ Q_N $ in Lemma~\ref{marginalconstructionlemma} simply by $ Q $.
Hence, we have the property that here is $ x_0\in\L_N $ such that
\begin{equation}\label{assumptionsingleentry}
Q(x_0,x_0)\ge\frac{(|\L_N|-1)^2}{N}.
\end{equation}

Now we define $ |\L_N|^2-|\L_N|=:\nu $ components of a vector in $ [0,1]^{\nu} $ which satisfy the  conditions (1)-(4) of \eqref{conditionspairmeasure} in Lemma~\ref{pairmeasurelem}. These coordinates define then according to Lemma~\ref{pairmeasurelem} uniquely a pair probability measure in $ \Pmf^{\ssup{N}}_{\L_N} $ when the single components are elements of $ \{0,1/N,\ldots,(N-1)/N,1\} $.
Now $ (|\L_N|-1)^2 $ components are defined as
\begin{equation}\label{deffirstcomponents}
Q_N^{\ssup{N}}(x,y):=\frac{\floor{NQ(x,y)}}{N}\quad \;\mbox{ for all }\; x,y\in\L_N\setminus\{x_0\},
\end{equation}
where $ \floor{x} $ is the largest integer smaller or equal to $ x\in\R $.
The remaining $ |\L_N|-1 $ coordinates are given as
\begin{equation}\label{defsecondcomponents}
Q_N^{\ssup{N,(1)}}(x):=\frac{\floor{NQ^{\ssup{1}}(x)}}{N}\quad, \;\mbox{ for all }\; x\in\L_N\setminus\{x_0\},
\end{equation}
where $ x_0\in\L_N $ as in Lemma~\ref{marginalconstructionlemma} and in  \eqref{assumptionsingleentry}.
We check that the conditions (1)-(4) in \eqref{conditionspairmeasure} of Lemma~\ref{pairmeasurelem} are satisfied, where we order the coordinates in an obvious way. Since $ Q_N^{\ssup{N,(1)}}(x)\le Q^{\ssup{1}}(x) $ for all $ x\in\L_N\setminus\{x_0\} $ we have $ \sum_{x\in\L_N\setminus\{x_0\}}Q_N^{\ssup{N,(1)}}(x)\le 1 $, i.e. (1) of \eqref{conditionspairmeasure} is satisfied.
For any $ x,y\in\R_+ $ one has $ \floor{x+y}\ge \floor{x}+\floor{y} $. Therefore,
$ Q^{\ssup{1}}(x)\ge\sum_{y\in\L_N\setminus\{x_0\}}Q(x,y) $ and $ Q^{\ssup{1}}(x)=Q^{\ssup{2}}(x)\ge \sum_{y\in\L_N\setminus\{x_0\}}Q(y,x) $ for all $ x\in\L_N\setminus\{x_0\} $ imply that
$$
Q_N^{\ssup{N,(1)}}(x)\ge \sum_{y\in\L_N\setminus\{x_0\}}Q_N^{\ssup{N}}(x,y)\;\mbox{ and }\; Q_N^{\ssup{N,(1)}}(x)\ge \sum_{y\in\L_N\setminus\{x_0\}}Q_N^{\ssup{N}}(y,x)
$$ for all $ x\in\L_N\setminus\{x_0\} $. Hence, also (2) and (3) of \eqref{conditionspairmeasure} are satisfied.
For $ x,y\in\L_N\setminus\{x_0\} $ we have $Q_N^{\ssup{N}}(x,y) \ge Q_N(x,y) -\frac{1}{N}=Q(x,y)-\frac{1}{N}. $ Notice further that from the proof of Lemma~\ref{marginalconstructionlemma} and our assumption \eqref{assumptionsingleentry} we have
\begin{equation}\label{esttimationlastpoint}
\begin{aligned}
Q_N(x_0,x_0)&=1-\sum_{x\in\L_N\setminus\{x_0\}}2Q^{\ssup{1}}(x)+\sum_{x,y\in\L_N\setminus\{x_0\}}Q(x,y)=Q^{\ssup{1}}(x_0)-\sum_{x\in\L_N\setminus\{x_0\}}Q(x,x_0)\\
&=Q(x_0,x_0) \ge \frac{(|\L_N|-1)^2}{N}.
\end{aligned}
\end{equation}
Hence, an estimation and application of \eqref{esttimationlastpoint} gives
$$
\begin{aligned}
\sum_{x\in\L_N\setminus\{x_0\}}&\Big(2Q_N^{\ssup{N,(1)}}(x)-\sum_{y\in\L_N\setminus\{x_0\}}Q_N^{\ssup{N}}(x,y)\Big)\le \sum_{x\in\L_N\setminus\{x_0\}}\Big(2Q^{\ssup{1}}(x)-\sum_{y\in\L_N\setminus\{x_0\}}Q(x,y)\Big)\\&+\frac{(|\L_N|-1)^2}{N}\le 1,
\end{aligned}
$$ and therefore (4) of \eqref{conditionspairmeasure}. We define the remaining $ 2|\L_N|-1 $ entries of the function $ Q_N^{\ssup{N}} $, see the proof of Lemma~\ref{pairmeasurelem}, as follows
\begin{equation}\label{defremainingentries}
\begin{aligned}
Q_N^{\ssup{N}}(x,x_0)&=Q_N^{\ssup{N,(1)}}(x)-\sum_{y\in\L_N\setminus\{x_0\}}Q_N^{\ssup{N}}(x,y)\\ Q_N^{\ssup{N}}(x_0,x)&=Q_N^{\ssup{N,(1)}}(x)-\sum_{y\in\L_N\setminus\{x_0\}}Q_N^{\ssup{N}}(y,x)\quad\mbox{ for } x\in\L_N\setminus\{x_0\},\\
Q_N^{\ssup{N}}(x_0,x_0)&=1-\Big(\sum_{x\in\L_N\setminus\{x_0\}}2Q_N^{\ssup{N,(1)}}(x)-\sum_{x,y\in\L_N\setminus\{x_0\}}Q_N^{\ssup{N}}(x,y)\Big).
\end{aligned}
\end{equation}

Then the components in \eqref{deffirstcomponents} and  in \eqref{defremainingentries} define uniquely a pair probability measure $ Q_N^{\ssup{N}}\in\Pmf^{\ssup{N}}_{\L_N} $ for $ N\in\N $.

\noindent {\bf Weak convergence of $ (Q_N^{\ssup{N}})_{N\in\N} $}

\noindent For $ x,y\in\L_N\setminus\{x_0\} $ we have
$$
\big|Q_N^{\ssup{N}}(x,y)-Q(x,y)\big|\le\frac{1}{N},
$$
and from this we get
\begin{equation}\label{est1QN}
\begin{aligned}
Q_N^{\ssup{N}}(x,x_0)-Q(x,x_0)&\le\big|Q_N^{\ssup{N,(1)}}(x)-Q^{\ssup{1}}(x)\big|+\sum_{y\in\L_N\setminus\{x_0\}}\big|Q(x,y)-Q_N^{\ssup{N}}(x,y)\big|\\ &\le \frac{1}{N}+\sum_{y\in\L_N\setminus\{x_0\}}\big|Q(x,y)-Q_N^{\ssup{N}}(x,y)\big|\le\frac{|\L_N|}{N}
\end{aligned}
\end{equation}
and analogously
\begin{equation}\label{est2QN}
Q_N^{\ssup{N}}(x_0,x)-Q(x_0,x)\le \frac{|\L_N|}{N}\quad\mbox{ for any }\;x\in\L_N\setminus\{x_0\}.
\end{equation}
Moreover, with \eqref{estfixedpoint1} we get
\begin{equation}\label{est3QN}
\begin{aligned}
Q_N^{\ssup{N}}(x_0,x_0)-Q(x_0,x_0)&\le |Q_N^{\ssup{N}}(x_0,x_0)-Q_N(x_0,x_0)|+|Q_N(x_0,x_0)-Q(x_0,x_0)|\\
&\le \frac{|\L_N|^2}{N}+\sum_{(x,y)\in (\L_N\times\L_N)^{\rm c}} Q(x,y),
\end{aligned}
\end{equation}
and therefore
\begin{equation}\label{estfinalconstruction}
\sum_{x,y\in\L_N}\big|Q_N^{\ssup{N}}(x,y)-Q(x,y)\big|\le \frac{2|\L_N|^2}{N} +\sum_{(x,y)\in (\L_N\times\L_N)^{\rm c}}Q(x,y),
\end{equation}
and the assertion follows.

\noindent {\bf Step 2: Lower bound via G\"artner-Ellis Theorem}
We are going to use the G\"artner-Ellis Theorem to deduce that
$$
\liminf\limits_{N\to\infty}\frac{1}{N}\log \P_{Q_{N}^{\ssup{N}},N}^{\beta}(L_N\in A)\ge I^{\ssup{Q}}_{\beta}(\mu).
$$
For doing this, we evaluate first the logarithmic moment generating function for a given $ F\in\Ccal_{\rm b}(D_{\beta}) $, where we take the dual pairing $ \Pmf(D_{\beta}) $ with $ \Ccal_{\rm b}(D_{\beta}) $ (\cite[Lemma 3.2.3]{DS01}).
$$
\begin{aligned}
\L(F)&:=\lim\limits_{N\to\infty}\frac{1}{N}\log\E_{Q_{N}^{\ssup{N}},N}^{\beta}\big({\rm e}^{N\langle F,L_N\rangle}\big)=\lim\limits_{N\to\infty}\frac{1}{N}\log\Bigl(\prod\limits_{x,y\in\L_N}\E_{x,y}^{\beta}\big({\rm e}^{F(\xi)}\big)^{NQ_{N}^{\ssup{N}}(x,y)}\Bigr)\\
&=\lim\limits_{N\to\infty}\sum\limits_{x,y\in\L_N}Q_{N}^{\ssup{N}}(x,y)\log\E_{x,y}^{\beta}\big({\rm e}^{F(\xi)}\big)\\
&=\lim\limits_{N\to\infty}\sum\limits_{x,y\in\L_N}Q_{N}^{\ssup{N}}(x,y)\log\E_{x,y}^{\beta}\big({\rm e}^{F(\xi)}\big)=\sum\limits_{x,y\in\Z^d}Q(x,y)\log\E_{x,y}^{\beta}\big({\rm e}^{F(\xi)}\big),
\end{aligned}
$$
because the limit and the sum can be interchanged, and recall that $ Q_{N}\to Q $ (weakly) in sense of probability measures as $ N\to\infty $. Hence, $ \Lambda(F) $ exists, and
$$
\Lambda(F)=\sum\limits_{x,y\in\Z^d}Q(x,y)\log\E_{x,y}^{\beta}\big({\rm e}^{F(\xi)}\big)\;\mbox { for all }\; F\in\Ccal_{\rm b}(D_{\beta}).
$$
Also, it is easily seen that $ \L $ is lower semi continuous and G\^ateax differentiable. Thus, \cite[4.5.27]{DZ98} together with the exponential tightness for the sequence 
$ (\P_{Q_{N}^{\ssup{N}},N}^{\beta}\circ L_N^{-1})_{N\in\N} $, which is derived in Lemma \ref{exptightnessprod}, gives
$$
\liminf\limits_{N\to\infty}\P_{Q_{N}^{\ssup{N}},N}^{\beta}(L_N\in A)\ge - I^{\ssup{Q}}_{\beta}(\mu),
$$
where
$$
I^{\ssup{Q}}_{\beta}(\mu)=\sup\limits_{F\in\Ccal_{\rm b}(D_{\beta})}\big\{\langle F,\mu\rangle-\L(F)\big\},\quad\mu\in\Pmf(D_{\beta}).
$$

\noindent {\bf Step 3: Estimation for the relative entropy}
We are going to show \eqref{assertionlower2}, i.e.
$$
\liminf\limits_{N\to\infty} -H(Q^{\ssup{N}}_N|Q^{\ssup{N,(1)}}_N\otimes m_{\L_N})\ge -H(Q|Q^{\ssup{1}}\otimes m).
$$
We are going to estimate the relative entropy of the pair probability measure $ Q_N^{\ssup{N}} $ with respect to $ Q_N^{\ssup{N,(1)}}\otimes m_N $. We intend to bound this relative entropy from above by the relative entropy of the restriction $ Q|_{\L_N} $ of $ Q $ onto $ \L_N\times\L_N $ with respect to the measure $ Q|_{\L_N}^{\ssup{1}}\otimes m_N $ plus some error terms. The latter relative entropy is known to converge to the relative entropy of $ Q $ with respect to $ \overline{Q}\otimes m $, see \cite[Lemma~4.4.15]{DS01}. We do this most conveniently with the help of the entropy
$$
H(Q_N^{\ssup{N}})=-\sum_{x,y\in\L_N}Q_N^{\ssup{N}}(x,y)\log Q_N^{\ssup{N}}(x,y) 
$$ of the probability measure $ Q_N^{\ssup{N}} $, and a well-know entropy estimation, which we cite and prove in Lemma~\ref{entropyestimation} in the appendix. We write 
\begin{equation}\label{entropyestimation1}
\begin{aligned}
H(Q_N^{\ssup{N}}|&Q_N^{\ssup{N,(1)}}\otimes m_N)= -H(Q_N^{\ssup{N}})-\sum_{x,y\in\L_N}Q_N^{\ssup{N}}(x,y)\log Q_N^{\ssup{N,(1)}}(x)m_N(y)\\
\le & -H(Q|_{\L_N})+|H(Q|_{\L_N})-H(Q_N^{\ssup{N}})|-\sum_{x,y\in\L_N}Q_N^{\ssup{N}}(x,y)\log Q_N^{\ssup{N,(1)}}(x)\otimes m_N(y)\\
\le & -H(Q|_{\L_N}) -\sum_{x,y\in\L_N}Q_N^{\ssup{N}}(x,y)\log Q_N^{\ssup{N,(1)}}(x)m_N(y) +\frac{4|\L_N|^2}{N}\log 4N,
\end{aligned}
\end{equation}
where we applied Lemma~\ref{entropyestimation} with our assumption that $ \frac{4|\L_N|^2}{N}\le \frac{1}{2} $, the estimate
\begin{equation}\label{errorentropy}
\sum_{x,y\in\L_N}|Q_N^{\ssup{N}}(x,y)-Q|_{\L_N}(x,y)|\le \frac{4|\L_N|^2}{N},
\end{equation} which follows from \eqref{est1QN},\eqref{est2QN} and \eqref{est3QN}, and a look at our previous assumption \eqref{assumptionsingleentry}. Hence the entropy difference is bounded by $ 4|\L_N|^2\log 4N/N$. Note
$$
-\sum_{x,y\in\L_N}Q_N^{\ssup{N}}(x,y)\log Q_N^{\ssup{N,(1)}}(x)m_N(y)=-\sum_{\heap{x,y\in\L_N,}{Q^{\ssup{1}}(x)\ge \frac{1}{N}}}Q_N^{\ssup{N}}(x,y)\log Q_N^{\ssup{N,(1)}}(x)m_N(y).
$$
Further
\begin{equation}\label{entropyestimation2}
\begin{aligned}
-\sum_{\heap{x,y\L_N,}{Q^{\ssup{1}}(x)\ge \frac{1}{N}}}&Q_N^{\ssup{N}}(x,y)\log Q_N^{\ssup{N,(1)}}(x)m_N(y)\le 
-\sum_{\heap{x,y\in\L_N,}{Q^{\ssup{1}}(x)\ge \frac{1}{N}}}Q|_{\L_N}(x,y)\log Q_N^{\ssup{N,(1)}}(x)m_N(y)\\
& \le -\sum_{\heap{x,y\in\L_N,}{Q^{\ssup{1}}(x)\ge \frac{1}{N}}}Q|_{\L_N}(x,y)\log Q^{\ssup{1}}|_{\L_N}(x)m_N(y)
+\Big|\sum_{\heap{x\in\L_N,}{Q^{\ssup{1}}(x)\ge \frac{1}{N}}}Q|_{\L_N}^{\ssup{1}}(x)\log\frac{Q_N^{\ssup{N,(1)}}(x)}{Q|_{\L_N}^{\ssup{1}}(x)}\Big|\\
\le & -\sum_{x,y\in\L_N} Q|_{\L_N}(x,y)\log Q|_{\L_N}^{\ssup{1}}(x)m_N(y) +\frac{4|\L_N|^3}{N},
\end{aligned}
\end{equation}
because of
$$
Q^{\ssup{N,(1)}}_N(x_0)-Q^{\ssup{1}}(x)\le \frac{4|\L_N|^3}{N}.
$$

\noindent Together with \eqref{entropyestimation1} and \eqref{entropyestimation2} we get finally the desired upper bound

\begin{equation}\label{entropyestimation3}
\begin{aligned}
H(Q_N^{\ssup{N}}|Q_N^{\ssup{N,(1)}}\otimes m_N)&\le -H(Q|_{\L_N})-\sum_{x,y\in\L_N}Q|_{\L_N}(x,y)\log Q|_{\L_N}^{\ssup{N,(1)}}(x)m_N(y)
+\Ocal(N^{\eps})\\
&=H(Q|_{\L_N}|Q|_{\L_N}^{\ssup{1}}\otimes m_N)+\Ocal(N^{\eps}).
\end{aligned}
\end{equation}
Clearly the right hand side of \eqref{entropyestimation3} converges to $ H(Q|Q^{\ssup{1}}\otimes m) $ as $ N\to\infty $.
Hence the assertion \eqref{assertionlower2} follows.

\noindent{\bf Finish of the proof of the proposition}

\noindent Now, \eqref{assertionlower1} and \eqref{assertionlower2} show that the left hand side of \eqref{proplower} is not smaller than 
$$ 
-H(Q|Q^{\ssup{1}}\otimes m)-I^{\ssup{Q}}_{\beta}(\mu). 
$$ 
Since this is the case for any $ \mu\in A $ and any pair measure $ Q\in\widetilde{\Pmf}(\Z^d\times\Z^d) $, the proof is finished. 

                                                      \hfill $ \hfill \Box $
\end{proofsect}

\noindent{\bf Finish of the proof of the lower bound of Theorem \ref{mainthm1}}

\noindent Now, Proposition \ref{mainlowerboundproposition} gives the desired lower bound for Theorem \ref{mainthm1} when we apply it to the derived inequality in \eqref{estlower3}. Note that
$$
\liminf\limits_{N\to\infty}\frac{1}{N}\log \Big(m(\L_N)^N{\rm e}^{-C|\L_N|^2\log N}\Big) =0
$$ 
due to our assumption that $ |\L_N|^3/N=N^{-\eps} $.
Thus,
$$
\liminf\limits_{N\to\infty}\log\P^{\ssup{\rm sym}}_{N,\beta}(L_N\in A)\ge -\inf\limits_{\mu\in A} I^{\ssup{\rm sym}}_{\beta}(\mu).
$$       \hfill $ \Box $

\subsubsection{Proof of the upper bound of theorem \ref{mainthm1}}\label{secupperboundmain}
For the upper bound we start with a finite box $ \L\subset\Z^d $. Later we will perform the limit $ \L\uparrow\Z^d $. The main task is to estimate the probability of the two events and to apply the combinatoric scheme introduced in the proof of the lower bound. Here, we are faced with the problem that a random walk may start in $ \L $ and terminate in $ \L $ or the complement $ \L^{\rm c} $, or start outside $ \L $ and terminate outside or inside the box $ \L $. All events have to be estimated. To start, fix a closed set $ F \subset \Pmf(D_{\beta}) $, a box $ \L \subset\Z^d $ and $ \eps>0 $. 

\noindent We split our sum over the $ N $ initial points of the random walks into a sum over the initial points in the box $ \L $ and over the  complement $ \L^{\rm c} $ for each single random walk. Thus we write
\begin{equation}
\sum\limits_{x_1\in\Z^d}\cdots\sum\limits_{x_N\in\Z^d}=\sum\limits_{a\in\{1,{\rm c}\}^N}\sum\limits_{x_1\in \L^{a_1}}\cdots\sum\limits_{x_N\in\L^{a_N}}.
\end{equation}
The estimate the probability $ \P^{\beta}_{N,\beta}(L_N\in F) $ from above, we write the probability as the sum of the following two events. The first event is the probability with additional indicator that the number of random walks not starting in $ \L $ is more than $ \eps N $, whereas the second event is the complement, i.e. the same probability with indicator that the number of random walks starting in $ \L $ is greater or equal than $ (1-\eps)N $. Thus formally
\begin{equation}
\sum\limits_{x_1\in\Z^d}\cdots\sum\limits_{x_N\in\Z^d}=\sum\limits_{\heap{a\in\{1,{\rm c}\}^N,}{\sharp\{i\colon a_i={\rm c}\}\ge \eps N}}\sum\limits_{x_1\in \L^{a_1}}\cdots\sum\limits_{x_N\in\L^{a_N}}+\sum\limits_{\heap{a\in\{1,{\rm c}\}^N,}{\sharp\{i\colon a_i=1\}\ge (1-\eps)N}}\sum\limits_{x_1\in \L^{a_1}}\cdots\sum\limits_{x_N\in\L^{a_N}}.
\end{equation}
The probability for the first event can be bounded from above by the factor $ 2^Nm(\L^{\rm c})^{\eps N} $. For the probability of the second event we have to sum over all possible subsets $ I\subset\{1,\ldots,N\} $ with $ |I|\ge (1-\eps)N $, hence
\begin{equation}\label{upperboundmain1}
\begin{aligned}
&\P^{\ssup{\rm sym}}_{N,\beta}(L_N\in F)=\frac{1}{N!}\sum\limits_{\s\in\Sym_N}\sum\limits_{a\in\{1,{\rm c}\}^N}\sum\limits_{x_1\in \L^{a_1}}\cdots\sum\limits_{x_N\in\L^{a_N}}\prod\limits_{i=1}^Nm(x_i)\Bigl(\bigotimes\limits_{i=1}^N\P^{\beta}_{x_i,x_{\s(i)}}\Bigr)(L_N\in F)\\
&\le 2^Nm(\L^{\rm c})^{\eps N} \\
&+\frac{1}{N!}\sum\limits_{\s\in\Sym_N}\sum\limits_{\heap{I\subset\{1,\ldots,N\}}{|I|>1-\eps N}}\sum\limits_{\heap{a\in\{1,{\rm c}\}^N\colon}{I=\{i\colon a_i=1\}}}\sum\limits_{x_1\in \L^{a_1}}\cdots\sum\limits_{x_N\in\L^{a_N}}\prod\limits_{i=1}^Nm(x_i)\Bigl(\bigotimes\limits_{i=1}^N\P^{\beta}_{x_i,x_{\s(i)}}\Bigr)(L_N\in F).
\end{aligned}
\end{equation}
We consider only those random walks in the product of random walk measures in \eqref{upperboundmain1}, which are conditioned to start and terminate in the given box $ \L $. The contributions of the remaining random walks are estimated from above by one. For any given permutation $ \s\in\Sym_N $ and subset $ I\subset\{1,\ldots,N\} $ with $ |I|>(1-\eps) N $ we define the subset $ I_{\s}=\{i\in \{1,\ldots,N\}\colon\s(i)\in I\} $. This means that we replace the product measure $ \bigotimes_{i=1}^N\P^{\beta}_{x_i,x_{\s(i)}} $ by the product measure $ \bigotimes_{i\in I_{\s}\cap I}\P^{\beta}_{x_i,x_{\s(i)}} $. To perform this, we have to replace the empirical path measure $ L_N $ by the empirical path measure $ L_{I_{\s}\cap I} $, given as
$$
L_{I_{\s}\cap I}=\frac{1}{|I_{\s}\cap I|}\sum\limits_{i\in I_{\s}} \delta_{\xi^{\ssup{i}}}. 
$$ From $ |I|>(1-\eps) N $ we get that $ |I_{\s}\cap I|\ge (1-2\eps)N $. 
We need some technical preliminaries to estimate the error of this replacement. First note that $ D_{\beta} $ equipped with the Skorokhod topology is Polish \cite{DS01}. Next we need a metric for the probability measures on the Polish space $ D_{\beta} $.
Recall the L\'{e}vy metric $ \d $ on the Polish space $ \Pmf(D_{\beta}) $ \cite{DS01}, defined for any two probability measures $ \mu,\nu\in\Pmf(D_{\beta}) $ as
\begin{equation}
\d(\mu,\nu)=\inf\limits_{\delta >0}\{\mu(\Gamma)\le \nu(\Gamma^{\delta})+\delta\,\mbox{ and }\, \nu(\Gamma)\le\mu(\Gamma^{\delta})+\delta\,\mbox{ for all }\, \Gamma=\overline{\Gamma}\subset D_{\beta}\} 
\end{equation} 
where $ \Gamma^{\delta}=\{\mu\in\Pmf(D_{\beta})\colon\dist(\mu,F)\le\delta\} $ is the closed $\delta$-neighbourhood of $ F $. (By $\dist(\mu,A)=\inf_{\nu\in A}\d(\mu,\nu)$ we denote the distance to a set $A\subset \Pmf(D_{\beta}) $). 
Note,  $ NL_N(A) $ is the number of random walk paths in any closed set $ A\subset D_{\beta} $ and therefore 
$$
NL_N(F)\le |I_{\s}\cap I|L_{I_{\s}\cap I} +2\eps N,
$$ which implies $ L_N(F)\le L_{I_{\s}\cap I}(F)+2\eps $. Thus, $ \d(L_N,L_{I_{\s}\cap I})\le 2\eps $, and we get using $ |I|>(1-\eps) N $
\begin{equation}
\Bigl(\bigotimes\limits_{i=1}^N\P^{\beta}_{x_i,x_{\s(i)}}\Bigr)(L_N\in F)\le \Bigl(\bigotimes\limits_{i\in I_{\s}\cap I}\P^{\beta}_{x_i,x_{\s(i)}}\Bigr)(L_N\in F^{2\eps}).
\end{equation}

We insert this in \eqref{upperboundmain1} and execute all the $ N-|I| $ summations over $ \L^{a_j} $ with $ j\notin I $ since they do not contribute anymore. All these contributions are estimated from above by one, and hence we are left with the $ |I| $ sums over those $ x_i $ with $ i\in I $, i.e. $ x_i\in\L $ for $ i\in I $. Hence, we write $ \overline{x}=(x_i)_{i\in I} $ for a configuration $ \overline{x}\in\L^I $ and get from \eqref{upperboundmain1}

\begin{equation}\label{estuppern1}
\begin{aligned}
\P^{\ssup{\rm sym}}_{N,\beta}(L_N\in F) & \le 2^Nm(\L^{\rm c})^{\eps N}\\
& + \frac{1}{N!}\sum\limits_{\s\in\Sym_N}\sum\limits_{\heap{I\subset\{1,\ldots,N\}}{|I|>(1-\eps) N}}\sum_{\overline{x}\in\L^I}\prod\limits_{i\in I}m(x_i)\Bigl(\bigotimes\limits_{i\in I_\sigma\cap I}\P^{\beta}_{x_i,x_{\s(i)}}\Bigr)(L_{I_\sigma\cap I}\in F^{2\eps}).
\end{aligned}
\end{equation}
In the following we put $ |I_\s\cap I|=n $ and recall $ (1-2\eps)N<n\le N $.
We observe that the product measure $ $ does not depend on the full information of the permutation $ \sigma\in\Sym_N $. It depends only on the frequencies of indices $ i\in I $ such that $ x_i=x $ and $ x_{\s(i)}=y $  for any $ x,y\in\L $. These frequencies can be expressed with some pair probability measure 
$$ Q\in\Pmf^{\ssup{n}}_\L:=\Pmf(\L\times\L)\cap\frac{1}{n}\N^{\L\times \L}; 
$$ i.e. for example there are $ nQ(x,y) $ indices $ i\in I_\s\cap I $ such that $ x_i=x $ and $ x_{\s(i)}=y $. Hence, we need to count those permutations that satisfy the constraint and further that 
$$ 
Q^{\ssup{1}}=L(\overline{x})=\frac{1}{n}\sum_{i\in I_\s\cap I}\delta_{x_i}\;\mbox{ and }\; Q^{\ssup{2}}=L(\s(\overline{x}))=\frac{1}{n}\sum_{i\in I_\s\cap I}\delta_{x_{\s(i)}}.
$$
Note that here the marginals $ Q^{\ssup{1}} $ and $ Q^{\ssup{2}} $ are not necessarily equal. But as $ n>(1-2\eps)N $ we can estimate the difference of these marginals with the metric $ \d $ on $ \Pmf(\L\times\L) $. The marginal differs at most by $ 2\eps $, i.e. $ \d(Q^{\ssup{1}}, Q^{\ssup{2}})\le 2\eps $. Define the set
$$
\Pmf^{\ssup{n,\eps}}_\L=\{Q\in \Pmf^{\ssup{n}}(\L\times\L)\colon \d(Q^{\ssup{1}}, Q^{\ssup{2}})\le 2\eps\}.
$$ 
We rewrite the right hand side of \eqref{estuppern1} with a sum on pair probability measures in the set $ \Pmf^{\ssup{n,\eps}}_\L $ and an additional sum over the subsets $ \widetilde I\subset I $ with $ |\widetilde I|=n>(1-2\eps)N $. 
\begin{equation}\label{estuppern2}
 \begin{aligned}
&\P^{\ssup{\rm sym}}_{N,\beta}(L_N\in F)\le 2^Nm(\L^{\rm c})^{\eps N}\\
& \quad + \frac{1}{N!}\sum\limits_{\s\in\Sym_N}\sum_{\heap{\widetilde I\subset I\subset\{1,\ldots,N\},}{n=|\widetilde I|\ge (1-2\eps)N}}\sum_{Q\in\Pmf^{\ssup{n,\eps}}_\L}\prod_{x\in\L}m(x)^{nQ^{\ssup{1}}(x)}\Big(\bigotimes_{x,y\in\L}\big(\P_{x,y}^\beta\big)^{nQ(x,y)}\Big)(L_n\in F^{2\eps})\prod_{i\in I\setminus\widetilde I}m(x_i)\\
& \quad \times \sum_{\overline{x}\in\L^I}\sum_{\s\in\Sym_N}\frac{\1\{\widetilde I=I_\s\cap I\}}{N!}\1_{\{\s\in\Sym_N\colon \forall\, x,y\in\L\colon\sharp\{i\in\widetilde I\colon x_i=x,x_{\s(i)}=y\}=nQ(x,y)\}}(\sigma)\\
&\quad \le 2^Nm(\L^{\rm c})^{\eps N}+\sum_{(1-2\eps)N\le n\le N}\sum_{Q\in\Pmf^{\ssup{n},\eps}_\L}\prod_{x\in\L}m(x)^{nQ^{\ssup{1}}(x)}\Big(\bigotimes_{x,y\in\L}\big(\P_{x,y}^\beta\big)^{nQ(x,y)}\Big)(L_n\in F^{2\eps})\\
&\times \sum_{\heap{\widetilde I\subset I\subset\{1,\ldots,N\},}{n=|\widetilde I|}}\sum_{\overline{x}\in\L^I}\frac{1}{N!}\Sym_N(\overline{x},\widetilde I,Q)\prod_{i\in I\setminus\widetilde I}m(x_i),
\end{aligned}
\end{equation}
where we introduced the set
$$
\Sym_N(\overline{x},\widetilde I,Q)=\{\s\in\Sym_N\colon \forall\, x,y\in\L\colon\sharp\{i\in\widetilde I\colon x_i=x,x_{\s(i)}=y\}=nQ(x,y)\}
$$
of permutations admissible with $ Q $ and the configuration $ \overline{x}\in\L^I $ on the index set $ \tilde I\subset I $.

\noindent {\bf Counting:} We estimate now the cardinality of the set $ \Sym_N(\overline{x},\widetilde I,Q) $. We fix a configuration $ \overline x\in\L^I$ and an index set $ \widetilde I\subset I\subset\{1,\ldots,N\} $ with $ |\widetilde I|=n, (1-2\eps)N\le n\le N $. We evaluate the cardinality of the set $ \Sym_N(\overline{x},\widetilde I,Q) $ with an additional sum over those configurations $ \omega\in\L^{\widetilde I} $ for which $ L_{\widetilde I}(\omega)=Q^{\ssup{2}} $, here $ L_{\widetilde I}(\omega) $ is the empirical measure of the configuration $ \omega $. This gives
\begin{equation}\label{estuppern3}
\begin{aligned}
\sharp \Sym_N(\overline{x},\widetilde I,Q)=&\sum_{\heap{\omega\in\L^{\widetilde I},}{L(\omega)=Q^{\ssup{2}}}}\1_{\{\forall\,x,y\in\L\colon\sharp\{i\in\widetilde I\colon x_i=x, \omega_i=y\}=nQ(x,y)\}}(\omega)\\
&\quad\times\sum_{\s\in\Sym_N}\1_{\{\omega_i=x_{\s(i)},\forall i\in\widetilde I\}}(\sigma).
\end{aligned}
\end{equation}
The last term is easily estimated as
$$
\begin{aligned}
\sum_{\s\in\Sym_N}\1_{\{\omega_i=x_{\s(i)},\forall i\in\widetilde I\}}(\sigma)&= \sum_{\widehat I\subset I, |\widehat I|=|\widetilde I|} \big|\{\s\colon\widetilde I\to \widehat I\;\mbox{bijective}\,\colon\omega_i=x_{\s(i)}\forall \,i\in\widetilde I\}\big| \\ & \times\big|\{\s\colon\{1,\ldots,N\}\setminus\widetilde I\to\{1,\ldots,N\}\setminus\widehat I \;\mbox{bijective}\;\}\\
&\le \binom{|I|}{n}\prod_{x\in\L}(nQ^{\ssup{2}}(x))!(N-n)!.
\end{aligned}
$$
The first term in \eqref{estuppern3} is given as in the lower bound via a counting of Euler trails (\cite{A01}). Therefore
$$
\sum_{\heap{\omega\in\L^{\widetilde I},}{L(\omega)=Q^{\ssup{2}}}}\1_{\{\forall\,x,y\in\L\colon\sharp\{i\in\widetilde I\colon x_i=x, \omega_i=y\}=nQ(x,y)\}}(\omega) \le \frac{\prod_{x\in\L}(nQ^{\ssup{1}}(x))!}{\prod_{x,y\in\L}(nQ(x,y))!}.
$$
Note that the two previous estimations do not depend on the configurations $ \overline x\in\L^I $ as long as 
$$ 
Q^{\ssup{1}}=L_{\widetilde I}(\overline{x})=\frac{1}{n}\sum_{i\in \widetilde I}\delta_{x_i}.
$$
The number of all these configurations is clearly equal $ n!/\prod_{x\in\L}(n Q^{\ssup{1}}(x))! $.
We split the sum over the configurations $ \overline x\in\L^I $ in the last line of \eqref{estuppern2} into a sum on $ (x_i)_{i\in\widetilde I}\in\L^{\widetilde I} $ and $ (x_i)_{i\in I\setminus\widetilde I}\in \L^{I\setminus\widetilde I} $, and we estimate the term $ \prod_{i\in  I\setminus\widetilde I}m(x_i) $ from above by one. Hence, the last line of \eqref{estuppern2} can be estimated as
\begin{equation}
\begin{aligned}
\sum_{\heap{\widetilde I\subset I\subset\{1,\ldots,N\},}{n=|\widetilde I|}}&\sum_{\overline{x}\in\L^I}\frac{1}{N!}\sharp \Sym_N(\overline{x},\widetilde I,Q)\prod_{i\in I\setminus\widetilde I}m(x_i)  \le \sum_{\heap{\widetilde I\subset I\subset\{1,\ldots,N\},}{n=|\widetilde I|}}\sharp\Big\{(x_i)_{i\in\widetilde I}\in\L^{\widetilde I}\colon L((x_i)_{i\in\widetilde I})=Q^{\ssup{1}}\Big\}\\
& \times\binom{|I|}{n}\frac{\prod_{x\in\L}(nQ^{\ssup{1}}(x))!\prod_{x\in\L}(nQ^{\ssup{2}}(x))!}{\prod_{x,y\in\L}(nQ(x,y))!}\frac{(N-n)!}{N!}\\
&\le\sum_{l=n}^N\binom{N}{l}\binom{l}{n}^2\frac{n!(N-n)!}{N!}\frac{\prod_{x\in\L}(nQ^{\ssup{2}}(x))!}{\prod_{x,y\in\L}(nQ(x,y))!}\\
& \le N\binom{N}{n}^2\frac{\prod_{x\in\L}(nQ^{\ssup{2}}(x))!}{\prod_{x,y\in\L}(nQ(x,y))!}\le N^{|\L|^2+\widetilde C}{\rm e}^{2C\eps N}\frac{\prod_{x\in\L}Q^{\ssup{2}}(x)^{n Q^{\ssup{2}}(x)}}{\prod_{x,y\in\L}Q(x,y)^{nQ(x,y)}},
\end{aligned}
\end{equation}
where we used that there is a constant $ C>0 $ such that $ \binom{N}{n}\le {\rm e}^{C\eps N} $ for all $ n $ with $ (1-2\eps)N< n\le N $, and where we used Stirling's formula and some constant $  \widetilde C $. Inserting all this in the estimation \eqref{estuppern2} gives 
\begin{equation}\label{estuppern4}
\begin{aligned}
\P^{\ssup{\rm sym}}_{N,\beta}(L_N\in F)\le 
N^{|\L|^2+\widetilde C}{\rm e}^{2C\eps N}\sum_{n>(1-2\eps)N}^N\sum_{Q\in\Pmf^{\ssup{n,\eps}}_\L}{\rm e}^{-nH(Q|m\otimes Q^{\ssup{2}})}\P_{Q,n}^\beta(L_n\in F^{2\eps})\quad +2^Nm(\L^{\rm c})^{\eps N}.
\end{aligned}
\end{equation}

\noindent {\bf Large deviations:}

\noindent We now show for $ \L\subset \Z^d $ fixed and any $ \eps>0 $ that
\begin{equation}\label{estuppern5}
\begin{aligned}
\lim_{n\to\infty}&\frac{1}{n}\log \Big(\sum_{Q\in\Pmf^{\ssup{n,\eps}}_\L}{\rm e}^{-nH(Q|m\otimes Q^{\ssup{2}})}\P_{Q,n}^\beta(L_n\in F^{2\eps})\Big)\\ &\le
-\inf_{\mu\in F^{2\eps}}\inf_{Q\in \Pmf^{\eps}_\L}\Big\{H(Q|m\otimes Q^{\ssup{2}})+\sup_{F\in\Ccal_{\rm b}(D_\beta)}\Big\{\langle F,\mu\rangle+\sum_{x,y\in\Z^d}Q(x,y)\log\E_{x,y}^\beta\Big({\rm e}^{F(\xi)}\Big)\Big\}\Big\},
\end{aligned}
\end{equation}
where $ \Pmf^{\ssup{\eps}}_\L=\{Q\in\Pmf(\L\times\L)\colon \d(Q^{\ssup{1}},Q^{\ssup{2}})\le 2\eps\} $ is the set of pair probability measure whose marginals differs by at most $ 2\eps $, and where the probability measure $ Q \in \Pmf^{\eps}_\L $ is trivially extended to $ \Z^d\times\Z^d $ by zero outside of $ \L\times\L $. To see this, consider the logarithmic moment generating function of the distribution of $ L_n $ under the probability measure $ \P_{Q,n}^\beta $,
\begin{equation}
\L^{\ssup{Q}}_n(\Phi)=\log\E_{Q,n}^\beta\big({\rm e}^{n\langle F,L_n\rangle}\big)=n\sum_{x,y\in\L}Q(x,y)\log\E_{x,y}^\beta\big({\rm e}^{F(\xi)}\big)
\end{equation}
for any $ Q\in\Pmf^{\ssup{n,\eps}}_\L $ and any $ F\in\Ccal_{\rm b}(D_\beta) $. Now let $ Q_n\in\Pmf^{\ssup{n,\eps}}_\L $ be maximal for the mapping $ Q\mapsto {\rm e}^{-nH(Q|m\otimes Q^{\ssup{2}})}\P_{Q,n}^\beta(L_n\in F^{2\eps}) $. Then, since the set $\Pmf^{\ssup{\eps}}_\L $ is compact, there is a pair measure $ Q\in\Pmf^{\ssup{\eps}}_\L $ with $ \lim_{n\to\infty}Q_n =Q $ weakly. Clearly the limit
$$
\L^Q(F)=\lim_{n\to\infty}\frac{1}{n}\L^{\ssup{Q}}_n(F)=\sum_{x,y\in\L}Q(x,y)\log\E_{x,y}^\beta\big({\rm e}^{ F(\xi)}\big)
$$
exists, and is lower semi continuous and G\^ateax differentiable. Now the G\"artner-Ellis theorem yields that
$$
\limsup_{n\to\infty}\frac{1}{n}\log\P_{Q_n,n}^\beta (L_n\in F^{2\eps})\le -\inf_{\mu\in F^{2\eps}}\sup_{F\in\Ccal_{\rm b}(D_\beta)}\Big\{\langle F,\mu\rangle+\sum_{x,y\in\Z^d}Q(x,y)\log\E_{x,y}^\beta\big({\rm e}^{F(\xi)}\big)\Big\},
$$
because we may assume that $ F^{2\eps} $ is compact. We can do so, because Lemma~\ref{lemma-exptight1} shows that there is a sequence of compact sets $ M_L\subset\Pmf(D_\beta) $ such that
$$
\lim_{L\to\infty}\limsup_{n\to\infty}\frac{1}{n}\log\Big(\sup_{Q\in \Pmf_\L^{\ssup{n}}}\P_{Q,n}^\beta(L_n\in  M_L^{\rm c})\Big)=-\infty.
$$
The cardinality of the set $ \Pmf^{\ssup{n,\eps}}_\L $ is clearly polynomial in $ n $, and by continuity of $ Q\mapsto H(Q|m\otimes Q^{\ssup{2}}) $, the assertion of \eqref{estuppern5} follows.

We are now in the position to perform the $ N\to\infty $ limit for the upper bound in \eqref{estuppern4}. We get
\begin{equation}
\begin{aligned}
\limsup_{N\to\infty}\frac{1}{N}\log\P^{\ssup{\rm sym}}_{N,\beta}(L_N\in F)&\le -\min\Big\{-2C\eps-\log 2-\eps\log m(\L^{\rm c}), \\
&\quad \inf_{\mu\in F^{2\eps}}\inf_{Q\in\Pmf^{\ssup{\eps}}_\L}\Big\{H(Q|m\otimes Q^{\ssup{2}})+I_\beta^{\ssup{Q}}(\mu)\Big\}\Big\},
\end{aligned}
\end{equation}
where we recall
$$
I_\beta^{\ssup{Q}}(\mu)=\sup_{F\in\Ccal_{\rm b}(D_\beta)}\Big\{\langle F,\mu\rangle+\sum_{x,y\in\Z^d}Q(x,y)\log\E_{x,y}^\beta\big({\rm e}^{F(\xi)}\big)\Big\}.
$$
The proof of the upper bound \eqref{secupperboundmain} of Theorem~\ref{mainthm1} is finished when we replace $ \L\subset \Z^d$ and $ \eps $ by sequences $ (\eps_N)_{N\in\N} $ and $ (\L_N)_{N\in\N} $ with $\eps_N\to 0 $ and $ \L_N\uparrow\Z^d $ as $ N\to\infty $ such that $ |\L_N|^2/N=N^{-\delta} $ for some $ \delta>0 $ and such that $ \eps_N\log m(\L_N^{\rm c})\to-\infty $ as N $\to\infty $ and use the following lemma. Recall that $ m_N\in\Pmf(\L_N) $ is the restriction of the initial distribution on the set $ \L_N $.

\begin{lemma}
Fix a closed set $ F\subset\Pmf(D_\beta) $. Then for any sequence $ (\eps_n)_{N\in\N} $ satisfying $ \eps_N\to 0 $ as  $ N \to\infty $  and any sequence $ (\L_N)_{N\in\N} $ with $ \L_N\uparrow\Z^d $ as $ N\to\infty $,

\begin{equation}\label{estupperlemma}
\begin{aligned}
\liminf_{N\to\infty}\inf_{\mu\in F^{2\eps}}\inf_{Q\in\Pmf^{\ssup{\eps}}_{\L_N}}\Big\{H(Q|m_N\otimes Q^{\ssup{2}})+I_\beta^{\ssup{Q}}(\mu)\Big\}\Big\}\\
\ge \inf_{\mu\in F}\inf_{Q\in\widetilde{\Pmf}(\Z^d\times\Z^d)}\Big\{H(Q|Q^{\ssup{1}}\otimes m)+I_\beta^{\ssup{Q}}(\mu) \Big\}.
\end{aligned}
\end{equation}
\end{lemma}

\begin{proofsect}{Proof}
Clearly, $ m_N\to m $ weakly as $ N\to\infty $. Now we pick approximating sequences of $ Q$'s and $\mu$'s and employ compactness arguments. Thus, for any $ N\in\N $ pick $ \mu_N\in F^{2\eps_N} $ and $ Q_N\in\Pmf^{\ssup{\eps_N}}_{\L_N} $ such that the sequences $ (H(Q_N|m_N\otimes Q^{\ssup{2}}_N)+I_\beta^{\ssup{Q_N}}(\mu_N))_{N\in\N} $ converges to the right hand side of \eqref{estupperlemma} and may therefore be assumed to be bounded. The sequence $ (H(Q_N^{\ssup{1}}|m_N))_{N\in\N} $ is bounded because of
$$
H(Q_N|m_N\otimes Q^{\ssup{2}}_N=H(Q_N^{\ssup{1}}|m_N)+H(Q_N|Q_N^{\ssup{1}}\otimes Q_N^{\ssup{2}}).
$$ 
As $ Q_N^{\ssup{1}} $ has support in $ \L_N $ we have $ H(Q_N^{\ssup{1}}|m_N)=H(Q_N^{\ssup{1}}|m) $, and thus the sequence $ (Q_N^{\ssup{1}})_{N\in\N} $ is tight due to the fact that the level sets of the relative entropy are compact (see Lemma~\cite[6.2.12]{DZ98}). As $ \d(Q_N^{\ssup{1}},Q_N^{\ssup{2}})\le 2\eps_N\to 0 $ as $ N\to\infty $, also the sequence $ (Q_N^{\ssup{2}})_{N\in\N} $ is tight. By boundedness of the sequence $ (H(Q_N|Q_N^{\ssup{1}}\otimes Q_N^{\ssup{1}}))_{N\in\N} $, also the set $ P:=\{Q_N\colon N\in\N\} $ is tight. Hence, according to Prohorov's theorem we may assume that $ Q_N\to Q $ as $ N\to\infty $ for some $ Q\in\Pmf(\Z^d\times\Z^d) $. Since both $ Q_N^{\ssup{1}}\to Q^{\ssup{1}} $ and $ Q_N^{\ssup{2}}\to Q^{\ssup{2}} $ as $ N\to\infty $, we get that $ Q\in\widetilde{\Pmf}(\Z^d\times\Z^d) $. 

For $ C>0 $ sufficiently large, the sequence $ (\mu_n)_{N\in\N} $ is contained in the set 
$$
\{\mu\in\Pmf(D_\beta)\colon \inf_{N\in\N} I_\beta^{\ssup{Q_N}}(\mu)\le C\}.
$$
Now it turns out that this set is relatively compact. We are going to prove this fact now. Note that it suffices to find a family of compact sets $ M_L\subset D_\beta , L>0 $, such that
$$
\lim_{L\to\infty}\inf_{Q\in P}\inf_{M_L^{\rm c}} I^{\ssup{Q}}=\infty.
$$
We prove this in the usual way with the exponential tightness and a lower bound for a large deviations principle. The sequence $ (L_N)_{N\in\N} $ is exponentially tight under the probability measure $ \P_{Q,N}^\beta $ , uniformly in $ Q\in  P $ (see Lemma~\ref{exptightnessprod}). Moreover, it is easy to see that it satisfies a large deviations principle with rate function $ I_\beta^{\ssup{Q}} $. Indeed, note that the logarithmic moment generating function of $ L_N $ under the probability measure $ \P_{Q_N,N}^\beta $ is easily shown to converge towards the function
$$
F\mapsto\sum_{x,y\in\Z^d}Q(x,y)\log\E_{x,y}^\beta\big({\rm e}^{F(\xi)}\big)\quad,F\in\Ccal_{\rm b}(D_\beta),
$$
whose Legendre-Fenchel transform is $ I_\beta^{\ssup{Q}} $. The G\"artner-Ellis theorem then provides the proof for the large deviations principle. For $ L\in\N $, pick a compact set $M_L\subset\Pmf(D_\beta) $ such that 
$$
\P_{Q_N,N}^\beta(L_N\in M_L^{\rm c})\le {\rm e}^{-NL}\quad\mbox{ for any }\; L,N\in\N,Q_N\in P.
$$
Now the lower bound in the mentioned large deviations principle gives us that
$$
\inf_{Q\in P}\inf_{M_L^{\rm c}} I_\beta^{\ssup{Q}}\ge -\liminf_{N\to\infty}\frac{1}{N}\log\P_{Q_N,N}^\beta(L_N\in M_L^{\rm c})\ge L,
$$
implying that the sequence $ (\mu_N)_{N\in\N} $ is tight. Therefore, we may assume that $ \mu_N\Rightarrow\mu $ as $ N\to\infty $ with some $ \mu\in F^1 $. Since $ \mu_N\in F^{2\eps_N} $ for any $ N\in\N $ and since $ \eps_N\to 0 $, we even have $ \mu\in F $, because $ F $ is closed. To finish now the proof of the lemma we employ the representation of the relative entropy as a Legendre transform (see \cite[Lemma~3.2.13]{DS01}.
This gives
$$
\begin{aligned}
\inf_{\mu\in F^{2\eps_N}}&\inf_{Q\in\Pmf^{\ssup{\eps_N}}_{\L_N}}\Big\{H(Q|m_N\otimes Q^{\ssup{2}})+I_\beta^{\ssup{Q}}(\mu)\Big\}\\
&\ge \langle g,Q_N\rangle -\log\langle {\rm e}^{g},m_N\otimes Q^{\ssup{2}}_N\rangle+\langle \Phi,\mu_N\rangle-\sum_{x,y\in\L_N}Q_N(x,y)\log\E_ {x,y}^\beta\Big({\rm e}^{F(\xi)}\Big),
\end{aligned}
$$
where $ g\in\Ccal_{\rm b}(\L_N\times\L_N) $ and $ F\in\Ccal_{\rm b}(D_\beta) $ are arbitrary.
Hence, we get
\begin{equation}\label{estuppern6}
\begin{aligned}
\liminf_{N\to\infty}&\inf_{\mu\in F^{2\eps_N}}\inf_{Q\in\Pmf^{\ssup{\eps_N}}_{\L_N}}\Big\{H(Q|m_N\otimes Q^{\ssup{2}})+I_\beta^{\ssup{Q}}(\mu)\Big\}\\
& \ge \langle g,Q\rangle -\log\langle {\rm e}^{g},Q^{\ssup{1}}\otimes m\rangle+\langle F,\mu\rangle-\sum_{x,y\in\Z^d}Q(x,y)\log\E_ {x,y}^\beta\big({\rm e}^{F(\xi)}\big).
\end{aligned}
\end{equation}
Since this holds for any $ g\in\Ccal_{\rm b}(\L_N\times\L_N) $ and any $ F\in\Ccal_{\rm b}(D_\beta) $, the left hand side of \eqref{estuppern6} is not smaller than $ H(Q|Q^{\ssup{1}}\otimes m)+I_\beta^{\ssup{Q}}(\mu) $. Therefore,
\begin{equation}
\mbox{ l.h.s. of }\;\eqref{estuppern6}\;\ge \inf_{\mu\in F}\inf_{Q\in\widetilde \Pmf(\Z^d\times\Z^d)} \Big\{H(Q|\overline{Q}\otimes m)+I_\beta^{\ssup{Q}}(\mu) \Big\},
\end{equation}
and the assertion of the lemma follows. \qed
\end{proofsect}

\subsection{Proof of Theorem~\ref{mainthm3}}\label{main3}
We prove Theorem~\ref{mainthm3} in the following. Denote by $ \Psi $ the continuous mapping
$$
\Psi\colon\Pmf(D_\beta)\to D([0,\beta];\R^d),\mu\mapsto \Psi(\mu)=\int_{D_\beta}\omega\,\mu(\d \omega).
$$
Note that $ Y_N=\Psi(L_N) $ and recall that $ D_\beta=D_\beta([0,\beta];\Z^d) $. The contraction principle \cite[Lemma~2.1.4]{DS01} ensures a large deviations principle for $ Y_N $ under the symmetrised measure $ \P_{N,\beta}^{\ssup{\rm sym}} $ with the rate function
\begin{equation}
\widehat I^{\ssup{\rm sym}}_\beta(\omega)=\inf_{Q\in\widetilde{\Pmf}(\Z^d\times\Z^d)} \Big\{H(Q|Q^{\ssup{1}}\otimes m)+\widehat I^{\ssup{Q}}_\beta(\omega)\Big\},
\end{equation}
where
\begin{equation}
\widehat I^{\ssup{Q}}_\beta(\omega)=\inf_{\heap{\mu\in\Pmf(D_\beta)\colon}{\Psi(\mu)=\omega}} I^{\ssup{Q}}_\beta(\mu).
\end{equation}
Therefore we need to show that $ \widehat I^{\ssup{\rm sym}}_\beta=\widetilde I^{\ssup{\rm sym}}_\beta $, and for that it suffices to show that $ \widehat I^{\ssup{Q}}_\beta=\widetilde I^{\ssup{Q}}_\beta $. If we consider the class of functions $ F\in\Ccal_{\rm b}(D_\beta) $ of the form $ F_f(\omega)=\int_0^\beta\d s \langle\omega_s,f_s\rangle_{\R^d} $ for $ f\in L^2([0,\beta];\R^d) $ we get
for $ \mu\in\Pmf(D_\beta) $ 
$$
\begin{aligned}
I^{\ssup{Q}}_\beta(\mu)&\ge \sup_{f\in L^2([0,\beta];\R^d)}\Big\{ \int_{D_\beta}\mu(\d \omega)\int_0^\beta\d s \langle \omega_s,f_s\rangle_{\R^d}-\sum_{x,y\in\Z^d}Q(x,y)\log\E_{x,y}^\beta\Big({\rm e}^{\int_0^\beta\d s \langle f_s,\xi_s\rangle_{\R^d}}\Big)\Big\}\\
&=\widetilde I^{\ssup{Q}}_\beta(\Psi(\mu)).
\end{aligned}
$$
If we now take the infimum over all probability measures $ \mu\in\Pmf(D_\beta) $ with $ \Psi(\mu)=\omega $ we get that $ \widehat I^{\ssup{Q}}_\beta\ge \widetilde I^{\ssup{Q}}_\beta $. To prove the complementary bound $ \widehat  I^{\ssup{Q}}_\beta\le \widetilde I^{\ssup{Q}}_\beta $ seems to cause major technical difficulties. We therefore proceed in an indirect way. We show that both $ \widehat I^{\ssup{Q}}_\beta $ and $ \widetilde I^{\ssup{Q}}_\beta $ are the rate function for the same large deviations principle. In the proof of Proposition~\eqref{mainlowerboundproposition} we have shown that the empirical path measures $ L_N $ satisfies a large deviations principle with rate function $ I^{\ssup{Q}}_\beta $ under the measure $ \P^\beta_{Q_N,N} $, where $ Q_N\in\Pmf_{\L_N}^{\ssup{N}} $ is the sequence of pair probability measure from step 1 in the proof of Proposition~\ref{mainlowerboundproposition}. According to the contraction principle the sequence $ (Y_N)_{N\in\N} $ satisfies, under the measure $ \P^\beta_{Q_N,N} $, a large deviations principle with rate function $ \widehat I^{\ssup{Q}}_\beta $.

Now we show directly that the sequence $ (Y_N)_{N\in\N} $ under the measure $ \P^\beta_{Q_N,N} $ satisfies a large deviations principle with rate function $ \widetilde I^{\ssup{Q}}_\beta $, which finishes the proof. Recall that $ Q_N\in\Pmf^{\ssup{N}}_{\L_N} $ with $ Q_N\to Q $ as $ N\to\infty $.
We perform this in the usual setting of the G\"artner-Ellis theorem. Let $ f\in L^2([0,\beta];\R^d)  $, the logarithmic moment generating function is then
$$
\begin{aligned}
\L_N(f)&=\log\E_{Q_N,N}^\beta\Big({\rm e}^{N\langle f,Y_N\rangle}\Big)=\log\Big(\prod_{x,y\in\L_N}\E_{x,y}^\beta\Big({\rm e}^{\int_0^\beta \d s\langle f_s,\xi_s\rangle_{\R^d}}\Big)^{NQ_N(x,y)}\Big)\\
&=N\sum_{x,y\in\L_N}Q_N(x,y)\log \E_{x,y}^\beta\Big({\rm e}^{\int_0^\beta \d s\langle f_s,\xi_s\rangle_{\R^d}}\Big),
\end{aligned}
$$
and hence
$$
\L(f)=\lim_{N\to\infty}\frac{1}{N}\L_N(f)=\sum_{x,y\in\Z^d}Q(x,y)\log \E_{x,y}^\beta\Big({\rm e}^{\int_0^\beta \langle f_s,\xi_s\rangle_{\R^d}\d s}\Big).
$$
It is easily seen that $ \L $ is lower semi continuous and G\^ateax differentiable. The Legendre-Fenchel transform is equal to $ \widetilde I^ {\ssup{Q}}_\beta $. According to Lemma~\ref{exptightnessprod}, the sequence $(Y_N)_{N\in\N} $ is exponentially tight under $ (\P^\beta_{Q_N,N})_{N\in\N} $. Hence, the G\"artner-Ellis theorem finishes the proof.\qed

\subsection{Proof of Theorem~\ref{mainthm2}}\label{main2}
In this subsection we prove Theorem~\ref{mainthm2}. We denote by $ \pi_s\colon D_\beta\to\Z^d , s\in[0,\beta] $, the canonical projection $ \pi_s(\omega)=\omega_s $ for $ \omega\in D_\beta $. Then the mapping
$$
T\colon\Pmf(D_\beta)\to\Pmf(\Z^d), \mu\mapsto T(\mu)=\frac{1}{\beta}\int_0^\beta\d s \mu\circ \pi_s^{-1}
$$
is continuous and $ Z_N=T(L_N) $. Then a large deviations principle for the mean $ Z_N $ of occupation local times with rate function
\begin{equation}\label{ratecontract2}
\widehat J^{\ssup{\rm sym}}_\beta(p)=\inf_{Q\in\widetilde{\Pmf}(\Z^d\times\Z^d)} \Big\{H(Q|Q^{\ssup{1}}\otimes m)+\widehat J^{\ssup{Q}}_\beta(p)\Big\},
\end{equation} where
\begin{equation}
\widehat J^{\ssup{Q}}_\beta(p)=\inf_{\heap{\mu\in\Pmf(D_\beta)\colon}{T(\mu)=p}} I^{\ssup{Q}}_\beta(\mu),
\end{equation} is given via the contraction principle \cite[Lemma~2.1.4]{DS01}.
Therefore we need to show that $ J^{\ssup{\rm sym}}_\beta =\widehat J^{\ssup{\rm sym}}_\beta $,
and for that it suffices to show that $ J^{\ssup{Q}}_\beta =\widehat J^{\ssup{Q}}_\beta $. 
We relax the set of functions over which we perform the supremum. For a fixed probability measure $ \mu\in\Pmf(D_\beta) $ we consider bounded continuous functions $ F\in\Ccal_{\rm b}(D_\beta) $ of the form $ F(\omega)=\frac{1}{\beta}\int_0^\beta \d s f(\omega_s) $ for any bounded function $ f\in\Bcal(\Z^d) $ and $ \omega\in D_\beta $.
Then for $ Q\in\widetilde\Pmf(\Z^d\times\Z^d) $
$$
\begin{aligned}
I^{\ssup{Q}}_\beta(\mu)&\ge \sup_{f\in\Bcal(\Z^d)}\Big\{\int_{D_\beta}\mu(\d \omega)\frac{1}{\beta}\int_0^\beta\d s f(\omega_s)-\sum_{x,y\in\Z^d}Q(x,y)\log\E_{x,y}^\beta\Big({\rm e}^{\frac{1}{\beta}\int_0^\beta f(\xi_s)\d s}\Big)\Big\}\\
&=J^{\ssup{Q}}(T(\mu)).
\end{aligned}
$$
If we now take the infimum over all probability measures $ \mu\in\Pmf(D_\beta) $ with $ T(\mu)=p $ we get that $ \widehat J^{\ssup{Q}}_\beta\ge J^{\ssup{Q}}_\beta $. To prove the complementary bound $ \widehat J^{\ssup{Q}}_\beta\le J^{\ssup{Q}}_\beta $ seems to cause major technical difficulties. We therefore proceed in an indirect way. We show that both $ \widehat J^{\ssup{Q}}_\beta $ and $ J^{\ssup{Q}}_\beta $ are the rate function for the same large deviations principle.
In the proof of Proposition~\eqref{mainlowerboundproposition} we have shown that the empirical path measures $ L_N $ satisfies a large deviations principle with rate function $ I^{\ssup{Q}}_\beta $ under the measure $ \P^\beta_{Q_N,N} $, where $ Q_N\in\Pmf_{\L_N}^{\ssup{N}} $ is the sequence of pair probability measure from step 1 in the proof of Proposition~\ref{mainlowerboundproposition}. According to the contraction principle the sequence $ (Z_N)_{N\in\N} $ satisfies, under the measure $ \P^\beta_{Q_N,N} $, a large deviations principle with rate function $ \widehat J^{\ssup{Q}}_\beta $.

Now we show directly that the sequence $ (Z_N)_{N\in\N} $ under the measure $ \P^\beta_{Q_N,N} $ satisfies a large deviations principle with rate function $ J^{\ssup{Q}}_\beta $, which finishes the proof. Recall that $ Q_N\in\Pmf^{\ssup{N}}_{\L_N} $ with $ Q_N\to Q $ as $ N\to\infty $.
We perform this in the usual setting of the G\"artner-Ellis theorem. Let $ f\in\Bcal(\Z^d) $ any bounded function, the logarithmic moment generating function is then
$$
\begin{aligned}
\L_N(f)&=\log\E_{Q_N,N}^\beta\Big({\rm e}^{N\langle f,Z_N\rangle}\Big)=\log\Big(\prod_{x,y\in\L_N}\E_{x,y}^\beta\Big({\rm e}^{\int_0^\beta f(\xi_s)\d s}\Big)^{NQ_N(x,y)}\Big)\\
&=N\sum_{x,y\in\L_N}Q_N(x,y)\log \E_{x,y}^\beta\Big({\rm e}^{\int_0^\beta f(\xi_s)\d s}\Big),
\end{aligned}
$$
and hence
$$
\L(f)=\lim_{N\to\infty}\frac{1}{N}\L_N(f)=\sum_{x,y\in\Z^d}Q(x,y)\log \E_{x,y}^\beta\Big({\rm e}^{\int_0^\beta f(\xi_s)\d s}\Big).
$$
It is easily seen that $ \L $ is lower semi continuous and G\^ateax differentiable. The Legendre-Fenchel transform is equal to $ J^ {\ssup{Q}}_\beta $. According to Lemma~\ref{exptightnessprod}, the sequence $(Z_N)_{N\in\N} $ is exponentially tight under $ (\P^\beta_{Q_N,N})_{N\in\N} $. Hence, the G\"artner-Ellis theorem finishes the proof.\qed

\subsection{Exponential tightness}\label{productldp-proof} 
We prove in this subsection the exponential tightness of the distributions of the empirical path measure $ L_N $ under $ \P_{N,\beta}^{\ssup{\rm sym}} $ and under $ \P_{Q,N}^\beta $ for any pair measure $ Q\in\Pmf_{\L_N}^{\ssup{N}} $. For the first result we use the compactification given by the initial distribution $ m $ of the random walks.

\begin{lemma}\label{lemma-exptight1}
The empirical path measures $ (L_N)_{N\in\N} $ are exponentially tight under the symmetrised measure $ P_{N,\beta}^{\ssup{\rm sym}} $.
\end{lemma}
\begin{proofsect}{Proof}
For $ l\in\N $, choose a subset  $ \L_l\subset\Z^d $ such that $ m(\L_l^{\rm c})\le {\rm e}^{-l^2}. $ Furthermore, choose $ \delta_l > 0 $ so small that
$$
\sup\limits_{x,y\in \L_l}\P_{x,y}^{\beta}\Bigl(\sup\limits_{t-\delta_l\le t^{\prime}\le t^{\prime\prime}\le t+\delta_l}|\xi_{t^{\prime}}-\xi_t|\wedge|\xi_{t^{\prime\prime}}-\xi_t|+\sup\limits_{0\le t\le\delta_l}|\xi(t)-\xi(0)|+\sup\limits_{\beta-\delta_l\le t\le\beta}|\xi(t)-\xi(\beta)| > \frac{1}{l}\Bigr)\le {\rm e}^{-l^2},
$$ where we write $ \xi=\xi^{\ssup{1}} $ for a single random walk. Consider the set $$
\begin{aligned}
A_l =\big\{\omega\in D_{\beta}\colon \omega(0)&\in \L_l, \omega(\beta)\in \L_l, \sup\limits_{t-\delta_l\le t^{\prime}\le t^{\prime\prime}\le t+\delta_l}|\omega_{t^{\prime}}-\omega_t|\wedge|\omega_{t^{\prime\prime}}-\omega_t|\le\frac{1}{3l}, \\&\quad \sup\limits_{0\le t\le\delta_l}|\omega(t)-\omega(0)|\le\frac{1}{3l}, \sup\limits_{\beta-\delta_l\le t\le\beta}|\omega(t)-\omega(\beta)|\le \frac{1}{3l}\big\}.
\end{aligned}
$$

\noindent According to a well-known characterisation of compact subsets in $ D_{\beta} $, compare e.g. \cite[Lemma~2.1]{Dor96} or \cite{Par67}, $ A_l $ is relative compact in $ D_{\beta}$ with respect to the Skorokhod topology. 
Now put $ K_l:=\{\mu\in\Pmf(D_{\beta}):\mu(\overline{A}_l^{\rm c})\le \frac{1}{l}\} $ and note that $ K_l $ is closed by Portmanteau's theorem. Let $ M\in\N $ be given and consider $ K_M:=\bigcap\limits_{l=M}^{\infty} K_l $. The set $ K_M $ is tight and by Prohorov's theorem $ \overline{K_M} $ is compact. We shall show that $ \P_{N,\beta}^{\ssup{\rm sym}}(L_N\in K^{\rm c}_M)\le {\rm e}^{-MN} $. Observe that 

$$
\begin{aligned}
&\{L_N\in K^{\rm c}_l\}\subset\Big\{\,\sharp\{i\in\{1,\ldots,N\}\colon \xi^{\ssup{i}}\in A^{\rm c}_l\} >\frac{N}{l}\Big\}\\&\quad \subset \Big\{\sharp\{i\colon \xi^{\ssup{i}}_0\in \L_l^{\rm c}\}\ge \frac{N}{3l}\Big\}\cup\Big\{\sharp\{i\colon \xi^{\ssup{i}}_{\beta}\in \L_l^{\rm c}\}\ge \frac{N}{3l}\Big\}\cup\Big\{\sharp\{i\colon \xi^{\ssup{i}}_0\in \L_l, \xi^{\ssup{i}}_{\beta}\in \L_l, \\&\quad\sup\limits_{t-\delta_l\le t^{\prime}\le t^{\prime\prime}\le t+\delta_l}|\xi_{t^{\prime}}-\xi_t|\wedge|\xi_{t^{\prime\prime}}-\xi_t|>\frac{1}{3l},\sup\limits_{0\le t\le\delta_l}|\xi(t)-\xi(0)|>\frac{1}{3l},\\&\quad\sup\limits_{\beta-\delta_l\le t\le\beta}|\xi(t)-\xi(\beta)| >\frac{1}{3l}\}\ge \frac{N}{3l}\Big\}.
\end{aligned}
$$

Clearly,
\begin{equation}
\begin{aligned}
\P_{N,\beta}^{\ssup{\rm sym}}&(\sharp\{i\colon \xi^{\ssup{i}}_{\beta}\in \L_l^{\rm c}\}\ge \frac{N}{3l})=\P_{N,\beta}^{\ssup{\rm sym}}(\sharp\{i\colon \xi^{\ssup{i}}_{0}\in \L_l^{\rm c}\}\ge \frac{N}{3l})\\
& \le \sum\limits_{\heap{I\subset\{1,\ldots,N\}:}{|I|\ge \frac{N}{3l}}}\frac{1}{N!}\sum\limits_{\s\in\Sym_N}\sum\limits_{x\in(\Z^d)^N}\prod\limits_{i=1}^N m(x_i)\bigotimes\limits_{i=1}^N\P_{x_i,x_{\s(i)}}(\forall\; i\in I\colon \xi^{\ssup{i}}_0\in \L^{\rm c}_l)\\
&\le\sum\limits_{|I|\ge \frac{N}{3l}} m(\L_l^{\rm c})^{|I|}\le {\rm e}^{-lN/3}2^N.
\end{aligned}
\end{equation}
Furthermore,
\begin{equation}
\begin{aligned}
\P_{N,\beta}^{\ssup{\rm sym}}\Bigl(&\sharp\Bigl\{i\colon \xi^{\ssup{i}}_0\in \L_l,\xi^{\ssup{i}}_{\beta}\in \L_l,   \sup\limits_{t-\delta_l\le t^{\prime}\le t^{\prime\prime}\le t+\delta_l}|\xi_{t^{\prime}}-\xi_t|\wedge|\xi_{t^{\prime\prime}}-\xi_t|>\frac{1}{3l},\\
& \sup\limits_{0\le t\le\delta_l}|\xi(t)-\xi(0)|>\frac{1}{3l},\sup\limits_{\beta-\delta_l\le t\le\beta}|\xi(t)-\xi(\beta)| >\frac{1}{3l}\Bigr)\\
& \le \sum\limits_{|I|\ge \frac{N}{3l}}\frac{1}{N!}\sum\limits_{\s\in\Sym_N}\sum\limits_{x\in(\Z^d)^N}\prod\limits_{i=1}^N m(x_i)\bigotimes\limits_{i=1}^N\P_{x_i,x_{\s(i)}}^{\beta}\Bigl(\forall\; i\in I\colon \xi^{\ssup{i}}_0\in \L_l,\xi^{\ssup{i}}_{\beta}\in \L_l,\\&
\sup\limits_{t-\delta_l\le t^{\prime}\le t^{\prime\prime}\le t+\delta_l}|\xi_{t^{\prime}}-\xi_t|\wedge|\xi_{t^{\prime\prime}}-\xi_t|>\frac{1}{3l},\\
& \sup\limits_{0\le t\le\delta_l}|\xi(t)-\xi(0)|>\frac{1}{3l},\sup\limits_{\beta-\delta_l\le t\le\beta}|\xi(t)-\xi(\beta)| >\frac{1}{3l}\Bigr)\\
&\le \sum\limits_{|I|\ge \frac{N}{3l}}\sup\limits_{(y_i)_{i\in I}\in \L_l^{I}}\sum_{x_i\in A_l,i\in I}\prod\limits_{i\in I} m(x_i)\prod_{i\in I}\P^{\beta}_{x_,y_i}\Bigl(   \sup\limits_{t-\delta_l\le t^{\prime}\le t^{\prime\prime}\le t+\delta_l}|\xi_{t^{\prime}}-\xi_t|\wedge|\xi_{t^{\prime\prime}}-\xi_t|>\frac{1}{3l},\\
& \sup\limits_{0\le t\le\delta_l}|\xi(t)-\xi(0)|>\frac{1}{3l},\sup\limits_{\beta-\delta_l\le t\le\beta}|\xi(t)-\xi(\beta)| >\frac{1}{3l}\Bigr)\\
&\le \sum\limits_{|I|\ge \frac{N}{3l}}{\rm e}^{-lN/3}\le {\rm e}^{-lN/3} 2^N.
\end{aligned}
\end{equation} Hence, 
$$ 
\P_{N,\beta}^{\ssup{\rm sym}}(L_N\in K_M^{\rm c})\le\sum\limits_{l=M}^{\infty}\P_{N,\beta}^{\ssup{\rm sym}}(L_N\in K^{\rm c}_l)\le 32^N\sum\limits_{l=M}^{\infty}{\rm e}^{-lN/3}\le 6\times 2^N{\rm e}^{-NM/3}\le {\rm e}^{-NM/5}
$$ for all large $ N $ if $ M> 24 $. This ends the proof.
\hfill $\Box $\end{proofsect}

The following exponential tightness is due to the product structure of the probability measure $ \P_{Q,N}^\beta $ for any $ Q\in\Pmf_{\L_N}^{\ssup{N}} $. Here we have a product of not necessarily identical distributed random walks.

\begin{lemma}\label{exptightnessprod}
Let $ (Q_N)_{N\in\N} $ be any sequence of pair measures $ Q_N\in\Pmf_{\L_N}^{\ssup{N}} $ on $ \L_N\times\L_N $ with $ \L_N\uparrow\Z^d $ as $ N\to\infty $. Then the family of distributions of $ L_N $ (respectively of $Y_N$ and of $ Z_N$) under $ \P_{Q_N,N}^\beta $ is exponentially tight.
\end{lemma}

\begin{proofsect}{Proof}
We will prove the case for the empirical path measure $ L_N $. The proofs for $ Y_N $ and $ Z_N $ follow analogously.
Our proof is an adaptation of the proof for the i.i.d. case (see \cite[Lemma~6.2.6]{DZ98}). From the previous Lemma~\ref{lemma-exptight1} we have a compact set $ A_l\subset D_\beta $ and a subset $ \L_l\subset\Z^d $ such that
\begin{equation}
\sup_{x,y\in\L_l}\P_{x,y}^{\beta}(\xi\in A_l^{\rm c})\le {\rm e}^{-2l^2}({\rm e}^l-1).
\end{equation}
The set $ M_l=\{\nu\in\Pmf(D_\beta)\colon \nu(A_l^{\rm c})\le 1/l\} $ is closed by Portmanteau's theorem. For $ M\in\N $ define $ K_M:=\bigcap_{l=M}^\infty M_l $. By Prohorov's theorem, each $ M_l $ is a relative compact subset of $ \Pmf(D_\beta) $. Then we derive via Chebycheff's inequality for any $ Q_N\in\Pmf_{\L_N}^{\ssup{N}} $
\begin{equation}
\begin{aligned}
\P_{Q_N,N}^\beta(L_N\notin M_l)&=\P_{Q_N,N}^\beta\Big(L_N(A_l^{\rm c}) >\frac{1}{l}\Big)\le {\rm e}^{-2Nl}\E_{Q_N,N}^\beta\Big({\rm e}^{2Nl^2L_N(A_l^{\rm c})}\Big)\\
& = {\rm e}^{-2Nl}\E_{Q_N,N}^\beta\Big(\exp\Big(2l^2\sum_{i=1}^N\1\{\xi^{\ssup{i}}\in A_l^{\rm c}\}\Big)\Big)\\
& = {\rm e}^{-2Nl}\prod_{x,y\in\L_N}\E_{x,y}^\beta\Big(\exp\Big(2l^2\1\{\xi^{\ssup{1}}\in A_l^{\rm c}\}\Big)\Big)^{N Q_N(x,y)}\\
& \le {\rm e}^{-2Nl}\prod_{x,y\in\L_N}\E_{x,y}^\beta\Big(1+{\rm e}^{2l^2}({\rm e}^{-2l^2}({\rm e}^l-1))\Big)^{N Q_N(x,y)}\le {\rm e}^{-Nl}.
\end{aligned}
\end{equation}
Therefore,

\begin{equation}
\P_{Q_N,N}^\beta(L_N\notin K_M)\le \sum_{l=M}^{\infty}\P_{Q_N,N}^\beta(L_N\notin M_l)\le \sum_{l=M}^{\infty}{\rm e}^{-Nl}\le 2{\rm e}^{-MN/2},
\end{equation}
which implies the exponential tightness.

\qed
\end{proofsect}
\section{Appendix}\label{appendix-sec}
\subsection{A Lemma for pair measures}

In this appendix we provide a lemma for a unique characterisation of a pair measure with equal marginals via a vector in the coordinate space. 
Let $ E\in\N $ and $ \nu:=E^2-E $ in the following. The following lemma characterises pair probability measures $ Q\in\Pmf(\{1,\ldots,E\}^2 ) $ which have equal first and second marginal. Recall the definition for the first and second marginal, respectively $ Q^{\ssup{1}}(j)=\sum_{k=1}^EQ(j,k) $ and $ Q^{\ssup{2}}(j)=\sum_{k=1}^EQ(k,j) $ for any $ j\in\{1,\ldots,E\}$.

\begin{lemma}\label{pairmeasurelem}
Let $ x=(x^{\ssup{1}},\ldots,x^{\ssup{\nu}})\in[0,1]^{\nu} $ be a vector with the following properties
\begin{equation}
\begin{aligned}\label{conditionspairmeasure}
(1) &\quad \sum_{k=0}^{E-2} x^{\ssup{kE+1}} \le 1;\\
(2) &\quad \sum_{j=kE+2}^{(k+1)E}x^{\ssup{j}} \le x^{\ssup{kE+1}} \;\mbox{ for }\; k=0,\ldots, E-2;\\
(3) &\quad \sum_{j=0}^{E-2}x^{\ssup{k+2+jE}}\le x^{\ssup{kE+1}} \;\mbox{ for }\; k=0,\ldots, E-2;\\
(4) &\quad 1 \ge \sum_{k=0}^{E-2}\Big(2x^{\ssup{kE+1}}-\sum_{j=0}^{E-2}x^{\ssup{k+2+jE}}\Big).
\end{aligned}
\end{equation}

Then $ x $ defines uniquely a pair probability measure on $ \{1,\ldots,E\}^2 $ with equal first and second marginal.
\end{lemma}

\begin{proofsect}{Proof}
Let $ x\in[0,1]^{\nu} $ with the properties (1)-(4) in \eqref{conditionspairmeasure} be given. We define $ (E-1)^2 $ entries of a function $ Q\colon\{1,\ldots,E\}^2\to[0,1] $ and $ E $ entries of a function $ Q^{\ssup{1}}\colon\{1,\ldots,E\}\to[0,1] $ in the following way:
\begin{equation}\label{Q1}
\begin{array}{cccc}
Q^{\ssup{1}}(1):=x^{\ssup{1}}; & Q(1,1):=x^{\ssup{2}}; & \ldots; & Q(1,E-1):=x^{\ssup{E}}\\
Q^{\ssup{1}}(2):=x^{\ssup{E+1}}; & Q(2,1):=x^{\ssup{E+2}}; & \ldots; & Q(2,E-1):=x^{\ssup{2E}}\\
\cdot & \cdot & \cdot &\cdot\\
\cdot & \cdot & \cdot &\cdot\\
\cdot & \cdot & \cdot &\cdot\\
\cdot & \cdot & \cdot &\cdot\\
Q^{\ssup{1}}(E-1):=x^{\ssup{(E-1)^2}}; & Q(E-1,1):=x^{\ssup{(E-1)^2+1}}; & \ldots ;& Q(E-1,E-1):=x^{\ssup{\nu}}
\end{array};
\end{equation}
respectively the $ E $ entries
\begin{equation}\label{Q2}
Q^{\ssup{1}}(j):=x^{\ssup{(j-1)E+1}}\;\mbox{ for }\; j=1,\ldots, E-1\;\mbox{ and }\; Q^{\ssup{1}}(E):=1-\sum_{j=1}^{E-1}Q^{\ssup{1}}(j).
\end{equation}
The remaining $ (2E-1) $ entries of the function $ Q$ are functions of the given entries in \eqref{Q1} and \eqref{Q2} and are given as
\begin{equation}\label{Q3}
Q(j,E):=Q^{\ssup{1}}(j)-\sum_{k=1}^{E-1}Q(j,k)\;\mbox{ and }\; Q(E,j):=Q^{\ssup{1}}(j)-\sum_{k=1}^{E-1}Q(k,j)\;\mbox{ for }\; j=1,\ldots, E-1;
\end{equation}
and
\begin{equation}\label{Q4}
Q(E,E):=1-\sum_{\heap{(j,k)\in\{1,\ldots,\}^2,}{(x,y)\not= (E,E)}} Q(j,k).
\end{equation}
The entries of the function $ Q $ defined \eqref{Q2} and \eqref{Q3} are elements in $ [0,1] $ because of the properties (1), (2) and (3) of \eqref{conditionspairmeasure}. Property (4) of \eqref{conditionspairmeasure} gives that $ Q(E,E)\in[0,1] $, because of 
$$ 
1-\sum_{\heap{(j,k)\in\{1,\ldots,\}^2,}{(x,y)\not= (E,E)}} Q(j,k)=1-\Big(\sum_{j=1}^{E-1}2Q^{\ssup{1}}(j)-\sum_{j,k=1}^{E-1}Q(k,j)\Big).
$$ 
Thus $ Q(j,k)\in[0,1] $ for all $ j,k\in\{1,\ldots,E\}$ and $ Q^{\ssup{1}}(j)\in[0,1], j\in \{1,\ldots,E\}$ and $ \sum_{j=1}^E Q^{\ssup{1}}(j)=1.$ 

Clearly,
$$
\sum_{j,k\in\{1,\ldots,E\}}Q(j,k)=\sum_{k,j=1}^{E-1}Q(j,k)+\sum_{j=1}^{E-1}\Big(2Q^{\ssup{1}}(j)-\sum_{k=1}^{E-1}\big( Q(j,k)+Q(k,j)\big)\Big)+Q(E,E)=1.
$$
For any $ k=1,\ldots, E-1 $ we have from the second equality in \eqref{Q3}
$$
Q^{\ssup{2}}(k)=Q(E,k)+\sum_{j=1}^{E-1} Q(j,k)=Q^{\ssup{1}}(k).
$$
From \eqref{Q2}, the first line of \eqref{Q3} and  \eqref{Q4} we get 
$$
\begin{aligned}
Q^{\ssup{2}}(E)&=Q(E,E)+\sum_{j=1}^{E-1}Q(j,E)=\\
&=1-\Big(\sum_{j=1}^{E-1} 2Q^{\ssup{1}}(j)-\sum_{j,k=1}^{E-1}Q(k,j)\Big)+\sum_{j=1}^{E-1}\Big(Q^{\ssup{1}}(j)-\sum_{k=1}^{E-1}Q(j,k)\Big)\\
&=1-\sum_{j=1}^{E-1}Q^{\ssup{1}}(j)=Q^{\ssup{1}}(E).
\end{aligned}
$$

Hence, altogether the defined function $ Q\colon\{1,\ldots,E\}^2\to[0,1] $ is a probability measure on $ \{1,\ldots,E\}^2 $ with equal first and second marginal.
\hfill $ \Box $
\end{proofsect}

\subsection{Entropy estimation}
The following lemma gives an entropy estimation. As we are not aware of any reference for this we include the proof of this lemma.

\begin{lemma}[{\bf Entropy estimation}]\label{entropyestimation}
Let $ E $ be a finite set and $ P,Q\in\Pmf(E) $ two sub-probability measures such that 
\begin{equation}\label{assumption1}
\sum_{x\in E}\big|P(x)-Q(x)\big|\le \alpha\le\frac{1}{2}\quad\;\mbox{ for }\; \alpha\in(0,\frac{1}{2}].
\end{equation}
Then the difference in the entropies is bounded as
\begin{equation}
\big| H(P)-H(Q)\big|\le -\alpha\log\frac{\alpha}{|E|}.
\end{equation}
\end{lemma}

\begin{proofsect}{Proof}
We write $ \Delta(x)=|P(x)-Q(x)|, x\in E $. Since the function $ f(x)=-x\log x $ is concave and $ f(0)=f(1)=0 $, we have for every $ 0\le x\le 1-\Delta, 0\le \Delta\le\frac{1}{2} $ the estimation
$$
|f(x)-f(x+\Delta)|\le\max\{f(\Delta),f(1-\Delta)\}=-\Delta\log\Delta.
$$ 

Hence for $ 0\le\alpha\le\frac{1}{2} $ we get
$$
\begin{aligned}
\big| H(P)-H(Q)\big|&\le \sum_{x\in E}\big|f(P(x))-f(Q(x))\big|\\
\le &-\sum_{x\in E}\Delta(x)\log\Delta(x)=\alpha\Big(-\sum_{x\in E}\frac{\Delta(x)}{\alpha}\log \frac{\Delta(x)}{\alpha}-\log\alpha\Big)\\
\le & \alpha\log |E|-\alpha\log\alpha,
\end{aligned}
$$
where we used for the last estimation the usual entropy estimate.
\qed
\end{proofsect}

\noindent






\end{document}